\documentclass[3p,times]{elsarticle}

\usepackage{ecrc}
\usepackage{placeins}
\usepackage{amsmath}
\usepackage{graphicx,subfigure}
\usepackage{epstopdf}
\volume{00}

\firstpage{1}

\journalname{Acta Astronautica}

\runauth{M. Zebenay T. Boge, and D. Chourkoun}

\CopyrightLine{2011}{Published by Elsevier Ltd.}

\usepackage{amssymb}
\usepackage[figuresright]{rotating}

\begin{document}

\begin{frontmatter}


 \author{M. Zebenay\corref{cor1}\fnref{a}}
 \ead{melak.zebenay@dlr.de}
 %\fntext[label2]{test2}
 \cortext[cor1]{This work was presented at the 5th EUCASS - European Conference of AeroSpace Sciences, Munich , July 5-8, 2013}

\dochead{}
%% e.g. \dochead{17th International Conference on Dynamical Processes in Excited States of Solids}

\title{Analytical and experimental stability investigation of a hardware-in-the-loop satellite docking simulator}

%\author[a]{M. Zebenay}
\author[a]{T. Boge}
\author[b]{R. Krenn}
\author[c]{D. Choukroun}
 \address[a]{German Aerospace Center, German Space Operation Center, 82234 Wessling, Germany}
  \address[b]{German Aerospace Center, Institute of System Dynamics and Control, 82234 Wessling, Germany}
 \address[b]{Delft University of Technology, Faculty of Aerospace Engineering, 2629 HS, Delft, The Netherlands}

%\begin{abstract}
%%% Text of abstract
%\end{abstract}
\begin{abstract}
\noindent The European Proximity Operation Simulator (EPOS) of the DLR-German Aerospace Center is a robotics-based simulator that aims at validating and verifying a satellite docking phase. The generic concept features a robotics tracking system working in closed loop with a force/torque feedback signal. Inherent delays in the tracking system combined with typical high stiffness at contact challenge the stability of the closed-loop system. The proposed concept of operations is hybrid: the feedback signal is a superposition of a measured value and of a virtual value that can be tuned in order to guarantee a desired behavior. This paper is concerned with an analytical study of the system's closed-loop stability, and with an experimental validation of the hybrid concept of operations in one dimension (1D). The robotics simulator is modeled as a second-order loop-delay system and closed-form expressions for the critical delay and associated frequency are derived as a function of the satellites' mass and the contact dynamics stiffness and damping parameters. A numerical illustration sheds light on the impact of the parameters on the stability regions. A first-order Pade approximation provides additional means of stability investigation. Experiments were performed and tests results are described for varying values of the mass and the damping coefficients. The empirical determination of instability is based on the coefficient of restitution and on the observed energy. There is a very good agreement between the critical damping values predicted by the analysis and observed during the tests. The contact duration shows also a very good fit between analysis and experiment. In addition, results from a 1D contact experiment carried on an air-floating testbed are successfully emulated using the proposed hybrid docking simulator. This illustrates the flexibility of the hybrid simulator, where various contact dynamics can be emulated without changing any hardware elements.
\end{abstract} 

\end{frontmatter}
\renewcommand{\include}{\input}

\newtheorem{prop}{Proposition}[section]
\newtheorem{defi}{Definition}[section]

\newcommand{\beq}{\begin{equation}}
\newcommand{\eeq}{\end{equation}}
\newcommand{\bdm}{\begin{displaymath}}
\newcommand{\edm}{\end{displaymath}}

\newcommand{\ever}{\ensuremath{\widetilde{\bf \theta}}}
\newcommand{\mever}{\ensuremath{\overline{\ever}}}
\newcommand{\sever}{\ensuremath{\sigma\ever}}

\newcommand{\mlmx}{\ensuremath{\overline{\lambda_{max}}}}
\newcommand{\slmx}{\ensuremath{\sigma_{\lmx}}}

\newcommand{\ger}{\ensuremath{\tilde{\bf{y}}}}

\newcommand{\Zer}{\ensuremath{\widetilde{\bf{Z}}}}

\newcommand{\sa}{\ensuremath{\sin{\alpha}}}
\newcommand{\ca}{\ensuremath{\cos{\alpha}}}
\newcommand{\sib}{\ensuremath{\sin{\beta}}}
\newcommand{\cb}{\ensuremath{\cos{\beta}}}

\newcommand{\Bfr}{\ensuremath{\mathcal{B}}}
\newcommand{\Bfrt}{\ensuremath{ \Bfr^{o} }}
\newcommand{\Befr}{\ensuremath{  \widehat{\Bfr} }}
\newcommand{\Bfrk}{\ensuremath{ \Bfr_{_k} }}
\newcommand{\Bfri}{\ensuremath{ \Bfr_{_i} }}
\newcommand{\Ifr}{\ensuremath{\mathcal{I}}}
\newcommand{\Rfr}{\ensuremath{\mathcal{R}}}
\newcommand{\Sfr}{\ensuremath{\mathcal{S}}}
\newcommand{\Tfr}{\ensuremath{\mathcal{T}}}

\newcommand{\eqdef}{\ensuremath{{\;\stackrel{\triangle}{=}\;}}}

\newcommand{\adjoint}{\ensuremath{ \,\mbox{adj} }}
\newcommand{\arcsec}{\ensuremath{ \,\mbox{arcsec} }}
\newcommand{\arcmn}{\ensuremath{ \,\mbox{arcmn} }}
\newcommand{\arcos}{\ensuremath{ \,\mbox{arccos} }}
\newcommand{\cov}{\ensuremath{ \,\mbox{cov}   }}
\newcommand{\const}{\ensuremath{ \,\mbox{const}   }}
\newcommand{\dghr}{\ensuremath{ \,\mbox{deg/hr} }}
\newcommand{\degree}{\ensuremath{ \,\mbox{deg} }}
\newcommand{\diagonal}{\ensuremath{ \,\mbox{diag} }}
\newcommand{\dt}{\ensuremath{ \,\Delta\!t   }}
\newcommand{\Dt}{\ensuremath{ \Delta\!t   }}
\newcommand{\dts}{\ensuremath{ \,\text{d}\hspace{-.25ex}t }}
\newcommand{\dst}{\ensuremath{ \,\text{d}\hspace{-.25ex}t }}
\newcommand{\du}{\ensuremath{ \,\text{d}\!u }}
\newcommand{\dvs}{\ensuremath{ \,\text{d}\!v }}
\newcommand{\dw}{\ensuremath{ \,\text{d}\!w }}
\newcommand{\dx}{\ensuremath{ \,\text{d}\!x }}
\newcommand{\dxt}{\ensuremath{ \,\text{d}\!\xt }}
\newcommand{\dy}{\ensuremath{ \,\text{d}\!y }}
\newcommand{\dz}{\ensuremath{ \,\text{d}\!z }}
\newcommand{\ds}{\ensuremath{ \,\text{d}\!s }}
\newcommand{\dalfa}{\ensuremath{ \,\text{d}\!\alpha }}
\newcommand{\dbeta}{\ensuremath{ \,\text{d}\!\beta}}
\newcommand{\detat}{\ensuremath{ \,\text{d}\!\etat}}
\newcommand{\dbetapr}{\ensuremath{ \,\text{d}\!\betapr}}
\newcommand{\eqr}{Eq.~\eqref }
\newcommand{\hertz}{\ensuremath{ \,\mbox{Hz} }}
\newcommand{\halfdt}{\ensuremath{ \,\frac{\dt}{2}\,   }}
\newcommand{\half}{\ensuremath{ \,\frac{1}{2}\,   }}
\newcommand{\hour}{\ensuremath{ \,\mbox{hr} }}
\newcommand{\kron}{\ensuremath{ \otimes }}
\newcommand{\mdg}{\ensuremath{ \,\mbox{mdeg} }}
\newcommand{\mdghr}{\ensuremath{ \,\mbox{mdeg/hr} }}
\newcommand{\integer}{\ensuremath{ \mathbb{N} }}
\newcommand{\Order}{\ensuremath{ \,\mathcal{O} }}
\newcommand{\Proof}{ \emph{Proof} }
\newcommand{\quarter}{\ensuremath{ \,\frac{1}{4}\,   }}
\newcommand{\radian}{\ensuremath{ \,\mbox{rad} }}
\newcommand{\rank}{\ensuremath{ \,\mbox{rank} }}
\newcommand{\rdsc}{\ensuremath{ \,\mbox{rad/sec} }}
\newcommand{\remo}{ \emph{Remark 1}: }
\newcommand{\remtw}{ \emph{Remark 2}: }
\newcommand{\remth}{ \emph{Remark 3}: }
\newcommand{\remf}{ \emph{Remark 4}: }
\newcommand{\real}{\ensuremath{ \mathbb{R} }}
\newcommand{\realn}{\ensuremath{ \real^n }}
\newcommand{\realm}{\ensuremath{ \real^m }}
\newcommand{\realplus}{\ensuremath{ \real^{+} }}
\newcommand{\realmn}{\ensuremath{ \real^{mn} }}
\newcommand{\realpq}{\ensuremath{ \real^{pq} }}
\newcommand{\realfour}{\ensuremath{ \real^4 }}
\newcommand{\realthree}{\ensuremath{ \real^3 }}
\newcommand{\realithifo}{\ensuremath{ \real^{\ith \ifo} }}
\newcommand{\realioith}{\ensuremath{ \real^{\io \ith} }}
\newcommand{\realioitw}{\ensuremath{ \real^{\io \itw} }}
\newcommand{\realthreebythree}{\ensuremath{ \real^{3 \times 3} }}
\newcommand{\realmbyn}{\ensuremath{ \real^{m \times n} }}
\newcommand{\realnbym}{\ensuremath{ \real^{n \times m} }}
\newcommand{\realmbym}{\ensuremath{ \real^{m \times m} }}
\newcommand{\realnbyn}{\ensuremath{ \real^{n \times n} }}
\newcommand{\realnubynu}{\ensuremath{ \real^{\nu \times \nu} }}
\newcommand{\realnbyp}{\ensuremath{ \real^{n \times p} }}
\newcommand{\realnx}{\ensuremath{ \real^{n_{x}} }}
\newcommand{\realny}{\ensuremath{ \real^{n_{y}} }}
\newcommand{\realnu}{\ensuremath{ \real^{\nu} }}
\newcommand{\realpbyq}{\ensuremath{ \real^{p \times q} }}
\newcommand{\realpbym}{\ensuremath{ \real^{p \times m} }}
\newcommand{\realnbyq}{\ensuremath{ \real^{n \times q} }}
\newcommand{\realmpbynq}{\ensuremath{ \real^{mp \times nq} }}
\newcommand{\realmnbypq}{\ensuremath{ \real^{mn \times pq} }}
\newcommand{\realpqbymn}{\ensuremath{ \real^{pq \times mn} }}
\newcommand{\realmnbymn}{\ensuremath{ \real^{mn \times mn} }}
\newcommand{\realiobyitw}{\ensuremath{ \real^{\io \times \itw} }}
\newcommand{\realitwbyio}{\ensuremath{ \real^{\itw \times \io} }}
\newcommand{\realitwbyifo}{\ensuremath{ \real^{\itw \times \ifo} }}
\newcommand{\realithbyio}{\ensuremath{ \real^{\ith \times \io} }}
\newcommand{\realithbyifo}{\ensuremath{ \real^{\ith \times \ifo} }}
\newcommand{\realioitwbyithifo}{\ensuremath{ \real^{\io \itw \times \ith \ifo} }}
\newcommand{\second}{\ensuremath{ \,\mbox{sec} }}
\newcommand{\trace}{\ensuremath{ \,\emph{tr} }}
\newcommand{\variance}{\ensuremath{ \,\mbox{var}   }}
\newcommand{\vecop}{\ensuremath{ \,\mbox{vec}   }}
\newcommand{\vctr}{\ensuremath{ \,\mbox{vec}   }}

\newcommand{\inm}{\ensuremath{{Q}}}
\newcommand{\onm}{\ensuremath{{R}}}
\newcommand{\pdm}{\ensuremath{P^d }}
\newcommand{\pttm}{\ensuremath{ P^{\xi} }}
\newcommand{\rttm}{\ensuremath{ \mathcal{Y}^{\xi} }}
\newcommand{\onttm}{\ensuremath{ \onm^{\xi} }}
\newcommand{\kttm}{\ensuremath{ K^{\xi} }}

% VECTOR SYMBOLS --------------------------------

\newcommand{\av}{\ensuremath{ \,{\bf a}  }}

\newcommand{\bv}{\ensuremath{ \,{\bf b}  }}
\newcommand{\bi}{\ensuremath{ \,{\bf b}_{_i}  }}
\newcommand{\bko}{\ensuremath{ \,{\bf b}_{_{k+1}}  }}
\newcommand{\bkoi}{\ensuremath{ \bko^i }}
\newcommand{\bk}{\ensuremath{ \,{\bf b}_{_k}  }}
\newcommand{\bki}{\ensuremath{ \,{\bf b}_{_{k+i}}  }}
\newcommand{\bz}{\ensuremath{ \,{\bf b}_{_0}  }}
\newcommand{\bkt}{\ensuremath{  \mathbf{b}_{k}^{o}  }}
\newcommand{\bkot}{\ensuremath{  \mathbf{b}_{k+1}^{o}  }}
\newcommand{\bt}{\ensuremath{ {\bv^{o}}   }}
\newcommand{\bit}{\ensuremath{\mathbf{b}_{i}^{o}  }}
\newcommand{\bo}{\ensuremath{ {\bf b}_{_1}  }}
\newcommand{\btwo}{\ensuremath{ {\bf b}_{_2}  }}
\newcommand{\bn}{\ensuremath{ {\bf b}_{_n}  }}
\newcommand{\berv}{\ensuremath{\mathbf{\delta b}}}
\newcommand{\bers}{\ensuremath{{\sigma_b}}}
\newcommand{\bq}{\ensuremath{ \bv_{_q} }}
\newcommand{\bqt}{\ensuremath{ \bq^{o} }}
\newcommand{\bkone}{\ensuremath{ \,{\bf b}_{_{k,1}}  }}
\newcommand{\bktwo}{\ensuremath{ \,{\bf b}_{_{k,2}}  }}
\newcommand{\bkN}{\ensuremath{ \,{\bf b}_{_{k,N}}  }}
\newcommand{\btkone}{\ensuremath{ \bkone^{o} }}
\newcommand{\btktwo}{\ensuremath{ \bktwo^{o} }}
\newcommand{\btkN}{\ensuremath{ \bkN^{o} }}
\newcommand{\btk}{\ensuremath{ \bk^{o} }}
\newcommand{\btko}{\ensuremath{ \bko^{o} }}
\newcommand{\bkoone}{\ensuremath{ \,{\bf b}_{_{k+1,1}}  }}
\newcommand{\bkotwo}{\ensuremath{ \,{\bf b}_{_{k+1,2}}  }}
\newcommand{\bkoN}{\ensuremath{ \,{\bf b}_{_{k+1,N}}  }}
\newcommand{\btkoone}{\ensuremath{ \bkoone^{o} }}
\newcommand{\btkotwo}{\ensuremath{ \bkotwo^{o} }}
\newcommand{\btkoN}{\ensuremath{ \bkoN^{o} }}

\newcommand{\bital}{\ensuremath{ \emph{\bv} }}

\newcommand{\dbv}{\ensuremath{{\,\boldsymbol{\delta b}}}}
\newcommand{\dbi}{\ensuremath{ \dbv_{_i}  }}
\newcommand{\dbj}{\ensuremath{ \dbv_{_j}  }}
\newcommand{\dbk}{\ensuremath{ \dbv_{_k}  }}
\newcommand{\dbl}{\ensuremath{ \dbv_{_l}  }}
\newcommand{\dbki}{\ensuremath{ \dbv_{_{k+i}}  }}
\newcommand{\dbko}{\ensuremath{ \dbv_{_{k+1}}  }}
\newcommand{\dbkoi}{\ensuremath{ {\dbko^i} }}
\newcommand{\dbkoj}{\ensuremath{ {\dbko^j} }}

\newcommand{\cv}{\ensuremath{ {\bf c}  }}
\newcommand{\ck}{\ensuremath{ \cv_{_{k}} }}
\newcommand{\cko}{\ensuremath{ \cv_{_{k+1}} }}
\newcommand{\cz}{\ensuremath{ \cv_{_{0}} }}
\newcommand{\cotw}{\ensuremath{ \cv_{_{12}} }}
\newcommand{\ctwo}{\ensuremath{ \cv_{_{21}} }}

\newcommand{\dv}{\ensuremath{{\bf d}}}
\newcommand{\dtr}{\ensuremath{{ {\dv^{o}} }}}
\newcommand{\dko}{\ensuremath{ \dv_{_{k+1}} }}
\newcommand{\dk}{\ensuremath{ \dv_{_{k}} }}
\newcommand{\dkone}{\ensuremath{ \dv_{_{k,1}} }}
\newcommand{\dktwo}{\ensuremath{ \dv_{_{k,2}} }}
\newcommand{\dkthr}{\ensuremath{ \dv_{_{k,3}} }}
\newcommand{\dkj}{\ensuremath{ \dv_{_{k,j}} }}

\newcommand{\dumykk}{\ensuremath{  {\widehat{\left( \cdot \right)}}_{_{k/k}}    }}
\newcommand{\dumykok}{\ensuremath{  {\widehat{\left( \cdot \right)}}_{_{k+1/k}}    }}
\newcommand{\dumyerkk}{\ensuremath{  {\widetilde{\left( \cdot \right)}}_{_{k/k}}    }}
\newcommand{\dumyerkok}{\ensuremath{  {\widetilde{\left( \cdot \right)}}_{_{k+1/k}}    }}

\newcommand{\dev}{\ensuremath{{\,\boldsymbol{\delta e}}}}
\newcommand{\dqv}{\ensuremath{{\,\boldsymbol{\delta q}}}}
\newcommand{\dqvs}{\ensuremath{\delta q}}
\newcommand{\dzv}{\ensuremath{ {\boldsymbol{\delta}} \zv }}
\newcommand{\dzko}{\ensuremath{ \dzv_{_{k+1}}  }}

\newcommand{\dfit}{\ensuremath{ \phi_{_{t}} }}
\newcommand{\dfix}{\ensuremath{ \phi_{_{\xv}} }}
\newcommand{\dfiu}{\ensuremath{ \phi_{_{u}} }}

\newcommand{\Dqv}{\ensuremath{\boldsymbol{\Delta q} }}
\newcommand{\Dqkk}{\ensuremath{\Dqv_{_{k/k}} }}
\newcommand{\Dqkok}{\ensuremath{ \Dqv_{_{k+1/k}} }}
\newcommand{\Dqz}{\ensuremath{ \Dqv^0 }}
\newcommand{\Dqzkok}{\ensuremath{ \Dqz_{_{k+1/k}} }}
\newcommand{\Dqi}{\ensuremath{ \Dqv^i }}
\newcommand{\Dqikok}{\ensuremath{ \Dqi_{_{k+1/k}} }}

\newcommand{\ev}{\ensuremath{ \,{\bf e}  }}
\newcommand{\etz}{\ensuremath{ \ev^0 }}
\newcommand{\ekt}{\ensuremath{ {\ek^{o}} }}
\newcommand{\eko}{\ensuremath{ \ev_{_{k+1}} }}
\newcommand{\ee}{\ensuremath{ \widehat{\ev}  }}
\newcommand{\eek}{\ensuremath{ \ee_{_{k}} }}
\newcommand{\eeko}{\ensuremath{ \ee_{_{k+1}} }}
\newcommand{\ekok}{\ensuremath{ \ee_{_{k+1/k}}  }}
\newcommand{\eekok}{\ensuremath{ \ee_{_{k+1/k}}  }}
\newcommand{\ekoko}{\ensuremath{ \ee_{_{k+1/k+1}}  }}
\newcommand{\eekoko}{\ensuremath{ \ee_{_{k+1/k+1}}  }}
\newcommand{\eekolk}{\ensuremath{ \ee_{_{k+l-1/k}}  }}
\newcommand{\eeklk}{\ensuremath{ \ee_{_{k+l/k}}  }}

\newcommand{\eone}{\ensuremath{ \ev_{_1} }}
\newcommand{\etwo}{\ensuremath{ \ev_{_2} }}
\newcommand{\ethr}{\ensuremath{ \ev_{_3} }}
\newcommand{\ethree}{\ensuremath{ \ev_{_3} }}
\newcommand{\efor}{\ensuremath{ \ev_{_4} }}
\newcommand{\evi}{\ensuremath{ \ev_{_i} }}
\newcommand{\ek}{\ensuremath{ \ev_{_{k}} }}
\newcommand{\el}{\ensuremath{ \ev_{_{l}} }}
\newcommand{\ei}{\ensuremath{ \ev_{_{i}} }}
\newcommand{\ej}{\ensuremath{ \ev_{_{j}} }}
\newcommand{\eny}{\ensuremath{ \ev_{_{n_y}} }}
\newcommand{\enu}{\ensuremath{ \ev_{_{\nu}} }}

\newcommand{\etav}{\ensuremath{ \boldsymbol{\eta} }}
\newcommand{\etavk}{\ensuremath{ \etav_{_{k}} }}

\newcommand{\eulv}{\ensuremath{ \,\boldsymbol{\theta} }}

\newcommand{\epsv}{\ensuremath{ \,\boldsymbol{\epsilon} }}
\newcommand{\epsk}{\ensuremath{ \epsv_{_k} }}
\newcommand{\epsl}{\ensuremath{ \epsv_{_l} }}
\newcommand{\epski}{\ensuremath{ \epsv_{_{k+i}} }}
\newcommand{\epsi}{\ensuremath{  \epsv_{_i} }}
\newcommand{\epsko}{\ensuremath{ \epsv_{_{k+1}} }}
\newcommand{\epsok}{\ensuremath{ \epsv_{_{k-1}} }}
\newcommand{\epsktt}{\ensuremath{  \epsv_{k,\,} }}
\newcommand{\epsbar}{\ensuremath{ \bar{\epsv}  }}
\newcommand{\epsbark}{\ensuremath{ \epsbar_{_k}  }}
\newcommand{\epskbar}{\ensuremath{ \epsbar_{_k}  }}
\newcommand{\epse}{\ensuremath{ \widehat{\epsv} }}
\newcommand{\epseokk}{\ensuremath{ \epse_{_{k-1/k}} }}
\newcommand{\epsekk}{\ensuremath{ \epse_{_{k/k}} }}

\newcommand{\epsxtk}{\ensuremath{ \epsk^{16} }}

\newcommand{\fv}{\ensuremath{ {\bf f}  }}
\newcommand{\fk}{\ensuremath{ \fv_{_{k}} }}

\newcommand{\fe}{\ensuremath{  \widehat{\fv} }}
\newcommand{\fkk}{\ensuremath{ \fe_{_{k/k}} }}

\newcommand{\ftld}{\ensuremath{  \widetilde{\fv} }}
\newcommand{\ftldkk}{\ensuremath{  \ftld_{_{k/k}} }}

\newcommand{\fik}{\ensuremath{{ {\bf \varphi}_{_k}}}}

\newcommand{\gv}{\ensuremath{ \,\boldsymbol{g} }}
\newcommand{\gk}{\ensuremath{ \gv_{_k} }}
\newcommand{\gko}{\ensuremath{ \gv_{_{k+1}} }}

\newcommand{\gev}{\ensuremath{  \widehat{\gv} }}
\newcommand{\gkk}{\ensuremath{ \gev_{_{k/k}} }}
\newcommand{\gkko}{\ensuremath{ \gev_{_{k/k+1}} }}
\newcommand{\gkoko}{\ensuremath{ \gev_{_{k+1/k+1}} }}

\newcommand{\gamav}{\ensuremath{ \boldsymbol{\gamma} }}

\newcommand{\hv}{\ensuremath{ \,{\bf h}  }}
\newcommand{\hko}{\ensuremath{ \hv_{_{k+1}} }}

\newcommand{\inv}{\ensuremath{{{\bf w}}}}

\newcommand{\kv}{\ensuremath{ {\bf k}  }}
\newcommand{\kvko}{\ensuremath{ \kv_{_{k+1}} }}
\newcommand{\kvj}{\ensuremath{ \kv^{j} }}
\newcommand{\kvjko}{\ensuremath{ \kvj_{_{k+1}} }}

\newcommand{\minv}{\ensuremath{\overline{\inv}}}
\newcommand{\monv}{\ensuremath{\overline{\onv}}}
\newcommand{\mberv}{\ensuremath{\mathbf{\mu_b}}}
\newcommand{\mepsk}{\ensuremath{ \mv_{_{\eps_{k}}}    }}

\newcommand{\mv}{\ensuremath{ \,{\bf m}  }}

\newcommand{\nv}{\ensuremath{ \,{\bf n}  }}
\newcommand{\no}{\ensuremath{  \nv_{_1} }}
\newcommand{\nt}{\ensuremath{   \nv_{_2}  }}
\newcommand{\nepsk}{\ensuremath{ \nv_{_{\eps_{k}}}    }}
\newcommand{\nwk}{\ensuremath{ \nv_{_{w_{k}}}    }}

\newcommand{\nuv}{\ensuremath{ \,{\boldsymbol \nu}  }}
\newcommand{\nuko}{\ensuremath{ \nuv_{_{k+1}}  }}
\newcommand{\nukl}{\ensuremath{ \nuv_{_{k+l}}  }}
\newcommand{\nukok}{\ensuremath{ \nuv_{_{k+1}}  }}
\newcommand{\nuklk}{\ensuremath{ \nuv_{_{k+l}}  }}
\newcommand{\nuknk}{\ensuremath{ \nuv_{_{k+n}}  }}
\newcommand{\nub}{\ensuremath{  {\bar{\nuv}} }}
\newcommand{\nubN}{\ensuremath{ {\nub^{^N}}  }}
\newcommand{\nubl}{\ensuremath{ {\nub^{^l}} }}
\newcommand{\nubol}{\ensuremath{ {\nub^{^{l-1}}}  }}
\newcommand{\nubz}{\ensuremath{ {\nub^{^{0}}}  }}
\newcommand{\nusq}{\ensuremath{  {\nuv\nuv^{_T}} }}
\newcommand{\nusqb}{\ensuremath{  \bar{\nusq} }}
\newcommand{\nusqbN}{\ensuremath{ \nusqb^{^N} }}
\newcommand{\nui}{\ensuremath{ \nuv_{_{i}}  }}

\newcommand{\Ov}{\ensuremath{ \,{\bf{0}}  }}
\newcommand{\Onev}{\ensuremath{ {\bf{1}}  }}

\newcommand{\omg}{\ensuremath{ \,\boldsymbol{\omega} }}
\newcommand{\omgv}{\ensuremath{ \,\boldsymbol{\omega} }}
\newcommand{\omk}{\ensuremath{  \omgv_{_k}  }}
\newcommand{\omkt}{\ensuremath{ {\omgv_{_k}^{o}}  }}
\newcommand{\omok}{\ensuremath{ \omgv^0_{_k} }}
\newcommand{\ome}{\ensuremath{  \widehat{\omgv} }}
\newcommand{\omekk}{\ensuremath{  \ome_{_{k/k}}  }}
\newcommand{\omgt}{\ensuremath{ {\omgv^{o}}  }}

\newcommand{\onv}{\ensuremath{{{\bf v}}}}

\newcommand{\pv}{\ensuremath{{\bf p}}}
\newcommand{\pz}{\ensuremath{ \pv_{_{0}} }}
\newcommand{\pk}{\ensuremath{ \pv_{_{k}} }}
\newcommand{\pko}{\ensuremath{ \pv_{_{k+1}} }}
\newcommand{\pok}{\ensuremath{ \pv_{_{k-1}} }}

\newcommand{\qt}{\ensuremath{{\bf q}}}
\newcommand{\qv}{\ensuremath{ {\bf q} }}
\newcommand{\qunit}{\ensuremath{ {\bf 1}_{_q}  }}
\newcommand{\qinv}{\ensuremath{ \qv^{^{-1}} }}
\newcommand{\qtz}{\ensuremath{ {\bf q}^0  }}
\newcommand{\qto}{\ensuremath{ {\bf q}^1  }}
\newcommand{\qtw}{\ensuremath{ {\bf q}^2  }}
\newcommand{\qth}{\ensuremath{ {\bf q}^3  }}
\newcommand{\qone}{\ensuremath{ \qv_{_1}  }}
\newcommand{\qtwo}{\ensuremath{ \qv_{_2}  }}
\newcommand{\qthr}{\ensuremath{ \qv_{_3}  }}
\newcommand{\qk}{\ensuremath{  {\bf q}_{_{k}} }}
\newcommand{\qak}{\ensuremath{{  {\bf q}_{_k}}}}
\newcommand{\qko}{\ensuremath{ \qv_{_{k+1}} }}
\newcommand{\qok}{\ensuremath{ \qv_{_{k-1}} }}
\newcommand{\qz}{\ensuremath{{  {\bf q}_{_0}}}}
\newcommand{\qo}{\ensuremath{ \qv_{_{1}} }}
\newcommand{\qN}{\ensuremath{ \qv_{_N} }}
\newcommand{\qoN}{\ensuremath{ \qv_{_{N-1}} }}
\newcommand{\qNo}{\ensuremath{ \qv_{_{N+1}} }}
\newcommand{\qNast}{\ensuremath{ \qN^{\ast} }}

\newcommand{\qiko}{\ensuremath{ {\qko^i} }}
\newcommand{\qzko}{\ensuremath{ {\qko^0} }}
\newcommand{\qoko}{\ensuremath{ {\qko^1} }}
\newcommand{\qtwko}{\ensuremath{ {\qko^2} }}
\newcommand{\qthko}{\ensuremath{ {\qko^3} }}

\newcommand{\qkt}{\ensuremath{ \qk^{o} }}
\newcommand{\qkot}{\ensuremath{ \qko^{o} }}
\newcommand{\qzt}{\ensuremath{ \qv_0^{o} }}

\newcommand{\qe}{\ensuremath{ \,\widehat{\bf q}}}
\newcommand{\qev}{\ensuremath{ \,\widehat{\bf q}}}
\newcommand{\qek}{\ensuremath{ \qe_{_k} }}
\newcommand{\qeko}{\ensuremath{ \qe_{_{k+1}} }}
\newcommand{\qekk}{\ensuremath{   \qe_{_{k/k}} }}
\newcommand{\qeNN}{\ensuremath{   \qe_{_{N/N}} }}
\newcommand{\qeNoN}{\ensuremath{   \qe_{_{N+1/N}} }}
\newcommand{\qeKok}{\ensuremath{  \qe_{_{k/k-1}}  }}
\newcommand{\qeNon}{\ensuremath{  \qe_{_{N/N-1}}  }}
\newcommand{\qeoNoN}{\ensuremath{  \qe_{_{N-1/N-1}}  }}
\newcommand{\qeNoNo}{\ensuremath{   \qe_{_{N+1/N+1}} }}
\newcommand{\qekoko}{\ensuremath{ \qe_{_{k+1/k+1}} }}
\newcommand{\qeokok}{\ensuremath{ \qe_{_{k-1/k-1}} }}
\newcommand{\qekokoast}{\ensuremath{ \qekoko^{\ast} }}
\newcommand{\qekkast}{\ensuremath{ \qekk^{\ast} }}
\newcommand{\qekok}{\ensuremath{   \qe_{_{k+1/k}} }}
\newcommand{\qekonk}{\ensuremath{   \qe_{_{k/{k-1}}} }}
\newcommand{\qeklk}{\ensuremath{   \qe_{_{k+l/k}} }}
\newcommand{\qekolk}{\ensuremath{   \qe_{_{k+l-1/k}} }}
\newcommand{\qezz}{\ensuremath{    \qe_{_{0/0}} }}
\newcommand{\qeoz}{\ensuremath{    \qe_{_{1/0}} }}
\newcommand{\qeoo}{\ensuremath{     \qe_{_{1/1}} }}
\newcommand{\qzstar}{\ensuremath{{ {\star{\bf q}}_{_0} }}}
\newcommand{\qostar}{\ensuremath{{ {\star{\bf q}}_{_1} }}}
\newcommand{\qater}{\ensuremath{\tilde{\bf q}}}
\newcommand{\qaterkoko}{\ensuremath{ \qater_{_{k+1/k+1}}    }}
\newcommand{\qeki}{\ensuremath{{ \,\widehat{\bf q}_{_{k/i} }}}}
\newcommand{\qezzast}{\ensuremath{    {\qezz^{\ast}} }}

\newcommand{\qezkok}{\ensuremath{ {\qekok^0} }}
\newcommand{\qeonekok}{\ensuremath{ {\qekok^1} }}
\newcommand{\qetwkok}{\ensuremath{ {\qekok^2} }}
\newcommand{\qethkok}{\ensuremath{ {\qekok^3} }}
\newcommand{\qeikok}{\ensuremath{ {\qekok^i} }}

\newcommand{\qb}{\ensuremath{ \overline{\bf q}  }}
\newcommand{\qbar}{\ensuremath{ \overline{\bf q}  }}
\newcommand{\qbarz}{\ensuremath{  \qbar_{_0} }}
\newcommand{\qbaro}{\ensuremath{  \qbar_{_1} }}
\newcommand{\qbarN}{\ensuremath{  \qbar_{_N} }}
\newcommand{\qbaroN}{\ensuremath{  \qbar_{_{N-1}} }}
\newcommand{\qbark}{\ensuremath{  \qbar_{_k} }}
\newcommand{\qbarok}{\ensuremath{  \qbar_{_{k-1}} }}
\newcommand{\qbari}{\ensuremath{  \qbar_{_i} }}
\newcommand{\qbarKok}{\ensuremath{  \qbar_{_{k/k-1}}   }}
\newcommand{\qbarbar}{\ensuremath{      \bar{\qbar}  }}
\newcommand{\qbarbarKok}{\ensuremath{  \qbarbar_{_{k/k-1}} }}

\newcommand{\qkbar}{\ensuremath{{  \bar{\bf q}_{_{k}} }}}
\newcommand{\qzbar}{\ensuremath{{  \bar{\bf q}_{_0} }}}
\newcommand{\qobar}{\ensuremath{{  \bar{\bf q}_{_1} }}}
\newcommand{\qkobar}{\ensuremath{{  \bar{\bf q}_{_{k+1}} }}}

\newcommand{\qkk}{\ensuremath{ {\bf q}_{_{k/k}}  }}
\newcommand{\qerkk}{\ensuremath{ {\boldsymbol{\delta}} \qkk }}
\newcommand{\qkok}{\ensuremath{ {\bf q}_{_{k+1/k}}  }}
\newcommand{\qerkok}{\ensuremath{ {\boldsymbol{\delta}} \qkok }}
\newcommand{\qkoko}{\ensuremath{ {\bf q}_{_{k+1/k+1}}  }}
\newcommand{\qerkoko}{\ensuremath{ {\boldsymbol{\delta}} \qkoko }}
\newcommand{\qerkokoast}{\ensuremath{ \qerkoko^{\ast} }}
\newcommand{\qerkkast}{\ensuremath{ \qerkk^{\ast} }}
\newcommand{\qierkok}{\ensuremath{ {\qerkok^i} }}
\newcommand{\qzerkok}{\ensuremath{ {\qerkok^0} }}

\newcommand{\qtl}{\ensuremath{ \widetilde{\qv} }}
\newcommand{\qtlkk}{\ensuremath{ \qtl_{k/k} }}

\newcommand{\Qvo}{\ensuremath{  {\bf{e}}_{_1}  }}
\newcommand{\Qvtw}{\ensuremath{  {\bf{e}}_{_2}  }}
\newcommand{\Qvth}{\ensuremath{  {\bf{e}}_{_3}  }}
\newcommand{\Qv}{\ensuremath{  {\bf{e}}  }}
\newcommand{\Quatv}{\ensuremath{{\bf e}}}

\newcommand{\rv}{\ensuremath{ \,{\bf r}  }}
\newcommand{\ri}{\ensuremath{ \,{\bf r}_{_i}  }}
\newcommand{\rj}{\ensuremath{ \,{\bf r}_{_j}  }}
\newcommand{\rko}{\ensuremath{ \rv_{_{k+1}} }}
\newcommand{\rkoi}{\ensuremath{ {\rv_{_{k+1}}^i} }}
\newcommand{\rz}{\ensuremath{ \,{\bf r}_{_0}  }}
\newcommand{\ro}{\ensuremath{ \,{\bf r}_{_1}  }}
\newcommand{\rt}{\ensuremath{ \,{\bf r}_{_2}  }}
\newcommand{\rone}{\ensuremath{ \,{\bf r}_{_1}  }}
\newcommand{\rtwo}{\ensuremath{ \,{\bf r}_{_2}  }}
\newcommand{\rn}{\ensuremath{ \,{\bf r}_{_n}  }}
\newcommand{\rttv}{\ensuremath{ \,\bf{\xi}_{r} }}
\newcommand{\res}{\ensuremath{ \,\widehat{\bf e}}}
\newcommand{\resid}{\ensuremath{ \,{\bf r}}}
\newcommand{\rk}{\ensuremath{ {\bf r}_{_k}  }}
\newcommand{\rvq}{\ensuremath{ \rv_{_q} }}
\newcommand{\rvo}{\ensuremath{ \rv_{_1} }}
\newcommand{\rvtw}{\ensuremath{ \rv_{_2} }}
\newcommand{\rvth}{\ensuremath{ \rv_{_3} }}
\newcommand{\rqv}{\ensuremath{ \rv^q }}
\newcommand{\rqko}{\ensuremath{ \rqv_{_{k+1}} }}

\newcommand{\sv}{\ensuremath{{\bf s}}}
\newcommand{\st}{\ensuremath{ {\sv^{o}} }}
\newcommand{\sko}{\ensuremath{ \sv_{_{k+1}} }}

\newcommand{\tetav}{\ensuremath{ \,\boldsymbol{\theta} }}
\newcommand{\tetavko}{\ensuremath{ \,\boldsymbol{\theta}_{_{k+1}} }}
\newcommand{\tetaev}{\ensuremath{ \,\widehat{\tetav} }}
\newcommand{\tetaeWLS}{\ensuremath{ \tetaev^{_{_{WLS}}} }}
\newcommand{\tetaeWLSk}{\ensuremath{ \tetaeWLS_{_k} }}
\newcommand{\tetaeWLSko}{\ensuremath{ \tetaeWLS_{_{k+1}} }}

\newcommand{\uv}{\ensuremath{\mathbf{u}}}
\newcommand{\ukok}{\ensuremath{  \uv_{_{k+1/k}} }}
\newcommand{\uk}{\ensuremath{ \uv_{_{k}} }}
\newcommand{\uko}{\ensuremath{ \uv_{_{k+1}} }}
\newcommand{\ut}{\ensuremath{ \uv_{_{t}} }}
\newcommand{\ui}{\ensuremath{ \uv_{_{i}} }}
\newcommand{\ukast}{\ensuremath{ \uk^{\ast} }}
\newcommand{\uoN}{\ensuremath{ \uv_{_{N-1}} }}
\newcommand{\uoNast}{\ensuremath{ \uoN^{\ast} }}
\newcommand{\utwN}{\ensuremath{ \uv_{_{N-2}} }}
\newcommand{\ukoast}{\ensuremath{ \uko^{\ast} }}

\newcommand{\vv}{\ensuremath{ \mathbf{v} }}
\newcommand{\vk}{\ensuremath{  \vv_{_k} }}
\newcommand{\vko}{\ensuremath{  \vv_{_{k+1}} }}
\newcommand{\vkon}{\ensuremath{  \vko^n }}
\newcommand{\vkoi}{\ensuremath{  \vko^i }}
\newcommand{\vb}{\ensuremath{  \overline{\bf v} }}
\newcommand{\vbk}{\ensuremath{  \vb_{_k} }}
\newcommand{\vvc}{\ensuremath{  \vv^{c} }}
\newcommand{\vci}{\ensuremath{  \vvc_{_i} }}
\newcommand{\vi}{\ensuremath{  \vv_{_i} }}
\newcommand{\vj}{\ensuremath{  \vv_{_j} }}
\newcommand{\vkone}{\ensuremath{  \vv_{_{k,1}} }}
\newcommand{\vktwo}{\ensuremath{  \vv_{_{k,2}} }}
\newcommand{\vkN}{\ensuremath{  \vv_{_{k,N}} }}

\newcommand{\wv}{\ensuremath{  {\bf w} }}
\newcommand{\wk}{\ensuremath{  {\bf w}_{_k} }}
\newcommand{\wok}{\ensuremath{  \wv_{_{k-1}} }}
\newcommand{\wN}{\ensuremath{  {\bf w}_{_N} }}
\newcommand{\woN}{\ensuremath{ \wv_{_{N-1}}  }}
\newcommand{\wko}{\ensuremath{     \wv_{_{k+1}} }}
\newcommand{\wtwN}{\ensuremath{  \wv_{_{N-2}} }}
\newcommand{\wz}{\ensuremath{  {\bf w}_{_0}   }}
\newcommand{\wo}{\ensuremath{  {\bf w}_{_1}  }}
\newcommand{\wsxtk}{\ensuremath{  \wk^{16} }}
\newcommand{\wtenk}{\ensuremath{  \wk^{10} }}
\newcommand{\wnink}{\ensuremath{  \wk^{9} }}

\newcommand{\wbar}{\ensuremath{ \,\bar{\wv} }}
\newcommand{\wkbar}{\ensuremath{ \wbar_{_k} }}
\newcommand{\wbark}{\ensuremath{ \wbar_{_k} }}
\newcommand{\wbarok}{\ensuremath{ \wbar_{_{k-1}} }}
\newcommand{\wbarN}{\ensuremath{   \wbar_{_{N}} }}
\newcommand{\wbaroN}{\ensuremath{   \wbar_{_{N-1}} }}
\newcommand{\wkobar}{\ensuremath{     \wbar_{_{k+1}} }}
\newcommand{\wzbar}{\ensuremath{   \wbar_{_0} }}
\newcommand{\wbarz}{\ensuremath{   \wbar_{_0} }}
\newcommand{\wbartwk}{\ensuremath{  \wbar_{_{k-2}} }}
\newcommand{\wbartwN}{\ensuremath{  \wbar_{_{N-2}} }}
\newcommand{\wbarKok}{\ensuremath{  \wbar_{_{k/k-1}} }}
\newcommand{\wbarbar}{\ensuremath{  \bar{\wbar}  }}
\newcommand{\wbarbarokok}{\ensuremath{  \wbarbar_{_{k-1/k-1}}  }}

\newcommand{\wzstar}{\ensuremath{{  \star{\bf w}_{_0}}}}
\newcommand{\wostar}{\ensuremath{{  \star{\bf w}_{_1}}}}
\newcommand{\wstar}{\ensuremath{  \, {\wv}^{\star} }}
\newcommand{\wstarz}{\ensuremath{ \wstar_{_0}  }}
\newcommand{\wstaro}{\ensuremath{ \wstar_{_1}  }}
\newcommand{\wstarok}{\ensuremath{ \wstar_{_{k-1}}  }}
\newcommand{\wstarokqk}{\ensuremath{
 \wstarok \left( \qk \right) }}
\newcommand{\wstartwk}{\ensuremath{ \wstar_{_{k-2}}  }}
\newcommand{\wstarstar}{\ensuremath{ \wstar^{\star}  }}
\newcommand{\wstarstartwk}{\ensuremath{ \wstarstar_{_{k-2}}  }}

\newcommand{\wezz}{\ensuremath{{ \widehat{\bf w}_{_{0/0} }}}}
\newcommand{\weoz}{\ensuremath{{ \widehat{\bf w}_{_{1/0} }}}}
\newcommand{\weoo}{\ensuremath{{ \widehat{\bf w}_{_{1/1} }}}}
\newcommand{\wekk}{\ensuremath{{ \,\widehat{\bf w}_{_{k/k} }}}}
\newcommand{\wekok}{\ensuremath{{ \,\widehat{\bf w}_{_{k+1/k} }}}}
\newcommand{\weki}{\ensuremath{{ \,\widehat{\bf w}_{_{k/i} }}}}
\newcommand{\wekoK}{\ensuremath{{ \,\widehat{\bf w}_{_{k+1/k} }}}}
\newcommand{\weKok}{\ensuremath{{ \,\widehat{\bf w}_{_{k/k-1} }}}}
\newcommand{\weokk}{\ensuremath{{ \,\widehat{\bf w}_{_{k-1/k} }}}}

\newcommand{\wtv}{\ensuremath{{\bf \omega}^o}}
\newcommand{\wtm}{\ensuremath{{ \Omega}^o}}
\newcommand{\wm}{\ensuremath{{\Omega}}}
\newcommand{\werm}{\ensuremath{{\mathcal{E}}}}
\newcommand{\wers}{\ensuremath{{\sigma_{\varepsilon}}}}

\newcommand{\xettv}{\ensuremath{ \bf{\xi}_{\widehat{x}} }}
\newcommand{\xv}{\ensuremath{ {\bf x}  }}
\newcommand{\xedv}{\ensuremath{\widehat{\xv}^d }}
\newcommand{\xev}{\ensuremath{{\widehat{\xv}}}}
\newcommand{\xt}{\ensuremath{ \xv_{_t}  }}
\newcommand{\xk}{\ensuremath{ \xv_{_k}  }}
\newcommand{\xl}{\ensuremath{ \xv_{_l}  }}
\newcommand{\xlo}{\ensuremath{ \xv_{_{l+1}}  }}
\newcommand{\xN}{\ensuremath{ \xv_{_N}  }}
\newcommand{\xoN}{\ensuremath{ \xv_{_{N-1}}  }}
\newcommand{\xko}{\ensuremath{ \xv_{_{k+1}}  }}
\newcommand{\xvi}{\ensuremath{ \xv_{_i}  }}
\newcommand{\xz}{\ensuremath{ \xv_{_0}  }}
\newcommand{\xo}{\ensuremath{ \xv_{_1}  }}
\newcommand{\xtw}{\ensuremath{ \xv_{_2}  }}
\newcommand{\xsxtk}{\ensuremath{ \xk^{16} }}
\newcommand{\xsxtko}{\ensuremath{ \xko^{16} }}
\newcommand{\xtenk}{\ensuremath{ \xk^{10} }}
\newcommand{\xtenko}{\ensuremath{ \xko^{10} }}
\newcommand{\xninko}{\ensuremath{ \xko^{9} }}
\newcommand{\xnink}{\ensuremath{ \xk^{9} }}
\newcommand{\xekk}{\ensuremath{  \xev_{_{k/k}} }}
\newcommand{\xekok}{\ensuremath{  \xev_{_{k+1/k}} }}
\newcommand{\xezz}{\ensuremath{  \xev_{_{0/0}} }}
\newcommand{\xekoko}{\ensuremath{  \xev_{_{k+1/k+1}} }}
\newcommand{\xzv}{\ensuremath{ \xv^0 }}
\newcommand{\xkk}{\ensuremath{  \xev_{_{k/k}} }}
\newcommand{\xkok}{\ensuremath{  \xev_{_{k+1/k}} }}
\newcommand{\xkoko}{\ensuremath{  \xev_{_{k+1/k+1}} }}
\newcommand{\xtz}{\ensuremath{ \xv(\tz)  }}
\newcommand{\xtzs}{\ensuremath{ \xv(\tz+s)  }}
\newcommand{\xio}{\ensuremath{ \xv_{_{i+1}} }}

\newcommand{\xtld}{\ensuremath{  \widetilde{\xv} }}
\newcommand{\xtldkok}{\ensuremath{  \xtld_{_{k+1/k}} }}

\newcommand{\yv}{\ensuremath{{\bf y}}}
\newcommand{\yzv}{\ensuremath{  \yv^0 }}
\newcommand{\yttv}{\ensuremath{ \bf{\xi}_y }}
\newcommand{\Yv}{\ensuremath{ {\bf y}  }}
\newcommand{\ydv}{\ensuremath{\yv^d }}
\newcommand{\yk}{\ensuremath{ \yv_{_k} }}
\newcommand{\yko}{\ensuremath{ \yv_{_{k+1}} }}
\newcommand{\yz}{\ensuremath{ \yv_{_0} }}
\newcommand{\ykl}{\ensuremath{ \yv_{_{k+l}} }}

\newcommand{\ytz}{\ensuremath{ y(\tz)  }}
\newcommand{\ytzD}{\ensuremath{ y(\tz-\Delta)  }}
\newcommand{\ytzmD}{\ensuremath{ y(\tz-\Delta)  }}
\newcommand{\ytzsmD}{\ensuremath{ y(\tz+s-\Delta)  }}
\newcommand{\ysmD}{\ensuremath{ y(s-\Delta)  }}

\newcommand{\ye}{\ensuremath{  \widehat{\yv} }}
\newcommand{\ykk}{\ensuremath{ \ye_{_{k/k}} }}
\newcommand{\yokk}{\ensuremath{ \ye_{_{k-1/k}} }}
\newcommand{\ykko}{\ensuremath{ \ye_{_{k/k+1}} }}
\newcommand{\ykoko}{\ensuremath{ \ye_{_{k+1/k+1}} }}
\newcommand{\yzz}{\ensuremath{ \ye_{_{0/0}} }}

\newcommand{\ytld}{\ensuremath{  \widetilde{\yv} }}
\newcommand{\ytldko}{\ensuremath{  \ytld_{_{k+1}} }}
\newcommand{\ytldkk}{\ensuremath{  \ytld_{_{k/k}} }}

\newcommand{\Zv}{\ensuremath{ {\bf z}  }}
\newcommand{\Zeps}{\ensuremath{ {\bf z}_{_\epsilon}  }}
\newcommand{\zeps}{\ensuremath{ {\bf z}_{_\epsilon}  }}
\newcommand{\Zb}{\ensuremath{ {\bf z}_{_b}  }}
\newcommand{\zv}{\ensuremath{ {\bf z}  }}
\newcommand{\zb}{\ensuremath{ {\bf z}_{_b}  }}
\newcommand{\zk}{\ensuremath{ \zv_{_{k}}  }}
\newcommand{\zko}{\ensuremath{ \zv_{_{k+1}}  }}
\newcommand{\zkot}{\ensuremath{ \zv_{_{k+1}}^{o}  }}
\newcommand{\zer}{\ensuremath{ \widetilde{\zv} }}
\newcommand{\zerk}{\ensuremath{ \zer_{_k} }}
\newcommand{\zz}{\ensuremath{ \zv_{_{0}}  }}
\newcommand{\zo}{\ensuremath{ \zv_{_{1}}  }}
\newcommand{\zN}{\ensuremath{ \zv_{_{N}}  }}
\newcommand{\zbi}{\ensuremath{ {\zb^i} }}
\newcommand{\zbn}{\ensuremath{ {\zb^n} }}
\newcommand{\zev}{\ensuremath{{\widehat{\zv}}}}

% SCALARS SYMBOLS ---------------------------------------------

\newcommand{\ai}{\ensuremath{ \,a_{_i}  }}
\newcommand{\aj}{\ensuremath{ \,a_{_j}  }}
\newcommand{\ako}{\ensuremath{ \,a_{_{k+1}}  }}
\newcommand{\ak}{\ensuremath{ \,a_{_k}  }}
\newcommand{\az}{\ensuremath{ \,a_{_0}  }}
\newcommand{\ao}{\ensuremath{ \,a_{_1}  }}
\newcommand{\aone}{\ensuremath{ \,a_{_1}  }}
\newcommand{\atw}{\ensuremath{ \,a_{_2}  }}
\newcommand{\aij}{\ensuremath{ \,a_{_{ij}}  }}
\newcommand{\aoo}{\ensuremath{ \,a_{_{11}}  }}
\newcommand{\aon}{\ensuremath{ \,a_{_{1n}}  }}
\newcommand{\amo}{\ensuremath{ \,a_{_{m1}}  }}
\newcommand{\amm}{\ensuremath{ \,a_{_{mm}}  }}
\newcommand{\amn}{\ensuremath{ \,a_{_{mn}}  }}
\newcommand{\aijkl}{\ensuremath{ \,a_{_{ijkl}}  }}
\newcommand{\auik}{\ensuremath{ \,a_{_{ik}}^{u} }}

\newcommand{\asz}{\ensuremath{ {a}_{_0} }}
\newcommand{\aso}{\ensuremath{ {a}_{_1} }}
\newcommand{\ask}{\ensuremath{ {a}_{_k} }}
\newcommand{\asko}{\ensuremath{ {a}_{_{k+1}} }}
\newcommand{\asoN}{\ensuremath{ {a}_{_{N-1}} }}
\newcommand{\astwN}{\ensuremath{ {a}_{_{N-2}} }}
\newcommand{\asthN}{\ensuremath{ {a}_{_{N-3}} }}

\newcommand{\asbar}{\ensuremath{ \overline{a} }}
\newcommand{\asbbar}{\ensuremath{ \overline{\asbar} }}
\newcommand{\asqbar}{\ensuremath{ \overline{a^2} }}
\newcommand{\asqbbar}{\ensuremath{ \overline{\asqbar} }}
\newcommand{\asbartwNthN}{\ensuremath{ \asbar_{_{N-2/N-3}} }}
\newcommand{\asqbartwNthN}{\ensuremath{ \asqbar_{_{N-2/N-3}} }}
\newcommand{\asbaroNtwN}{\ensuremath{ \asbar_{_{N-1/N-2}} }}
\newcommand{\asqbaroNtwN}{\ensuremath{ \asqbar_{_{N-1/N-2}} }}
\newcommand{\asbaroNalfoN}{\ensuremath{ \asbar_{_{N-1/N-1-\alfa}} }}

\newcommand{\alfa}{\ensuremath{ \alpha  }}
\newcommand{\alfaz}{\ensuremath{ \alpha_{_{\!0}}  }}
\newcommand{\alfao}{\ensuremath{ \alpha_{_{\!1}}  }}
\newcommand{\alfatw}{\ensuremath{ \alpha_{_{\!2}}  }}
\newcommand{\alfk}{\ensuremath{ \alpha_{_{\!k}}  }}
\newcommand{\alfak}{\ensuremath{ \alpha_{_{\!k}}  }}
\newcommand{\alfaN}{\ensuremath{ \alpha_{_{\!N}}  }}
\newcommand{\alfako}{\ensuremath{ \alpha_{_{\!k+1}}  }}
\newcommand{\alfaj}{\ensuremath{ \alpha_{_{\!j}}  }}
\newcommand{\alfaijkl}{\ensuremath{ \alpha_{_{\!ijkl}}  }}
\newcommand{\alfai}{\ensuremath{ \alpha_{_{\!i}}  }}
\newcommand{\alfaoN}{\ensuremath{ \alpha_{_{\!N\!\!-1}}  }}
\newcommand{\alfatwN}{\ensuremath{ \alpha_{_{\!N\!-2}}  }}
\newcommand{\alfakop}{\ensuremath{ \alfako^{+} }}
\newcommand{\alfakom}{\ensuremath{ \alfako^{-} }}

\newcommand{\bsk}{\ensuremath{ {b}_{_k} }}
\newcommand{\bos}{\ensuremath{ \,b_{_1}  }}
\newcommand{\btws}{\ensuremath{ \,b_{_2}  }}
\newcommand{\bths}{\ensuremath{ \,b_{_3}  }}
\newcommand{\bjs}{\ensuremath{ \,b_{_j}  }}
\newcommand{\bts}{\ensuremath{ \,b^o  }}
\newcommand{\btjs}{\ensuremath{ \,\bjs^o  }}
\newcommand{\btos}{\ensuremath{ \,\bos^o  }}
\newcommand{\bttwos}{\ensuremath{ \,\btws^o  }}

\newcommand{\bulj}{\ensuremath{ \,b_{_{lj}}^{u} }}
\newcommand{\bij}{\ensuremath{ \,b_{_{ij}} }}

\newcommand{\betai}{\ensuremath{ \beta_{_i}  }}
\newcommand{\betak}{\ensuremath{ \beta_{_k}  }}
\newcommand{\betao}{\ensuremath{ \beta_{_{1}}  }}
\newcommand{\betatw}{\ensuremath{ \beta_{_{2}}  }}
\newcommand{\betamu}{\ensuremath{ \beta_{_{\mu}}  }}
\newcommand{\betaj}{\ensuremath{ \beta_{_{j}}  }}
\newcommand{\betal}{\ensuremath{ \beta_{_l}  }}
\newcommand{\betam}{\ensuremath{ \beta_{_m}  }}
\newcommand{\betan}{\ensuremath{ \beta_{_n}  }}
\newcommand{\betaminus}{\ensuremath{ \beta^{-} }}
\newcommand{\betaplus}{\ensuremath{ \beta^{+} }}
\newcommand{\betapr}{\ensuremath{ \beta' }}

\newcommand{\cs}{\ensuremath{ c }}
\newcommand{\cstwtw}{\ensuremath{ c_{_{22}} }}

\newcommand{\dsbar}{\ensuremath{ \overline{d} }}
\newcommand{\dsbaroNtwN}{\ensuremath{ \dsbar_{_{N-1/N-2}} }}
\newcommand{\dsbartwNthN}{\ensuremath{ \dsbar_{_{N-2/N-3}} }}
\newcommand{\dasbar}{\ensuremath{ \overline{(d+1)a } }}
\newcommand{\dasbartwNthN}{\ensuremath{ \dasbar_{_{N-2/N-3}} }}
\newcommand{\dasqbar}{\ensuremath{ \overline{(d+1)a^2 } }}
\newcommand{\dasqbartwNthN}{\ensuremath{ \dasqbar_{_{N-2/N-3}} }}
\newcommand{\dasbaralftwN}{\ensuremath{ \overline{ [d(\ysalfoN)+1] \astwN^2 } }}
\newcommand{\dsbaralfoNalftwN}{\ensuremath{ \dsbar_{_{N-1-\alfa/N-2-\alfa}} }}
\newcommand{\dasqbbar}{\ensuremath{ \overline{\overline{[d(\zsoN)+1]\astwN^2}} }}
\newcommand{\dasbbar}{\ensuremath{ \overline{\overline{[d(\zsoN)+1]\astwN}} }}
\newcommand{\dsbbar}{\ensuremath{ \overline{\overline{d(\zsoN)}} }}

\newcommand{\dbs}{\ensuremath{ \,\delta b }}
\newcommand{\dbos}{\ensuremath{ \dbs_{_1} }}
\newcommand{\dbtws}{\ensuremath{  \dbs_{_2} }}
\newcommand{\dbjs}{\ensuremath{  \dbs_{_j} }}

\newcommand{\dfi}{\ensuremath{ \;\delta\phi }}
\newcommand{\dfikk}{\ensuremath{ \;\delta \phi_{_{k/k}} }}
\newcommand{\dfizz}{\ensuremath{ \;\delta \phi_{_{0/0}} }}

\newcommand{\deltaok}{\ensuremath{ \,\delta_{_{k-1}}  }}
\newcommand{\deltatwk}{\ensuremath{ \,\delta_{_{k-2}}  }}
\newcommand{\deltaij}{\ensuremath{ \,\delta_{_{ij}}  }}
\newcommand{\deltajl}{\ensuremath{ \,\delta_{_{jl}}  }}
\newcommand{\deltakl}{\ensuremath{ \,\delta_{_{kl}}  }}
\newcommand{\dqes}{\ensuremath{ \,\delta q }}
\newcommand{\deltakol}{\ensuremath{ \,\delta_{_{k+1,l}}  }}
\newcommand{\deltakoi}{\ensuremath{ \,\delta_{_{k+1,i}}  }}
\newcommand{\deltai}{\ensuremath{ \delta_{_i}  }}

\newcommand{\dbone}{\ensuremath{ \delta \bos  }}
\newcommand{\dbtwo}{\ensuremath{ \delta \btws  }}
\newcommand{\dbthr}{\ensuremath{ \delta \bths  }}

\newcommand{\des}{\ensuremath{ \delta e  }}
\newcommand{\deone}{\ensuremath{ \des_{_1}  }}
\newcommand{\detwo}{\ensuremath{ \des_{_2}  }}
\newcommand{\dethr}{\ensuremath{ \des_{_3}  }}

\newcommand{\eos}{\ensuremath{ \,e_{_1}  }}
\newcommand{\etws}{\ensuremath{ \,e_{_2}  }}
\newcommand{\eths}{\ensuremath{ \,e_{_3}  }}

\newcommand{\etaz}{\ensuremath{ \eta_{_0}  }}
\newcommand{\etat}{\ensuremath{ \eta_{_t}  }}
\newcommand{\etak}{\ensuremath{ \eta_{_k}  }}
\newcommand{\etaok}{\ensuremath{ \eta_{_{k-1}}  }}
\newcommand{\etai}{\ensuremath{ \eta_{_i}  }}
\newcommand{\etaoz}{\ensuremath{ \eta_{_{1/0}}  }}
\newcommand{\etaoo}{\ensuremath{ \,\eta_{_{1/1}}  }}
\newcommand{\etae}{\ensuremath{\widehat{\eta}}}
\newcommand{\etakast}{\ensuremath{ \etak^{\ast}  }}
\newcommand{\etaokast}{\ensuremath{ \etaok^{\ast}  }}
\newcommand{\etaoN}{\ensuremath{ \,\eta_{_{N-1}}  }}
\newcommand{\etapr}{\ensuremath{ \widehat{\eta}^{-}  }}
\newcommand{\etapo}{\ensuremath{ \widehat{\eta}^{*}  }}

\newcommand{\eps}{\ensuremath{ \epsilon  }}
\newcommand{\epso}{\ensuremath{ \epsilon_{_1}  }}
\newcommand{\epsone}{\ensuremath{ \epsilon_{_1}  }}
\newcommand{\epstw}{\ensuremath{ \epsilon_{_2}  }}
\newcommand{\epstwo}{\ensuremath{ \epsilon_{_2}  }}
\newcommand{\epsth}{\ensuremath{ \epsilon_{_3}  }}
\newcommand{\epsthr}{\ensuremath{ \epsilon_{_3}  }}

\newcommand{\Etxk}{\ensuremath{ E_{_{t,\xv,k}} }}
\newcommand{\Esxk}{\ensuremath{ E_{_{s,\xv,k}} }}
\newcommand{\Etxtmdk}{\ensuremath{ E_{_{t,\xv;t-\Delta,k}} }}
\newcommand{\Etmdk}{\ensuremath{ E_{_{t-\Delta,k}} }}
\newcommand{\Etonxttwk}{\ensuremath{ E_{_{t_{_{1}},\xv;t_{_{2}},k}} }}
\newcommand{\Etk}{\ensuremath{ E_{_{t,k}} }}

\newcommand{\fsbbar}{\ensuremath{ \overline{\overline{f}} }}

\newcommand{\ftot}{\ensuremath{ f_{\tot} }}
\newcommand{\ftotbeta}{\ensuremath{ \ftot(\beta) }}
\newcommand{\ftauk}{\ensuremath{ f_{\tauk} }}
\newcommand{\ftokbeta}{\ensuremath{ \ftauk(\beta) }}
\newcommand{\ftototpr}{\ensuremath{ f_{\tot | \totpr} }}
\newcommand{\ftok}{\ensuremath{ f_{\tauk} }}
\newcommand{\ftokok}{\ensuremath{ f_{\toko | \tauk} }}

\newcommand{\fij}{\ensuremath{ \,\phi_{_j} }}
\newcommand{\fijp}{\ensuremath{ \fij' }}

\newcommand{\gamazz}{\ensuremath{ \gamma_{_{0/0}}  }}
\newcommand{\gamaoz}{\ensuremath{ \gamma_{_{1/0}}  }}
\newcommand{\gamaoo}{\ensuremath{ \gamma_{_{1/1}}  }}
\newcommand{\gamaoostar}{\ensuremath{ \star{\gamaoo}  }}
\newcommand{\gamakk}{\ensuremath{ \,\gamma_{_{k/k}}  }}
\newcommand{\gamaokok}{\ensuremath{ \,\gamma_{_{ {k-1}/{k-1} }}  }}
\newcommand{\gamakoK}{\ensuremath{ \,\gamma_{_{ {k+1}/{k} }}  }}
\newcommand{\gamaKok}{\ensuremath{  \,\gamma_{_{    k /{k-1} }}  }}
\newcommand{\gamabar}{\ensuremath{  \,\bar{\gamma}  }}
\newcommand{\gamabarKok}{\ensuremath{  \,\gamabar_{_{    k /{k-1} }}  }}
\newcommand{\gamabarbar}{\ensuremath{  \,\bar{\gamabar}  }}
\newcommand{\gamabarbarKok}{\ensuremath{  \,\gamabarbar_{_{    k /{k-1} }}  }}
\newcommand{\gamai}{\ensuremath{ \gamma_{_i}  }}
\newcommand{\gamaone}{\ensuremath{ \gamma_{_1}  }}
\newcommand{\gamatwo}{\ensuremath{ \gamma_{_2}  }}
\newcommand{\gamathr}{\ensuremath{ \gamma_{_3}  }}
\newcommand{\gamanu}{\ensuremath{ \gamma_{_\nu}  }}
\newcommand{\gamany}{\ensuremath{ \gamma_{_{n_y}}  }}

\newcommand{\kes}{\ensuremath{ \widehat{k} }}

\newcommand{\gNN}{\ensuremath{  g_{_{N/N}} }}

\newcommand{\io}{\ensuremath{  i_{_1} }}
\newcommand{\itw}{\ensuremath{  i_{_2} }}
\newcommand{\ith}{\ensuremath{  i_{_3} }}
\newcommand{\ifo}{\ensuremath{  i_{_4} }}

\newcommand{\Kkoo}{\ensuremath{  \Kk_{_{11}}  }}
\newcommand{\Kkotw}{\ensuremath{ \Kk_{_{12}}  }}
\newcommand{\Kkoth}{\ensuremath{ \Kk_{_{13}}  }}
\newcommand{\Kkof}{\ensuremath{  \Kk_{_{14}}  }}
\newcommand{\Kktwo}{\ensuremath{ \Kk_{_{21}}  }}
\newcommand{\Kktwtw}{\ensuremath{\Kk_{_{22}}  }}
\newcommand{\Kktwth}{\ensuremath{ \Kk_{_{23}}  }}
\newcommand{\Kktwf}{\ensuremath{ \Kk_{_{24}}  }}
\newcommand{\Kktho}{\ensuremath{  \Kk_{_{31}}  }}
\newcommand{\Kkthtw}{\ensuremath{ \Kk_{_{32}}  }}
\newcommand{\Kkthth}{\ensuremath{  \Kk_{_{33}}  }}
\newcommand{\Kkthf}{\ensuremath{ \Kk_{_{34}}  }}
\newcommand{\Kkfo}{\ensuremath{  \Kk_{_{41}}  }}
\newcommand{\Kkftw}{\ensuremath{ \Kk_{_{42}}  }}
\newcommand{\Kkfth}{\ensuremath{ \Kk_{_{43}}  }}
\newcommand{\Kkff}{\ensuremath{  \Kk_{_{44}}  }}
\newcommand{\Kkij}{\ensuremath{  {\Kk}_{_{ij}}  }}

\newcommand{\Kkooo}{\ensuremath{  \Kko_{_{11}}  }}
\newcommand{\Kkootw}{\ensuremath{ \Kko_{_{12}}  }}
\newcommand{\Kkooth}{\ensuremath{ \Kko_{_{13}}  }}
\newcommand{\Kkoof}{\ensuremath{  \Kko_{_{14}}  }}
\newcommand{\Kkotwo}{\ensuremath{ \Kko_{_{21}}  }}
\newcommand{\Kkotwtw}{\ensuremath{\Kko_{_{22}}  }}
\newcommand{\Kkotwth}{\ensuremath{ \Kko_{_{23}}  }}
\newcommand{\Kkotwf}{\ensuremath{ \Kko_{_{24}}  }}
\newcommand{\Kkotho}{\ensuremath{  \Kko_{_{31}}  }}
\newcommand{\Kkothtw}{\ensuremath{ \Kko_{_{32}}  }}
\newcommand{\Kkothth}{\ensuremath{  \Kko_{_{33}}  }}
\newcommand{\Kkothf}{\ensuremath{ \Kko_{_{34}}  }}
\newcommand{\Kkofo}{\ensuremath{  \Kko_{_{41}}  }}
\newcommand{\Kkoftw}{\ensuremath{ \Kko_{_{42}}  }}
\newcommand{\Kkofth}{\ensuremath{ \Kko_{_{43}}  }}
\newcommand{\Kkoff}{\ensuremath{  \Kko_{_{44}}  }}
\newcommand{\Kkoij}{\ensuremath{  {\Kko}_{_{ij}}  }}

\newcommand{\Koo}{\ensuremath{  K_{_{11}}  }}
\newcommand{\Kotw}{\ensuremath{ K_{_{12}}  }}
\newcommand{\Koth}{\ensuremath{ K_{_{13}}  }}
\newcommand{\Kof}{\ensuremath{  K_{_{14}}  }}
\newcommand{\Ktwo}{\ensuremath{ K_{_{21}}  }}
\newcommand{\Ktwtw}{\ensuremath{  K_{_{22}}  }}
\newcommand{\Ktwth}{\ensuremath{ K_{_{23}}  }}
\newcommand{\Ktwf}{\ensuremath{ K_{_{24}}  }}
\newcommand{\Ktho}{\ensuremath{  K_{_{31}}  }}
\newcommand{\Kthtw}{\ensuremath{ K_{_{32}}  }}
\newcommand{\Kthth}{\ensuremath{  K_{_{33}}  }}
\newcommand{\Kthf}{\ensuremath{ K_{_{34}}  }}
\newcommand{\Kfo}{\ensuremath{  K_{_{41}}  }}
\newcommand{\Kftw}{\ensuremath{ K_{_{42}}  }}
\newcommand{\Kfth}{\ensuremath{ K_{_{43}}  }}
\newcommand{\Kff}{\ensuremath{  K_{_{44}}  }}

\newcommand{\Ki}{\ensuremath{ K_{_{i}} }}

\newcommand{\lbd}{\ensuremath{{\lambda}}}
\newcommand{\lbdz}{\ensuremath{ \lambda^0  }}
\newcommand{\lbdo}{\ensuremath{ \lambda^1  }}
\newcommand{\lbdtw}{\ensuremath{ \lambda^2  }}
\newcommand{\lbdth}{\ensuremath{ \lambda^3  }}
\newcommand{\lmx}{\ensuremath{\lambda_{max}}}
\newcommand{\lkok}{\ensuremath{ \,\lambda_{_{k+1/k}} }}
\newcommand{\likok}{\ensuremath{ {\lkok^i} }}
\newcommand{\lzkok}{\ensuremath{ {\lkok^0} }}
\newcommand{\lokok}{\ensuremath{ {\lkok^1} }}
\newcommand{\ltwkok}{\ensuremath{ {\lkok^2} }}
\newcommand{\lthkok}{\ensuremath{ {\lkok^3} }}
\newcommand{\lmxk}{\ensuremath{  {\lmx}_{_k} }}
\newcommand{\lbdi}{\ensuremath{ \lambda_{_i}  }}
\newcommand{\lko}{\ensuremath{ \lambda_{_{k+1}} }}
\newcommand{\liko}{\ensuremath{ \lko^i }}
\newcommand{\lzko}{\ensuremath{ \lko^0 }}
\newcommand{\loko}{\ensuremath{ \lko^1 }}
\newcommand{\ltwko}{\ensuremath{ \lko^2 }}
\newcommand{\lthko}{\ensuremath{ \lko^3 }}

\newcommand{\mk}{\ensuremath{ m_{_k}  }}
\newcommand{\mko}{\ensuremath{ m_{_{k+1}}  }}
\newcommand{\dmk}{\ensuremath{ \delta \mk }}
\newcommand{\dmko}{\ensuremath{ \delta \mko }}
\newcommand{\mz}{\ensuremath{ m_{_0}  }}
\newcommand{\dmz}{\ensuremath{ \delta \mz }}
\newcommand{\mone}{\ensuremath{ m_{_1}  }}

\newcommand{\muko}{\ensuremath{ \mu_{_{k+1}}  }}
\newcommand{\mukoast}{\ensuremath{ {\muko^{\ast}}  }}
\newcommand{\muz}{\ensuremath{ \mu_{_0}  }}
\newcommand{\muezz}{\ensuremath{ \widehat{\mu}_{_{0/0}}  }}
\newcommand{\muk}{\ensuremath{ \,\mu_{_k}  }}
\newcommand{\mui}{\ensuremath{ \,\mu_{_i}  }}
\newcommand{\muo}{\ensuremath{ \,\mu_{_1}  }}
\newcommand{\mutw}{\ensuremath{ \,\mu_{_2}  }}
\newcommand{\muth}{\ensuremath{ \,\mu_{_3}  }}
\newcommand{\mufo}{\ensuremath{ \,\mu_{_4}  }}
\newcommand{\muN}{\ensuremath{ \mu_{N}  }}
\newcommand{\muNast}{\ensuremath{ \muN^{\ast}  }}
\newcommand{\muort}{\ensuremath{ \mu^{ort}  }}

\newcommand{\nk}{\ensuremath{ {n_{_k}} }}

\newcommand{\Nj}{\ensuremath{  \,N_{_j} }}

\newcommand{\nuoz}{\ensuremath{ \nu_{_{1/0}}  }}
\newcommand{\nuon}{\ensuremath{ \,\nu_{_1} }}
\newcommand{\nutw}{\ensuremath{  \,\nu_{_2} }}

\newcommand{\omo}{\ensuremath{ \,\omega_{_{1}}   }}
\newcommand{\omtw}{\ensuremath{ \,\omega_{_{2}}   }}
\newcommand{\omth}{\ensuremath{ \,\omega_{_{3}}   }}

\newcommand{\omot}{\ensuremath{ \,\omega_{_{1}}^{o}   }}
\newcommand{\omtwt}{\ensuremath{ \,\omega_{_{2}}^{o}   }}
\newcommand{\omtht}{\ensuremath{ \,\omega_{_{3}}^{o}   }}

\newcommand{\pinf}{\ensuremath{ p_{_{\infty}}  }}

\newcommand{\pij}{\ensuremath{ \,p_{_{ij}} }}
\newcommand{\pki}{\ensuremath{ \,p_{_{ki}} }}
\newcommand{\pji}{\ensuremath{ \,p_{_{ji}} }}
\newcommand{\pkj}{\ensuremath{ \,p_{_{kj}} }}
\newcommand{\pkk}{\ensuremath{ \,p_{_{kk}} }}
\newcommand{\pjk}{\ensuremath{ \,p_{_{jk}} }}
\newcommand{\pjl}{\ensuremath{ \,p_{_{jl}} }}
\newcommand{\pil}{\ensuremath{ \,p_{_{il}} }}
\newcommand{\plj}{\ensuremath{ \,p_{_{lj}} }}

\newcommand{\Pff}{\ensuremath{ P_{_{44}} }}
\newcommand{\Pfsv}{\ensuremath{ P_{_{47}} }}
\newcommand{\Pfni}{\ensuremath{ P_{_{49}} }}
\newcommand{\Psvsv}{\ensuremath{ P_{_{77}} }}
\newcommand{\Psvni}{\ensuremath{ P_{_{79}} }}
\newcommand{\Pnini}{\ensuremath{ P_{_{99}} }}

\newcommand{\pdf}{\ensuremath{ p_\bv \left( \bv,A \right)  }}

\newcommand{\qs}{\ensuremath{ \,\emph{q}  }}
\newcommand{\qst}{\ensuremath{ \qs^{o}  }}
\newcommand{\qsone}{\ensuremath{ \qs_{_1} }}
\newcommand{\qstwo}{\ensuremath{ \qs_{_2} }}
\newcommand{\qsthr}{\ensuremath{ \qs_{_3} }}
\newcommand{\qsz}{\ensuremath{ \qs^0 }}
\newcommand{\qskt}{\ensuremath{ \qsk^{o}  }}
\newcommand{\qsko}{\ensuremath{ \qs_{_{k+1}} }}
\newcommand{\qes}{\ensuremath{ \,\widehat{\qs}  }}
\newcommand{\qesk}{\ensuremath{ \qes_{_{k}} }}
\newcommand{\qesko}{\ensuremath{ \qes_{_{k+1}} }}
\newcommand{\qeskok}{\ensuremath{ \qes_{_{k+1/k}}  }}
\newcommand{\qeskoko}{\ensuremath{ \qes_{_{k+1/k+1}}  }}
\newcommand{\qeskolk}{\ensuremath{ \qes_{_{k+l-1/k}}  }}
\newcommand{\qesklk}{\ensuremath{ \qes_{_{k+l/k}}  }}
\newcommand{\quats}{\ensuremath{\,\emph{q}}}
\newcommand{\qeps}{\ensuremath{ \,q_{_\epsilon}  }}

\newcommand{\qij}{\ensuremath{ \,q_{_{ij}} }}
\newcommand{\qii}{\ensuremath{ \,q_{_{ii}} }}
\newcommand{\qkj}{\ensuremath{ \,q_{_{kj}} }}
\newcommand{\qki}{\ensuremath{ \,q_{_{ki}} }}
\newcommand{\qji}{\ensuremath{ \,q_{_{ji}} }}

\newcommand{\qsi}{\ensuremath{ \,q_{_{i}} }}
\newcommand{\qsk}{\ensuremath{ \,q_{_{k}} }}

\newcommand{\rl}{\ensuremath{ \,r^{\lambda}  }}
\newcommand{\rlko}{\ensuremath{ \rl_{_{k+1}}  }}
\newcommand{\rlkok}{\ensuremath{ \rl_{_{k+1/k}}  }}

\newcommand{\rsym}{\ensuremath{ r^{sym}  }}
\newcommand{\rtr}{\ensuremath{ r^{tr}  }}
\newcommand{\rtrko}{\ensuremath{ \rtr_{_{k+1}}  }}

\newcommand{\rsk}{\ensuremath{ {r}_{_k} }}
\newcommand{\ros}{\ensuremath{ \,r_{_1}  }}
\newcommand{\rtws}{\ensuremath{ \,r_{_2}  }}
\newcommand{\rths}{\ensuremath{ \,r_{_3}  }}

\newcommand{\rkos}{\ensuremath{ \, r_{_{k+1}} }}
\newcommand{\rkoso}{\ensuremath{ {\rkos}_{_1} }}
\newcommand{\rkostw}{\ensuremath{{\rkos}_{_2} }}
\newcommand{\rkosth}{\ensuremath{ {\rkos}_{_3} }}

\newcommand{\rooz}{\ensuremath{ \,\rho_{_{1/0}}  }}
\newcommand{\rozo}{\ensuremath{ \,\rho_{_{0/1}}  }}
\newcommand{\roko}{\ensuremath{ \,\rho_{_{k+1}}  }}
\newcommand{\rok}{\ensuremath{ \,\rho_{_{k}}  }}
\newcommand{\roast}{\ensuremath{ \rho^{\ast}  }}
\newcommand{\rokoast}{\ensuremath{ \roko^{\ast}  }}
\newcommand{\rokast}{\ensuremath{ \rok^{\ast}  }}
\newcommand{\rooo}{\ensuremath{  \,\rho_{_{1/1}}  }}
\newcommand{\roe}{\ensuremath{   \,\widehat{\rho}}}
\newcommand{\rookk}{\ensuremath{ \,\rho_{_{{k-1}/k}}  }}
\newcommand{\rotwkok}{\ensuremath{ \,\rho_{_{{k-2}/{k-1}}}  }}
\newcommand{\rookok}{\ensuremath{ \,\rho_{_{{k-1}/{k-1}}}  }}
\newcommand{\robar}{\ensuremath{  \,\bar{\rho}  }}
\newcommand{\robarokok}{\ensuremath{ \,\robar_{_{{k-1}/{k-1}}}  }}
\newcommand{\robarbar}{\ensuremath{  \,\bar{\robar}  }}
\newcommand{\robarbarokok}{\ensuremath{ \,\robarbar_{_{{k-1}/{k-1}}}  }}
\newcommand{\roi}{\ensuremath{ \,\rho_{_{i}}  }}

\newcommand{\skos}{\ensuremath{ s_{_{k+1}} }}

\newcommand{\sigeps}{\ensuremath{ \,\sigma_{_\epsilon}   }}
\newcommand{\sigb}{\ensuremath{ \,\sigma_{_b}   }}
% Change for the IsraConfPaper
%\newcommand{\sigeps}{\ensuremath{ \,\kappa_{_\epsilon}   }}
%\newcommand{\sigb}{\ensuremath{ \,\kappa_{_b}   }}

\newcommand{\sigk}{\ensuremath{ \,\sigma_{_{k}}   }}
\newcommand{\sigko}{\ensuremath{ \,\sigma_{_{k+1}}   }}
\newcommand{\dsigko}{\ensuremath{ \,\delta \sigma_{_{k+1}}   }}
\newcommand{\sigkot}{\ensuremath{ \sigko^{o} }}
\newcommand{\sigi}{\ensuremath{ \,\sigma_{_i}   }}
\newcommand{\sigone}{\ensuremath{ \,\sigma_{_1}   }}
\newcommand{\sigtwo}{\ensuremath{ \,\sigma_{_2}   }}
\newcommand{\sigtot}{\ensuremath{ \,\sigma_{_{tot}}   }}
\newcommand{\sigm}{\ensuremath{   \,\sigma_{_m}   }}
\newcommand{\sigbi}{\ensuremath{ {\sigb^i} }}
\newcommand{\sigbn}{\ensuremath{ {\sigb^n} }}
\newcommand{\sigoneone}{\ensuremath{ \,\sigma_{_{11}}   }}
\newcommand{\sigtwotwo}{\ensuremath{ \,\sigma_{_{22}}   }}
\newcommand{\sigthrthr}{\ensuremath{ \,\sigma_{_{33}}   }}
\newcommand{\sigforfor}{\ensuremath{ \,\sigma_{_{44}}   }}
\newcommand{\sigz}{\ensuremath{ \,\sigma^{o} }}
\newcommand{\sig}{\ensuremath{ \sigma }}
\newcommand{\sige}{\ensuremath{ \widehat{\sigma} }}
\newcommand{\sigsym}{\ensuremath{ \sigma^{sym} }}
\newcommand{\sigdist}{\ensuremath{ \sigma_{dst} }}

\newcommand{\ssko}{\ensuremath{ s_{_{k+1}}   }}

\newcommand{\sxsys}{\ensuremath{ [s,\xv(s),y(s)] }}
\newcommand{\sxk}{\ensuremath{ (s,\xv,k) }}
\newcommand{\sxys}{\ensuremath{ [s,\xv,y(s)] }}

\newcommand{\Soo}{\ensuremath{ \,S_{_{11}} }}
\newcommand{\Stwtw}{\ensuremath{  \,S_{_{22}} }}

\newcommand{\tz}{\ensuremath{ t_{_0}   }}
\newcommand{\tone}{\ensuremath{ t_{_1}   }}
\newcommand{\tk}{\ensuremath{ t_{_k}   }}
\newcommand{\tko}{\ensuremath{ t_{_{k+1}}   }}
\newcommand{\tktw}{\ensuremath{ t_{_{k+2}}   }}
\newcommand{\tok}{\ensuremath{ t_{_{k-1}}   }}
\newcommand{\tN}{\ensuremath{ t_{_N}   }}
\newcommand{\toN}{\ensuremath{ t_{_{N-1}}   }}
\newcommand{\tkN}{\ensuremath{ t_{_{k+N}}   }}
\newcommand{\tNo}{\ensuremath{ t_{_{N+1}}   }}
\newcommand{\ti}{\ensuremath{ t_{_{i}}   }}
\newcommand{\tj}{\ensuremath{ t_{_{j}}   }}
\newcommand{\tkl}{\ensuremath{ t_{_{k+l}}   }}
\newcommand{\tl}{\ensuremath{ t_{_{l}}   }}

\newcommand{\txj}{\ensuremath{ (t,\xv,j) }}
\newcommand{\tzxj}{\ensuremath{ (\tz,\xv,j) }}
\newcommand{\txk}{\ensuremath{ (t,\xv,k) }}
\newcommand{\tzxk}{\ensuremath{ (\tz,\xv,k) }}
\newcommand{\txi}{\ensuremath{ (t,\xv,i) }}
\newcommand{\txtyt}{\ensuremath{ [t,\xv(t),y(t)] }}
\newcommand{\tdtxtdtytdt}{\ensuremath{ [t+\dt,\xv(t+\dt),y(t+\dt)] }}
\newcommand{\txytdt}{\ensuremath{ [t,\xv(t),y(t+\dt)] }}
\newcommand{\txyt}{\ensuremath{ [t,\xv,y(t)] }}
\newcommand{\txys}{\ensuremath{ [t,\xv,y(s)] }}
\newcommand{\tztxtztytzt}{\ensuremath{ [\tz+t,\xv(\tz+t),y(\tz+t)] }}
\newcommand{\tztsxtztsytzts}{\ensuremath{ [\tz+t+s,\xv(\tz+t+s),y(\tz+t+s)] }}
\newcommand{\vxvyv}{\ensuremath{ [v,\xv(v),y(v)] }}

\newcommand{\tetaek}{\ensuremath{ \,\widehat{\Theta}_{_k} }}

\newcommand{\tot}{\ensuremath{ \tau_{t} }}
\newcommand{\tos}{\ensuremath{ \tau_{s} }}
\newcommand{\totpr}{\ensuremath{ \tau_{t'} }}
\newcommand{\tauk}{\ensuremath{ \tau_{k} }}
\newcommand{\toko}{\ensuremath{ \tau_{k+1} }}
\newcommand{\toes}{\ensuremath{ \widehat{\tau} }}

\newcommand{\uxuyu}{\ensuremath{ [u,\xv(u),y(u)] }}

\newcommand{\usk}{\ensuremath{ {u}_{_k} }}
\newcommand{\usko}{\ensuremath{ {u}_{_{k+1}} }}
\newcommand{\usoN}{\ensuremath{ {u}_{_{N-1}} }}
\newcommand{\ustwN}{\ensuremath{ {u}_{_{N-2}} }}
\newcommand{\usthN}{\ensuremath{ {u}_{_{N-3}} }}

\newcommand{\uskast}{\ensuremath{ \usk^{\ast} }}
\newcommand{\uskoast}{\ensuremath{ \usko^{\ast} }}
\newcommand{\usoNast}{\ensuremath{ \usoN^{\ast} }}
\newcommand{\ustwNast}{\ensuremath{ \ustwN^{\ast} }}
\newcommand{\usthNast}{\ensuremath{ \usthN^{\ast} }}

\newcommand{\vtr}{\ensuremath{  v^{tr}  }}
\newcommand{\vtrko}{\ensuremath{  \vtr_{_{k+1}}  }}

\newcommand{\Vkoioo}{\ensuremath{  \Vkoi_{_{11}}  }}
\newcommand{\Vkoiotw}{\ensuremath{ \Vkoi_{_{12}}  }}
\newcommand{\Vkoioth}{\ensuremath{ \Vkoi_{_{13}}  }}
\newcommand{\Vkoiof}{\ensuremath{  \Vkoi_{_{14}}  }}
\newcommand{\Vkoitwo}{\ensuremath{ \Vkoi_{_{21}}  }}
\newcommand{\Vkoitwtw}{\ensuremath{\Vkoi_{_{22}}  }}
\newcommand{\Vkoitwth}{\ensuremath{ \Vkoi_{_{23}}  }}
\newcommand{\Vkoitwf}{\ensuremath{ \Vkoi_{_{24}}  }}
\newcommand{\Vkoitho}{\ensuremath{  \Vkoi_{_{31}}  }}
\newcommand{\Vkoithtw}{\ensuremath{ \Vkoi_{_{32}}  }}
\newcommand{\Vkoithth}{\ensuremath{  \Vkoi_{_{33}}  }}
\newcommand{\Vkoithf}{\ensuremath{ \Vkoi_{_{34}}  }}
\newcommand{\Vkoifo}{\ensuremath{  \Vkoi_{_{41}}  }}
\newcommand{\Vkoiftw}{\ensuremath{ \Vkoi_{_{42}}  }}
\newcommand{\Vkoifth}{\ensuremath{ \Vkoi_{_{43}}  }}
\newcommand{\Vkoiff}{\ensuremath{  \Vkoi_{_{44}}  }}
\newcommand{\Vkoiij}{\ensuremath{  {\Vkoi}_{_{ij}}  }}

\newcommand{\vsk}{\ensuremath{ {v}_{_{k}} }}
\newcommand{\vsko}{\ensuremath{ {v}_{_{k+1}} }}

\newcommand{\Wkoo}{\ensuremath{  \Wk_{_{11}}  }}
\newcommand{\Wkotw}{\ensuremath{ \Wk_{_{12}}  }}
\newcommand{\Wkoth}{\ensuremath{ \Wk_{_{13}}  }}
\newcommand{\Wkof}{\ensuremath{  \Wk_{_{14}}  }}
\newcommand{\Wktwo}{\ensuremath{ \Wk_{_{21}}  }}
\newcommand{\Wktwtw}{\ensuremath{\Wk_{_{22}}  }}
\newcommand{\Wktwth}{\ensuremath{ \Wk_{_{23}}  }}
\newcommand{\Wktwf}{\ensuremath{ \Wk_{_{24}}  }}
\newcommand{\Wktho}{\ensuremath{  \Wk_{_{31}}  }}
\newcommand{\Wkthtw}{\ensuremath{ \Wk_{_{32}}  }}
\newcommand{\Wkthth}{\ensuremath{  \Wk_{_{33}}  }}
\newcommand{\Wkthf}{\ensuremath{ \Wk_{_{34}}  }}
\newcommand{\Wkfo}{\ensuremath{  \Wk_{_{41}}  }}
\newcommand{\Wkftw}{\ensuremath{ \Wk_{_{42}}  }}
\newcommand{\Wkfth}{\ensuremath{ \Wk_{_{43}}  }}
\newcommand{\Wkff}{\ensuremath{  \Wk_{_{44}}  }}
\newcommand{\Wkij}{\ensuremath{  {\Wk}_{_{ij}}  }}

\newcommand{\wij}{\ensuremath{  w_{_{ij}}  }}
\newcommand{\wsk}{\ensuremath{ {w}_{_{k}} }}

\newcommand{\xoo}{\ensuremath{ \,x_{_{11}}   }}
\newcommand{\xotw}{\ensuremath{ \,x_{_{12}}   }}
\newcommand{\xtwtw}{\ensuremath{ \,x_{_{22}}   }}
\newcommand{\xij}{\ensuremath{ \,x_{_{ij}}   }}
\newcommand{\xkl}{\ensuremath{ \,x_{_{kl}}   }}
\newcommand{\xbeta}{\ensuremath{ \,x_{_{\beta}}   }}
\newcommand{\xks}{\ensuremath{ \,x_{_k} }}
\newcommand{\xkso}{\ensuremath{ {\xks}_{_1} }}
\newcommand{\xkstw}{\ensuremath{ {\xks}_{_2} }}
\newcommand{\xksth}{\ensuremath{ {\xks}_{_3} }}
\newcommand{\xksf}{\ensuremath{ {\xks}_{_4} }}
\newcommand{\xksfv}{\ensuremath{ {\xks}_{_5} }}
\newcommand{\xkssx}{\ensuremath{ {\xks}_{_6} }}
\newcommand{\xkssv}{\ensuremath{ {\xks}_{_7} }}
\newcommand{\xksei}{\ensuremath{ {\xks}_{_8} }}
\newcommand{\xksni}{\ensuremath{ {\xks}_{_9} }}

\newcommand{\xes}{\ensuremath{ \,\widehat{x} }}
\newcommand{\xeso}{\ensuremath{ {\xes}_{_1} }}
\newcommand{\xestw}{\ensuremath{ {\xes}_{_2} }}
\newcommand{\xesth}{\ensuremath{ {\xes}_{_3} }}
\newcommand{\xesf}{\ensuremath{ {\xes}_{_4} }}
\newcommand{\xesfv}{\ensuremath{ {\xes}_{_5} }}
\newcommand{\xessx}{\ensuremath{ {\xes}_{_6} }}
\newcommand{\xessv}{\ensuremath{ {\xes}_{_7} }}
\newcommand{\xesei}{\ensuremath{ {\xes}_{_8} }}
\newcommand{\xesni}{\ensuremath{ {\xes}_{_9} }}
\newcommand{\xesten}{\ensuremath{ {\xes}_{_{10}} }}

\newcommand{\xsz}{\ensuremath{ {x}_{_0} }}
\newcommand{\xso}{\ensuremath{ {x}_{_1} }}
\newcommand{\xstw}{\ensuremath{ {x}_{_2} }}
\newcommand{\xsth}{\ensuremath{ {x}_{_3} }}
\newcommand{\xsf}{\ensuremath{ {x}_{_4} }}
\newcommand{\xsfv}{\ensuremath{ {x}_{_5} }}
\newcommand{\xssx}{\ensuremath{ {x}_{_6} }}
\newcommand{\xssv}{\ensuremath{ {x}_{_7} }}
\newcommand{\xsei}{\ensuremath{ {x}_{_8} }}
\newcommand{\xsni}{\ensuremath{ {x}_{_9} }}
\newcommand{\xsten}{\ensuremath{ {x}_{_{10}} }}
\newcommand{\xss}{\ensuremath{ x(s) }}
\newcommand{\xsko}{\ensuremath{ {x}_{_{k+1}} }}
\newcommand{\xsk}{\ensuremath{ {x}_{_{k}} }}
\newcommand{\xsN}{\ensuremath{ {x}_{_{N}} }}
\newcommand{\xsoN}{\ensuremath{ {x}_{_{N-1}} }}
\newcommand{\xstwN}{\ensuremath{ {x}_{_{N-2}} }}
\newcommand{\xsl}{\ensuremath{ {x}_{_{l}} }}
\newcommand{\xslo}{\ensuremath{ {x}_{_{l+1}} }}

\newcommand{\XX}{\ensuremath{ \mathcal{X} }}
\newcommand{\XXk}{\ensuremath{ \XX^k }}
\newcommand{\XXt}{\ensuremath{ \XX^t }}
\newcommand{\lXXk}{\ensuremath{ \XX_l^k }}
\newcommand{\XXoN}{\ensuremath{ \XX^{N-1} }}
\newcommand{\XXtwN}{\ensuremath{ \XX^{N-2} }}
\newcommand{\XXi}{\ensuremath{ \XX^i }}
\newcommand{\XXko}{\ensuremath{ \XX^{k+1} }}

\newcommand{\ysz}{\ensuremath{ {y}_{_0} }}
\newcommand{\yso}{\ensuremath{ {y}_{_1} }}
\newcommand{\ysk}{\ensuremath{ {y}_{_k} }}
\newcommand{\yson}{\ensuremath{ {y}_{_{\!n-1}} }}
\newcommand{\ysok}{\ensuremath{ {y}_{_{\!k-\!1}} }}
\newcommand{\ysko}{\ensuremath{ {y}_{_{k+1}} }}
\newcommand{\ysoN}{\ensuremath{ {y}_{_{\!N\!-1}} }}
\newcommand{\ystwN}{\ensuremath{ {y}_{_{\!N-\!2}} }}
\newcommand{\ysthN}{\ensuremath{ {y}_{_{\!N-\!3}} }}
\newcommand{\ysl}{\ensuremath{ {y}_{_{\!l}} }}
\newcommand{\yslo}{\ensuremath{ {y}_{_{\!l\!+1}} }}
\newcommand{\ysalfoN}{\ensuremath{ {y}_{_{\!\!N\!-1\!\!-\alfa}} }}
\newcommand{\ysalftwN}{\ensuremath{ {y}_{_{N-2-\alfa}} }}
\newcommand{\ysalfk}{\ensuremath{ {y}_{_{k-\alfa}} }}
\newcommand{\ysbetk}{\ensuremath{ {y}_{_{k-\beta}} }}
\newcommand{\ysbetko}{\ensuremath{ {y}_{_{k+1-\beta}} }}
\newcommand{\ysalfkk}{\ensuremath{ {y}_{_{k-\alfak}} }}
\newcommand{\ysalfkoko}{\ensuremath{ {y}_{_{k+1-\alfako}} }}
\newcommand{\ysalfoNoN}{\ensuremath{ {y}_{_{\!N-\!1-\alfaoN}} }}
\newcommand{\ysbetaloN}{\ensuremath{ {y}_{_{N-1-\betal}} }}
\newcommand{\ysalfkopko}{\ensuremath{ {y}_{_{k+1-\alfakop}} }}
\newcommand{\ysalfkomko}{\ensuremath{ {y}_{_{k+1-\alfakom}} }}

\newcommand{\ys}{\ensuremath{ y^s }}
\newcommand{\yt}{\ensuremath{ y^t }}
\newcommand{\ytko}{\ensuremath{ \yt_{_{k+1}} }}
\newcommand{\ytr}{\ensuremath{ y^{tr} }}
\newcommand{\ytrko}{\ensuremath{ \ytr_{_{k+1}}}}
\newcommand{\yss}{\ensuremath{ y(s) }}
\newcommand{\yst}{\ensuremath{ y_{_{t}} }}

\newcommand{\yse}{\ensuremath{ \widehat{y} }}
\newcommand{\yskk}{\ensuremath{ \yse_{_{k/k}} }}

\newcommand{\YY}{\ensuremath{ \mathcal{Y} }}
\newcommand{\YYk}{\ensuremath{ \YY^k }}
\newcommand{\YYt}{\ensuremath{ \YY^t }}
\newcommand{\YYtot}{\ensuremath{ \YY^{t-\tot} }}
\newcommand{\YYtos}{\ensuremath{ \YY^{t-\tos} }}
\newcommand{\YYok}{\ensuremath{ \YY^{k-1} }}
\newcommand{\YYoN}{\ensuremath{ \YY^{N-1} }}
\newcommand{\YYtwN}{\ensuremath{ \YY^{N-2} }}
\newcommand{\YYthN}{\ensuremath{ \YY^{N-3} }}
\newcommand{\kYYoN}{\ensuremath{ \YY_{k}^{N-1} }}
\newcommand{\lYYk}{\ensuremath{ \YY_l^k }}
\newcommand{\YYalfk}{\ensuremath{ \YY^{k-\alfa} }}
\newcommand{\YYalfoN}{\ensuremath{ \YY^{N-1-\alfa} }}
\newcommand{\YYalftwN}{\ensuremath{ \YY^{N-2-\alfa} }}
\newcommand{\YYalfkk}{\ensuremath{ \YY^{k-\alfak} }}
\newcommand{\YYalfkoko}{\ensuremath{ \YY^{k+1-\alfako} }}
\newcommand{\YYbetaik}{\ensuremath{ \YY^{k-\betai} }}

\newcommand{\zetak}{\ensuremath{ \zeta_{_k}  }}
\newcommand{\zij}{\ensuremath{ \,z_{_{ij}}  }}
\newcommand{\zalfa}{\ensuremath{ \,z_{_{\alpha}}  }}

\newcommand{\zsz}{\ensuremath{ {z}_{_{\!0}} }}
\newcommand{\zso}{\ensuremath{ {z}_{_{\!1}} }}
\newcommand{\zsk}{\ensuremath{ {z}_{_{\!k}} }}
\newcommand{\zsko}{\ensuremath{ {z}_{_{\!k\!+1}} }}
\newcommand{\zsl}{\ensuremath{ {z}_{_{\!l}} }}
\newcommand{\zslo}{\ensuremath{ {z}_{_{\!l+\!1}} }}
\newcommand{\zsoN}{\ensuremath{ {z}_{_{\!N\!-1}} }}
\newcommand{\zstwN}{\ensuremath{ {z}_{_{\!N\!-2}} }}
\newcommand{\zsN}{\ensuremath{ {z}_{_{\!N}} }}

\newcommand{\ZZk}{\ensuremath{ \mathcal{Z}^{k} }}
\newcommand{\ZZko}{\ensuremath{ \mathcal{Z}^{k+1} }}
\newcommand{\lZZk}{\ensuremath{ \mathcal{Z}_{l}^{k} }}
\newcommand{\ZZoN}{\ensuremath{ \mathcal{Z}^{N-1} }}
\newcommand{\ZZtwN}{\ensuremath{ \mathcal{Z}^{N-2} }}
\newcommand{\ZZi}{\ensuremath{ \mathcal{Z}^{i} }}
\newcommand{\ZZok}{\ensuremath{ \mathcal{Z}^{k-1} }}

% SCALAR FUNCTIONS SYMBOLS -----------------------------------------------------

\newcommand{\fx}{\ensuremath{ f\left(\xv\right)  }}
\newcommand{\fq}{\ensuremath{ f\left(\qv\right)  }}

% COST FUNCTIONS SYMBOLS --------------------------------------------------

\newcommand{\dhat}{\ensuremath{ \,\widehat{D}  }}
\newcommand{\dhato}{\ensuremath{ \dhat_{_1}  }}

\newcommand{\fNqN}{\ensuremath{ \fN \left( \qN \right) }}
\newcommand{\fkqk}{\ensuremath{ \fk \left( \qk \right) }}
\newcommand{\foNqoN}{\ensuremath{ \foN \left( \qoN \right) }}
\newcommand{\fzqz}{\ensuremath{ \fz \left( \qz\right) }}

\newcommand{\fstar}{\ensuremath{ \,{F}^{\star}  }}
\newcommand{\fstarz}{\ensuremath{ \fstar_{_0}  }}
\newcommand{\fstarzqz}{\ensuremath{ \fstarz \left( \qz \right)  }}
\newcommand{\fstarok}{\ensuremath{ \fstar_{_{k-1}}  }}
\newcommand{\fstarokqok}{\ensuremath{ \fstarok \left( \qok \right)  }}
\newcommand{\fstark}{\ensuremath{ \fstar_{_k}  }}
\newcommand{\fstarkqk}{\ensuremath{ \fstark \left( \qk \right)  }}
\newcommand{\fstaroktwk}{\ensuremath{ \fstar_{_{k-1,k-2}}  }}

\newcommand{\fhat}{\ensuremath{ \,\widehat{F}  }}
\newcommand{\fhatz}{\ensuremath{ \fhat_{_0}  }}
\newcommand{\fhatok}{\ensuremath{ \fhat_{_{k-1}}  }}
\newcommand{\fhatk}{\ensuremath{ \fhat_{_k}  }}
\newcommand{\fhatoktwk}{\ensuremath{ \fhat_{_{k-1,k-2}}  }}

\newcommand{\gz}{\ensuremath{ \,G_{_0}  }}
\newcommand{\gN}{\ensuremath{ \,G_{_N}  }}
\newcommand{\goN}{\ensuremath{ \,G_{_{N-1}}  }}
\newcommand{\gsk}{\ensuremath{ \,G_{_k}  }}
\newcommand{\gkqk}{\ensuremath{ \gsk \left( \qk \right)  }}
\newcommand{\goktwk}{\ensuremath{ \,G_{_{k-1,k-2}}  }}
\newcommand{\gokok}{\ensuremath{ \,G_{_{k-1,k-1}}  }}
\newcommand{\gNqN}{\ensuremath{ \gN \left( \qN \right) }}

\newcommand{\gstar}{\ensuremath{ \,{G}^{\star}  }}
\newcommand{\gstark}{\ensuremath{ \gstar_{_k}  }}
\newcommand{\gstarkqk}{\ensuremath{ \gstark \left( \qk \right)  }}

\newcommand{\ghat}{\ensuremath{ \,\widehat{G}  }}
\newcommand{\ghatk}{\ensuremath{ \ghat_{_k}  }}

\newcommand{\hN}{\ensuremath{ \,H_{_N}  }}
\newcommand{\hk}{\ensuremath{ \,H_{_k}  }}
\newcommand{\hokok}{\ensuremath{ \,H_{_{k-1,k-1}}  }}
\newcommand{\hKok}{\ensuremath{ \,H_{_{k,k-1}}  }}

\newcommand{\hstar}{\ensuremath{ \,\star{H}  }}
\newcommand{\hstarokok}{\ensuremath{ \hstar_{_{k-1,k-1}}  }}
\newcommand{\hstarKok}{\ensuremath{ \hstar_{_{k,k-1}}  }}

\newcommand{\hhat}{\ensuremath{ \,\widehat{H}  }}
\newcommand{\hhatz}{\ensuremath{ \hhat_{_0}  }}
\newcommand{\hhato}{\ensuremath{ \hhat_{_1}  }}
\newcommand{\hhatokok}{\ensuremath{ \hhat_{_{k-1,k-1}}  }}
\newcommand{\hhatKok}{\ensuremath{ \hhat_{_{k,k-1}}  }}

\newcommand{\iN}{\ensuremath{ \,I_{_N}  }}
\newcommand{\iKok}{\ensuremath{ \,I_{_{k,k-1}}  }}

\newcommand{\istar}{\ensuremath{ \,\star{I}  }}
\newcommand{\istarKok}{\ensuremath{ \istar_{_{k,k-1}}  }}

\newcommand{\ihat}{\ensuremath{ \,\widehat{I}  }}
\newcommand{\ihatKok}{\ensuremath{ \ihat_{_{k,k-1}}  }}

\newcommand{\Jko}{\ensuremath{ J_{_{k+1}}  }}
\newcommand{\Jzz}{\ensuremath{ \,J_{_{0/0}}  }}
\newcommand{\Joz}{\ensuremath{ \,J_{_{1/0}}  }}
\newcommand{\Joo}{\ensuremath{ \,J_{_{1/1}}  }}
\newcommand{\Je}{\ensuremath{  \,\widehat{J}}}
\newcommand{\Jezz}{\ensuremath{ \,\Je_{_{0/0}}  }}
\newcommand{\Jeoz}{\ensuremath{ \,\Je_{_{1/0}}  }}
\newcommand{\Jeoo}{\ensuremath{ \,\Je_{_{1/1}}  }}
\newcommand{\Jekk}{\ensuremath{ \,\Je_{_{k/k}}  }}
\newcommand{\Jeokok}{\ensuremath{ \,\Je_{_{k-1/k-1}}  }}
\newcommand{\Jstar}{\ensuremath{  \,\star{J}}}
\newcommand{\Jzzstar}{\ensuremath{ \,\Jstar_{_{0/0}}   }}
\newcommand{\Jozstar}{\ensuremath{ \,\Jstar_{_{1/0}}   }}
\newcommand{\Joostar}{\ensuremath{ \,\Jstar_{_{1/1}}   }}
\newcommand{\Jbar}{\ensuremath{  {  \,\bar{J}} }}
\newcommand{\Joobar}{\ensuremath{  \,\Jbar_{{1/1}} }}
\newcommand{\Joobarstar}{\ensuremath{ \,\Joobar^\star    }}
\newcommand{\Jeoobar}{\ensuremath{  \,\widehat{\Joobar} }}
\newcommand{\JNN}{\ensuremath{ \,J_{_{N/N}}  }}
\newcommand{\JNoN}{\ensuremath{ J_{_{N+1/N}}  }}
\newcommand{\JNoNo}{\ensuremath{ J_{_{N+1/N+1}}  }}
\newcommand{\Jc}{\ensuremath{ \,J_{_{c}}  }}
\newcommand{\Jo}{\ensuremath{ \,J_{_{o}}  }}

\newcommand{\JN}{\ensuremath{ J_{_{N}}  }}
\newcommand{\JoN}{\ensuremath{ J_{_{N-1}}  }}
\newcommand{\JtwN}{\ensuremath{ J_{_{N-2}}  }}
\newcommand{\Jk}{\ensuremath{ J_{_{k}}  }}

\newcommand{\Jast}{\ensuremath{ J^{\ast} }}
\newcommand{\JNast}{\ensuremath{ \JN^{\ast} }}
\newcommand{\JoNast}{\ensuremath{ \JoN^{\ast} }}
\newcommand{\JtwNast}{\ensuremath{ \JtwN^{\ast} }}
\newcommand{\Jkast}{\ensuremath{ \Jk^{\ast} }}
\newcommand{\Jkoast}{\ensuremath{ \Jko^{\ast} }}

% MATRIX SYMBOLS -----------------------------------------

\newcommand{\across}{\ensuremath{ \,\left[ \av \times \right] }}

\newcommand{\AAA}{\ensuremath{ \mathcal{A} }}
\newcommand{\AAz}{\ensuremath{ \AAA_{_{0}} }}
\newcommand{\AAone}{\ensuremath{ \AAA_{_{1}} }}
\newcommand{\AAoneone}{\ensuremath{ \AAA_{_{11}} }}
\newcommand{\AAonetwo}{\ensuremath{ \AAA_{_{12}} }}
\newcommand{\AAtwo}{\ensuremath{ \AAA_{_{2}} }}
\newcommand{\AAthr}{\ensuremath{ \AAA_{_{3}} }}
\newcommand{\AAfor}{\ensuremath{ \AAA_{_{4}} }}

\newcommand{\Aml}{\ensuremath{{ \,\widehat{A}{_{_{ML}}} }}}
\newcommand{\At}{\ensuremath{{ \,A \left( t \right)  }}}
\newcommand{\Ak}{\ensuremath{ A_{_{k}}  }}
\newcommand{\AoN}{\ensuremath{ A_{_{N-1}}  }}
\newcommand{\Ae}{\ensuremath{  \widehat{A} }}
\newcommand{\Akk}{\ensuremath{  \Ae_{_{k/k}} }}
\newcommand{\Aet}{\ensuremath{  \Ae \left( t \right) }}
\newcommand{\Aetp}{\ensuremath{   \Ae \left( t' \right) }}
\newcommand{\Ako}{\ensuremath{ \,A_{_{k+1}}  }}
\newcommand{\Atp}{\ensuremath{{ \,A \left( t' \right)  }}}
\newcommand{\Atrue}{\ensuremath{ \,A^{\mbox{o}}  }}
\newcommand{\Au}{\ensuremath{ \,A^{u} }}
\newcommand{\Aoo}{\ensuremath{ \,A^{11} }}
\newcommand{\Aotw}{\ensuremath{ \,A^{12} }}
\newcommand{\Aol}{\ensuremath{ \,A^{1l} }}
\newcommand{\Aoifo}{\ensuremath{ \,A^{1\ifo} }}
\newcommand{\Atwo}{\ensuremath{ \,A^{21} }}
\newcommand{\Atwtw}{\ensuremath{ \,A^{22} }}
\newcommand{\Atwl}{\ensuremath{ \,A^{2l} }}
\newcommand{\Atwifo}{\ensuremath{ \,A^{2\ifo} }}
\newcommand{\Ajo}{\ensuremath{ \,A^{j1} }}
\newcommand{\Ajtw}{\ensuremath{ \,A^{j2} }}
\newcommand{\Ajl}{\ensuremath{ \,A^{jl} }}
\newcommand{\Ajifo}{\ensuremath{ \,A^{j\ifo} }}
\newcommand{\Aitho}{\ensuremath{ \,A^{\ith1} }}
\newcommand{\Aithtw}{\ensuremath{ \,A^{\ith2} }}
\newcommand{\Aithl}{\ensuremath{ \,A^{\ith l} }}
\newcommand{\Aithifo}{\ensuremath{ \,A^{\ith \ifo} }}
\newcommand{\Aitwo}{\ensuremath{ \,A^{\itw1} }}
\newcommand{\Aitwtw}{\ensuremath{ \,A^{\itw2} }}
\newcommand{\Aitwl}{\ensuremath{ \,A^{\itw l} }}
\newcommand{\Aitwifo}{\ensuremath{ \,A^{\itw \ifo} }}
\newcommand{\Atld}{\ensuremath{  \widetilde{A} }}
\newcommand{\Atldkk}{\ensuremath{  \Atld_{_{k/k}} }}

\newcommand{\asak}{\ensuremath{ \Ak^T \Sko \Ak }}
\newcommand{\astlak}{\ensuremath{ \Ak^T \Stlko \Ak }}
\newcommand{\aqak}{\ensuremath{ \Ak^T \Qk \Ak }}
\newcommand{\asbk}{\ensuremath{ \Ak^T \Sko \Bk }}
\newcommand{\asaoN}{\ensuremath{ \AoN^T \SN \AoN }}
\newcommand{\asboN}{\ensuremath{ \AoN^T \SN \BoN }}
\newcommand{\bsoN}{\ensuremath{ \BoN^T \SN}}
\newcommand{\asoNm}{\ensuremath{ \AoN^T \SN }}

\newcommand{\BB}{\ensuremath{ \mathcal{B} }}
\newcommand{\BBone}{\ensuremath{ \BB_{_{1}}  }}

\newcommand{\Be}{\ensuremath{ \widehat{B} }}
\newcommand{\Bkk}{\ensuremath{{ \Be_{_{k/k}}  }}}
\newcommand{\Beps}{\ensuremath{{ \,B_{_\epsilon} }}}
\newcommand{\Bb}{\ensuremath{{ \,B_{_b} }}}
\newcommand{\Bt}{\ensuremath{{ \,B \left( t \right)  }}}
\newcommand{\Btp}{\ensuremath{{ \,B \left( t' \right)  }}}
\newcommand{\Btx}{\ensuremath{{ \,B \left( t , \xv\right)  }}}
\newcommand{\Bk}{\ensuremath{ B_{_{k}}  }}
\newcommand{\BoN}{\ensuremath{ B_{_{N-1}}  }}
\newcommand{\Bko}{\ensuremath{ \,B_{_{k+1}}  }}
\newcommand{\dBko}{\ensuremath{{ \,\delta B_{_{k+1}}  }}}
\newcommand{\Bkot}{\ensuremath{{ \,B_{_{k+1}}^{o}  }}}
\newcommand{\Bkkt}{\ensuremath{{ \,B_{_{k/k}}^{o} }}}
\newcommand{\Bkok}{\ensuremath{{ \,B_{_{k+1/k}}  }}}
\newcommand{\Bkoko}{\ensuremath{{ \,B_{_{k+1/k+1}}  }}}
\newcommand{\bicross}{\ensuremath{ \left[ {\bf b}_{_i} \times \right]  }}
\newcommand{\bcross}{\ensuremath{ \left[ \bv \times \right]  }}
\newcommand{\Bbn}{\ensuremath{ \Bb^n }}
\newcommand{\Bu}{\ensuremath{ \,B^{u} }}
\newcommand{\Bbi}{\ensuremath{{ \,B_{_{b_i}} }}}
\newcommand{\Btld}{\ensuremath{  \widetilde{B} }}
\newcommand{\Btldkk}{\ensuremath{  \Btld_{_{k/k}} }}

\newcommand{\bsbrk}{\ensuremath{ \Bk^T \Sko \Bk + \Rk }}
\newcommand{\bstlbk}{\ensuremath{ \Bk^T \Stlko \Bk}}
\newcommand{\bqbk}{\ensuremath{ \Bk^T \Qk \Bk}}
\newcommand{\bsak}{\ensuremath{ \Bk^T \Sko \Ak  }}
\newcommand{\bstlak}{\ensuremath{ \Bk^T \Stlko \Ak  }}
\newcommand{\bqak}{\ensuremath{ \Bk^T \Qk \Ak  }}
\newcommand{\bsbroN}{\ensuremath{ \BoN^T \SN \BoN + \RoN }}
\newcommand{\bsaoN}{\ensuremath{ \BoN^T \SN \AoN  }}

\newcommand{\CC}{\ensuremath{ \mathcal{C} }}
\newcommand{\CCone}{\ensuremath{ \CC_{_{1}}  }}

\newcommand{\Coo}{\ensuremath{{ \,C{_{_{11}}} }}}
\newcommand{\Cotw}{\ensuremath{ \cotw }}
\newcommand{\Ctwo}{\ensuremath{ \ctwo }}
\newcommand{\Ctwtw}{\ensuremath{ \cstwtw }}

\newcommand{\Cl}{\ensuremath{  \,\mbox{Cl} }}
\newcommand{\Clo}{\ensuremath{  \Cl_{_1} }}
\newcommand{\Cltw}{\ensuremath{  \Cl_{_2} }}
\newcommand{\Clth}{\ensuremath{  \Cl_{_3} }}
\newcommand{\Clf}{\ensuremath{  \Cl_{_4} }}
\newcommand{\Clfv}{\ensuremath{  \Cl_{_5} }}
\newcommand{\Clsx}{\ensuremath{  \Cl_{_6} }}
\newcommand{\Clsv}{\ensuremath{  \Cl_{_7} }}
\newcommand{\Clei}{\ensuremath{  \Cl_{_8} }}
\newcommand{\Clni}{\ensuremath{  \Cl_{_9} }}
\newcommand{\Clten}{\ensuremath{  \Cl_{_{10}} }}
\newcommand{\Clel}{\ensuremath{  \Cl_{_{11}} }}
\newcommand{\Cltwl}{\ensuremath{  \Cl_{_{12}} }}
\newcommand{\Clthr}{\ensuremath{  \Cl_{_{13}} }}
\newcommand{\Clfrt}{\ensuremath{  \Cl_{_{14}} }}
\newcommand{\Clfft}{\ensuremath{  \Cl_{_{15}} }}
\newcommand{\Clsxt}{\ensuremath{  \Cl_{_{16}} }}

\newcommand{\ckcross}{\ensuremath{ \left[ \ck \times \right]  }}

\newcommand{\DD}{\ensuremath{ \mathcal{D} }}
\newcommand{\DDone}{\ensuremath{ \DD_{_{1}}  }}

\newcommand{\dbicross}{\ensuremath{ \left[ \delta {\bf b}_{_i} \times \right]  }}
\newcommand{\dbjcross}{\ensuremath{ \left[ \delta {\bf b}_{_j} \times \right]  }}
\newcommand{\dbkocross}{\ensuremath{ \left[ \dbko \times \right]  }}
\newcommand{\dBcal}{\ensuremath{ \delta \mathcal{B} }}
\newcommand{\dBcali}{\ensuremath{ \dBcal_{_i} }}

\newcommand{\DX}{\ensuremath{  \,\Delta X }}
\newcommand{\DXkk}{\ensuremath{  \DX_{_{k/k}} }}
\newcommand{\DXkok}{\ensuremath{  \DX_{_{k+1/k}} }}
\newcommand{\DXkoko}{\ensuremath{  \DX_{_{k+1/k+1}} }}
\newcommand{\DXzz}{\ensuremath{  \DX_{_{0/0}} }}
\newcommand{\DXoneone}{\ensuremath{  \DX_{_{11}} }}
\newcommand{\DXonetwo}{\ensuremath{  \DX_{_{12}} }}
\newcommand{\DXonethr}{\ensuremath{  \DX_{_{13}} }}
\newcommand{\DXonefor}{\ensuremath{  \DX_{_{14}} }}
\newcommand{\DXtwoone}{\ensuremath{  \DX_{_{21}} }}
\newcommand{\DXtwotwo}{\ensuremath{  \DX_{_{22}} }}
\newcommand{\DXthrthr}{\ensuremath{  \DX_{_{33}} }}
\newcommand{\DXforfor}{\ensuremath{  \DX_{_{44}} }}
\newcommand{\Dx}{\ensuremath{ \,\boldsymbol{\Delta}\xv }}
\newcommand{\Dxkk}{\ensuremath{  \Dx_{_{k/k}} }}
\newcommand{\Dxkok}{\ensuremath{  \Dx_{_{k+1/k}} }}
\newcommand{\Dxkoko}{\ensuremath{  \Dx_{_{k+1/k+1}} }}

\newcommand{\Dk}{\ensuremath{  D_{_{k}} }}
\newcommand{\Dko}{\ensuremath{  D_{_{k+1}} }}
\newcommand{\Dz}{\ensuremath{  D_{_{0}} }}
\newcommand{\De}{\ensuremath{  \widehat{D} }}
\newcommand{\Dzz}{\ensuremath{  \De_{_{0/0}}  }}
\newcommand{\Dkk}{\ensuremath{  \De_{_{k/k}}  }}
\newcommand{\Dkok}{\ensuremath{  \De_{_{k+1/k}}  }}
\newcommand{\Dkoko}{\ensuremath{  \De_{_{k+1/k+1}}  }}
\newcommand{\Dast}{\ensuremath{  D^{\ast} }}
\newcommand{\Dastkoko}{\ensuremath{  \Dast_{_{k+1/k+1}} }}
\newcommand{\Dastkk}{\ensuremath{  \Dast_{_{k/k}} }}

\newcommand{\DEE}{\ensuremath{  \Delta \EE}}

\newcommand{\dkonecross}{\ensuremath{ \,\left[ \dkone \times \right] }}
\newcommand{\dktwocross}{\ensuremath{ \,\left[ \dktwo \times \right] }}
\newcommand{\dkthrcross}{\ensuremath{ \,\left[ \dkthr \times \right] }}

\newcommand{\Eij}{\ensuremath{ \, E^{ij} }}
\newcommand{\Ekl}{\ensuremath{ \, E^{kl} }}
\newcommand{\Eik}{\ensuremath{ \, E^{ik} }}
\newcommand{\Elj}{\ensuremath{ \, E^{lj} }}
\newcommand{\Eji}{\ensuremath{ \, E^{ji} }}
\newcommand{\Ejj}{\ensuremath{ \, E^{jj} }}
\newcommand{\Eoneone}{\ensuremath{ \, E^{11} }}
\newcommand{\Etwotwo}{\ensuremath{ \, E^{22} }}
\newcommand{\Ethrthr}{\ensuremath{ \, E^{33} }}
\newcommand{\Eforfor}{\ensuremath{ \, E^{44} }}
\newcommand{\Eforone}{\ensuremath{ \, E^{41} }}
\newcommand{\Efortwo}{\ensuremath{ \, E^{42} }}
\newcommand{\Eforthr}{\ensuremath{ \, E^{43} }}

\newcommand{\ekcross}{\ensuremath{ \,\left[ \ek \times \right] }}
\newcommand{\ektcross}{\ensuremath{ \,\left[ \ekt \times \right] }}
\newcommand{\ekocross}{\ensuremath{ \,\left[ \eko \times \right] }}
\newcommand{\ecross}{\ensuremath{ \,\left[ \ev \times \right] }}
\newcommand{\eecross}{\ensuremath{ \,\left[ \ee \times \right] }}
\newcommand{\eekcross}{\ensuremath{ \,\left[ \eek \times \right] }}
\newcommand{\eekocross}{\ensuremath{ \,\left[ \eeko \times \right] }}
\newcommand{\eekokcross}{\ensuremath{ \,\left[ \eekok \times \right] }}
\newcommand{\eekokocross}{\ensuremath{ \,\left[ \eekoko \times \right] }}
\newcommand{\eekolkcross}{\ensuremath{ \,\left[ \eekolk \times \right] }}
\newcommand{\eeklkcross}{\ensuremath{ \,\left[ \eeklk \times \right] }}
\newcommand{\eonecross}{\ensuremath{ \,\left[ \eone \times \right] }}
\newcommand{\etwocross}{\ensuremath{ \,\left[ \etwo \times \right] }}
\newcommand{\ethreecross}{\ensuremath{ \,\left[ \ethree \times \right] }}

\newcommand{\Epsk}{\ensuremath{{ \,\mathcal{E}_{_k}}}}
\newcommand{\Epsko}{\ensuremath{{ \,\mathcal{E}_{_{k+1}}}}}
\newcommand{\Eps}{\ensuremath{ \,\mathcal{E} }}
\newcommand{\epscross}{\ensuremath{ \,\left[ \eps \times \right] }}
\newcommand{\epskocross}{\ensuremath{ \,\left[ \epsko \times \right] }}
\newcommand{\epskcross}{\ensuremath{ \,\left[ \epsk \times \right] }}

\newcommand{\EE}{\ensuremath{ \mathcal{E} }}
\newcommand{\EEz}{\ensuremath{ \EE_{_{0}}  }}
\newcommand{\EEone}{\ensuremath{ \EE_{_{1}}  }}

\newcommand{\FF}{\ensuremath{ \mathcal{F} }}
\newcommand{\FFone}{\ensuremath{ \FF_{_{1}}  }}
\newcommand{\Ftt}{\ensuremath{ \,F_{_{\theta \theta}}  }}
\newcommand{\FFk}{\ensuremath{ \mathcal{F}_{_k}  }}
\newcommand{\Fi}{\ensuremath{ F_{_{i}} }}
\newcommand{\Fko}{\ensuremath{ F_{_{k+1}} }}

\newcommand{\Fbar}{\ensuremath{ \overline{F} }}
\newcommand{\Fbarbar}{\ensuremath{ \overline{\Fbar} }}
\newcommand{\Fbarbarbar}{\ensuremath{ \overline{\Fbarbar} }}
\newcommand{\Fbartilda}{\ensuremath{ \widetilde{\Fbar} }}

\newcommand{\Fik}{\ensuremath{   \,\Phi_{_k} }}
\newcommand{\FiN}{\ensuremath{   \,\Phi_{_N} }}
\newcommand{\Fiko}{\ensuremath{   \,\Phi_{_{k+1}} }}
\newcommand{\Fikt}{\ensuremath{   {\Phi_{_k}^{o}} }}
\newcommand{\Fiqk}{\ensuremath{   \,\Phi_{4_k} }}
\newcommand{\Fiz}{\ensuremath{   \,\Phi_{_0} }}
\newcommand{\dFik}{\ensuremath{  \,\Delta \Phi_{_k} }}
\newcommand{\Fitt}{\ensuremath{  \, \Phi \left( t' , t \right)   }}
\newcommand{\fikcross}{\ensuremath{ \,\left[ \fik \times \right] }}
\newcommand{\Fikol}{\ensuremath{ \, \Phi_{_{k+l-1}} }}
\newcommand{\Fiok}{\ensuremath{{   \,\Phi_{_{k-1}}  }}}
\newcommand{\FioN}{\ensuremath{   \,\Phi_{_{N-1}}  }}
\newcommand{\Fiktrue}{\ensuremath{{  \,\Phi^{o}_{_k}}}}
\newcommand{\Fie}{\ensuremath{   \,\widehat{\Phi} }}
\newcommand{\Fiekk}{\ensuremath{  \Fie_{_{k/k}}  }}
\newcommand{\Fieokok}{\ensuremath{  \Fie_{_{k-1/k-1}}  }}
\newcommand{\Fiezz}{\ensuremath{  \Fie_{_{0/0}}  }}

\newcommand{\Fisxtk}{\ensuremath{   \Fik^{16} }}
\newcommand{\Fitenk}{\ensuremath{   \Fik^{10} }}
\newcommand{\Finink}{\ensuremath{   \Fik^{9} }}

\newcommand{\fitm}{\ensuremath{{ \Phi}^o}}
\newcommand{\fim}{\ensuremath{{ \Phi}}}
\newcommand{\fierm}{\ensuremath{{\Delta \Phi}}}
\newcommand{\fikrm}{\ensuremath{\fim_{[2]_k} }}

\newcommand{\Gam}{\ensuremath{ \,\Gamma }}
\newcommand{\Gamk}{\ensuremath{ \Gam_{_{k}} }}
\newcommand{\Gamak}{\ensuremath{ \Gam_{_{k}} }}
\newcommand{\Gamae}{\ensuremath{ \widehat{\Gam} }}
\newcommand{\Gamekk}{\ensuremath{ \Gamae_{_{k/k}} }}
\newcommand{\Gamek}{\ensuremath{ \Gamae_{_{k}} }}

\newcommand{\Gamsxtk}{\ensuremath{ \Gamk^{16} }}
\newcommand{\Gamtenk}{\ensuremath{ \Gamk^{10} }}
\newcommand{\Gamnink}{\ensuremath{ \Gamk^{9} }}

\newcommand{\Gbar}{\ensuremath{ \overline{G} }}

\newcommand{\Gk}{\ensuremath{ G_{_{k}} }}
\newcommand{\Gko}{\ensuremath{ G_{_{k+1}} }}
\newcommand{\GWLS}{\ensuremath{ \,G^{_{_{WLS}}} }}
\newcommand{\GWLSko}{\ensuremath{ \GWLS_{_{k+1}} }}
\newcommand{\Gkos}{\ensuremath{ \Gko^s }}
\newcommand{\Gi}{\ensuremath{ G_{_{i}} }}
\newcommand{\GoN}{\ensuremath{ G_{_{N-1}} }}

\newcommand{\GG}{\ensuremath{ \mathcal{G} }}
\newcommand{\GGz}{\ensuremath{ \GG_{_{0}}  }}
\newcommand{\GGone}{\ensuremath{ \GG_{_{1}}  }}

\newcommand{\Hbar}{\ensuremath{ \overline{H} }}

\newcommand{\Ht}{\ensuremath{ H^{o} }}
\newcommand{\Hko}{ \,H_{_{k+1}} }
\newcommand{\Hkos}{\ensuremath{ \Hko^s }}
\newcommand{\Hk}{\ensuremath{ \,H_{_{k}} }}
\newcommand{\Hz}{\ensuremath{ \,H_{_{0}} }}
\newcommand{\Hzz}{\ensuremath{ \,H^{0}_{_{0}} }}
\newcommand{\Hzo}{\ensuremath{ \,H^{1}_{_{0}} }}
\newcommand{\Ho}{\ensuremath{ \,H_{_{1}} }}
\newcommand{\HN}{\ensuremath{ \,H_{_{N}} }}
\newcommand{\HNo}{\ensuremath{ \,H_{_{N+1}} }}
\newcommand{\Hktw}{\ensuremath{ \,H_{_{k+2}} }}
\newcommand{\Hkl}{\ensuremath{ \,H_{_{k+l}} }}
\newcommand{\Hkot}{\ensuremath{ \Hko^{o} }}
\newcommand{\Herko}{\ensuremath{ \,\Delta H_{_{k+1}} }}
\newcommand{\Hglobal}{\ensuremath{ \,\mathcal{H} }}
\newcommand{\Hglobaloo}{\ensuremath{ \Hglobal_{_{1/1}} }}
\newcommand{\HglobalNN}{\ensuremath{ \Hglobal_{_{N/N}} }}
\newcommand{\HglobalNoN}{\ensuremath{ \Hglobal_{_{N+1/N}} }}
\newcommand{\HglobalNoNo}{\ensuremath{ \Hglobal_{_{N+1/N+1}} }}
\newcommand{\Hglobalkk}{\ensuremath{ \Hglobal_{_{k/k}} }}
\newcommand{\Hglobalkok}{\ensuremath{ \Hglobal_{_{k+1/k}} }}
\newcommand{\Hglobalkoko}{\ensuremath{ \Hglobal_{_{k+1/k+1}} }}
\newcommand{\Hglobalzz}{\ensuremath{ \Hglobal_{_{0/0}} }}

\newcommand{\HH}{\ensuremath{\,\mathcal{H}}}
\newcommand{\HHko}{ \HH_{_{k+1}} }
\newcommand{\HHtr}{\ensuremath{ \hv^{tr} }}
\newcommand{\HHsym}{\ensuremath{ \HH^{sym} }}

\newcommand{\Ithree}{\ensuremath{ \,I_{_3}  }}
\newcommand{\Ifour}{\ensuremath{  \,I_{_4}  }}
\newcommand{\Isix}{\ensuremath{  \,I_{_6}  }}
\newcommand{\Inine}{\ensuremath{  \,I_{_9}  }}
\newcommand{\Imn}{\ensuremath{  \,I_{_{mn}}  }}
\newcommand{\Isixteen}{\ensuremath{  \,I_{_{16}}  }}
\newcommand{\In}{\ensuremath{  \,I_{_{n}}  }}
\newcommand{\Ip}{\ensuremath{  \,I_{_{p}}  }}
\newcommand{\Inu}{\ensuremath{ \,I_{_\nu}  }}

\newcommand{\Kk}{\ensuremath{ K_{_k} }}
\newcommand{\Kko}{\ensuremath{ K_{_{k+1}} }}
\newcommand{\Kepsk}{\ensuremath{{ K_{_{\epsk}} }}}
\newcommand{\Kbko}{\ensuremath{{ K_{_{b_{k+1}}} }}}
\newcommand{\Kbk}{\ensuremath{{ K_{_{b_{k}}} }}}
\newcommand{\Kkk}{\ensuremath{{ \,K{_{_{k/k}}}  }}}
\newcommand{\Kij}{\ensuremath{{ \,K{_{_{i/j}}}  }}}
\newcommand{\Kzz}{\ensuremath{{ \,K{_{_{0/0}}}  }}}
\newcommand{\KNN}{\ensuremath{{ \,K{_{_{N/N}}}  }}}
\newcommand{\Kkok}{\ensuremath{{ \,K_{_{k+1/k}}  }}}
\newcommand{\Kkoko}{\ensuremath{{ \,K_{_{k+1/k+1}}  }}}
\newcommand{\Kbar}{\ensuremath{ \overline{K} }}
\newcommand{\Kbarko}{\ensuremath{ \Kbar_{_{k+1}} }}

\newcommand{\Kt}{\ensuremath{ \,K^o  }}
\newcommand{\Kijt}{\ensuremath{{ \,K_{_{i/j}}^{o}  }}}
\newcommand{\Ktij}{\ensuremath{ \Kt_{_{i/j}}  }}
\newcommand{\Kzzt}{\ensuremath{ \Kt_{_{0/0}} }}
\newcommand{\Kkokt}{\ensuremath{ \Kt_{_{k+1/k}} }}
\newcommand{\Kkkt}{\ensuremath{ \Kt_{_{k/k}} }}
\newcommand{\Kkokot}{\ensuremath{ \Kt_{_{k+1/k+1}} }}
\newcommand{\Kkokap}{\ensuremath{ \Kkokt }}
\newcommand{\Kkkap}{\ensuremath{ \Kkkt }}
\newcommand{\Kkokoap}{\ensuremath{ \Kkokot }}
\newcommand{\Kzzap}{\ensuremath{ \Kzzt }}

\newcommand{\dK}{\ensuremath{ \,\delta K }}
\newcommand{\dKi}{\ensuremath{ \,\delta K_{_{i}} }}
\newcommand{\dKz}{\ensuremath{ \,\delta K_{_{0}} }}
\newcommand{\dKko}{\ensuremath{ \,\delta K_{_{k+1}} }}
\newcommand{\dKk}{\ensuremath{ \,\delta K_{_{k}} }}
\newcommand{\dKkot}{\ensuremath{ \delta K_{_{k+1}}^{o}  }}
\newcommand{\dKit}{\ensuremath{ \delta K_{_{i}}^{o}  }}
\newcommand{\dKt}{\ensuremath{  \dK^o }}
\newcommand{\dKti}{\ensuremath{ \dKt_{_i}  }}

\newcommand{\Kwerm}{\ensuremath{{{{\Delta K}^\varepsilon} }}}
\newcommand{\DKkokeps}{\ensuremath{ \Kwerm_{_{k+1/k}} }}
\newcommand{\Kweronem}{\ensuremath{ \Kwerm^{(1)}}}
\newcommand{\Kberm}{\ensuremath{{{{\Delta K}^b} }}}
\newcommand{\Ker}{\ensuremath{  \Delta K  }}
\newcommand{\Kerij}{\ensuremath{  \Ker_{_{ij}}  }}
\newcommand{\Kerkk}{\ensuremath{  \Ker_{_{k/k}}  }}
\newcommand{\Kerkok}{\ensuremath{  \Ker_{_{k+1/k}}  }}
\newcommand{\Kzerkok}{\ensuremath{  {\Kerkok^0} }}
\newcommand{\Kerkoko}{\ensuremath{  \Ker_{_{k+1/k+1}}  }}

\newcommand{\Ke}{\ensuremath{ \,\widehat{K} }}
\newcommand{\Kekk}{\ensuremath{ \Ke_{_{k/k}} }}
\newcommand{\Kjl}{\ensuremath{ K^{jl} }}
\newcommand{\Kjlko}{ \Kjl_{_{k+1}} }
\newcommand{\Kkojl}{ \Kjl_{_{k+1}} }

\newcommand{\KK}{\ensuremath{ \mathcal{K} }}
\newcommand{\KKko}{\ensuremath{ \KK_{_{k+1}} }}
\newcommand{\KKjl}{\ensuremath{ \KK^{jl} }}
\newcommand{\KKjlko}{\ensuremath{ \KKjl_{_{k+1}} }}
\newcommand{\KKj}{\ensuremath{ \KK^j }}
\newcommand{\KKjko}{\ensuremath{ \KKj_{_{k+1}} }}
\newcommand{\KKone}{\ensuremath{ \KK^1 }}
\newcommand{\KKtwo}{\ensuremath{ \KK^2 }}
\newcommand{\KKthree}{\ensuremath{ \KK^3 }}

\newcommand{\KW}{\ensuremath{ \,K^{_{_W} } }}
\newcommand{\KWko}{\ensuremath{ \KW_{_{k+1}} }}

\newcommand{\Lko}{\ensuremath{ \,L_{_{k+1}} }}
\newcommand{\Ltw}{\ensuremath{ \,L_{_{2}} }}
\newcommand{\LN}{\ensuremath{ \,L_{_{N}} }}
\newcommand{\Lone}{\ensuremath{ \,L_{_{1}} }}
\newcommand{\Lll}{\ensuremath{ \,L_{_{l}} }}
\newcommand{\Lol}{\ensuremath{ \,L_{_{l-1}} }}
\newcommand{\Lz}{\ensuremath{ \,L_{_{0}} }}

\newcommand{\Ll}{\ensuremath{  \,\mbox{Li} }}
\newcommand{\Llo}{\ensuremath{  \Ll_{_1} }}
\newcommand{\Lltw}{\ensuremath{  \Ll_{_2} }}
\newcommand{\Llth}{\ensuremath{  \Ll_{_3} }}
\newcommand{\Llf}{\ensuremath{  \Ll_{_4} }}
\newcommand{\Llfv}{\ensuremath{  \Ll_{_5} }}
\newcommand{\Llsx}{\ensuremath{  \Ll_{_6} }}
\newcommand{\Llsv}{\ensuremath{  \Ll_{_7} }}
\newcommand{\Llei}{\ensuremath{  \Ll_{_8} }}
\newcommand{\Llni}{\ensuremath{  \Ll_{_9} }}
\newcommand{\Llten}{\ensuremath{  \Ll_{_{10}} }}
\newcommand{\Llel}{\ensuremath{  \Ll_{_{11}} }}
\newcommand{\Lltwl}{\ensuremath{  \Ll_{_{12}} }}
\newcommand{\Llthr}{\ensuremath{  \Ll_{_{13}} }}
\newcommand{\Llfrt}{\ensuremath{  \Ll_{_{14}} }}
\newcommand{\Llfft}{\ensuremath{  \Ll_{_{15}} }}
\newcommand{\Llsxt}{\ensuremath{  \Ll_{_{16}} }}

\newcommand{\LL}{\ensuremath{ \mathcal{L} }}

\newcommand{\Lam}{\ensuremath{ \,\Lambda }}
\newcommand{\Lamko}{\ensuremath{ \Lam_{_{k+1}} }}
\newcommand{\Lame}{\ensuremath{ \widehat{\Lam} }}
\newcommand{\Lameko}{\ensuremath{ \Lame_{_{k+1}} }}
\newcommand{\Lami}{\ensuremath{ {\Lam^i} }}
\newcommand{\Lamiko}{\ensuremath{ {\Lami_{_{k+1}}} }}
\newcommand{\Lamj}{\ensuremath{ \Lam^j }}
\newcommand{\Lamjko}{\ensuremath{ {\Lamj_{_{k+1}}} }}

\newcommand{\Mk}{\ensuremath{ \,M_{_k} }}
\newcommand{\Mtw}{\ensuremath{ \,M_{_2} }}
\newcommand{\MN}{\ensuremath{ \,M_{_N} }}
\newcommand{\Mone}{\ensuremath{ \,M_{_1} }}
\newcommand{\Ml}{\ensuremath{ \,M_{_l} }}
\newcommand{\Mol}{\ensuremath{ \,M_{_{l-1}} }}
\newcommand{\Mz}{\ensuremath{ \,M_{_0} }}
\newcommand{\Mi}{\ensuremath{ M_{_{i}} }}

\newcommand{\Mbar}{\ensuremath{ \overline{M} }}

\newcommand{\Nol}{\ensuremath{ \,N_{_{l-1}} }}
\newcommand{\Ni}{\ensuremath{ N_{_{i}} }}

\newcommand{\Nbar}{\ensuremath{ \overline{N} }}

\newcommand{\onem}{\ensuremath{\,\widehat{R}}}
\newcommand{\Omg}{\ensuremath{ \,\Omega }}
\newcommand{\Omk}{\ensuremath{ \,\Omega_{_k} }}
\newcommand{\Omkt}{\ensuremath{ \,\Omega_{_k}^{o} }}
\newcommand{\omgcross}{\ensuremath{ \,\left[ \omg \times \right] }}
\newcommand{\omkcross}{\ensuremath{ \,\left[ \omk \times \right] }}
\newcommand{\omktcross}{\ensuremath{ \,\left[ \omkt \times \right] }}
\newcommand{\omekkcross}{\ensuremath{ \,\left[ \omekk \times \right] }}
\newcommand{\omgtcross}{\ensuremath{ \,\left[ \omgt \times \right] }}
\newcommand{\Ome}{\ensuremath{ \widehat{\Omg} }}
\newcommand{\Omekk}{\ensuremath{ \Ome_{_{k/k}}  }}

\newcommand{\Ofour}{\ensuremath{ \,O_{_4} }}
\newcommand{\Othree}{\ensuremath{ \,O_{_3} }}
\newcommand{\Osix}{\ensuremath{ \,O_{_6} }}
\newcommand{\Omn}{\ensuremath{ \,O_{_{mn}} }}
\newcommand{\Ombyn}{\ensuremath{ \,O_{_{m\times n}} }}
\newcommand{\Opbyq}{\ensuremath{ \,O_{_{p\times q}} }}
\newcommand{\Omnbypq}{\ensuremath{ \,O_{_{mn\times pq}} }}
\newcommand{\Oibyj}{\ensuremath{ \,O_{_{i\times j}} }}
\newcommand{\Oonebythr}{\ensuremath{ \,O_{_{1\times 3}} }}
\newcommand{\Otwth}{\ensuremath{ \,O_{_{2,3}} }}
\newcommand{\Otwbyth}{\ensuremath{ \,O_{_{2\times 3}} }}
\newcommand{\Otwfo}{\ensuremath{ \,O_{_{2,4}} }}
\newcommand{\Othrbyone}{\ensuremath{ \,O_{_{3\times 1}} }}
\newcommand{\Othrbytwo}{\ensuremath{ \,O_{_{3\times 2}} }}
\newcommand{\Othrbythr}{\ensuremath{ \,O_{_{3\times 3}} }}
\newcommand{\Othrbynin}{\ensuremath{ \,O_{_{3\times 9}} }}
\newcommand{\Oforbythr}{\ensuremath{ \,O_{_{4\times 3}} }}

\newcommand{\Piz}{\ensuremath{ \Pi{_{_{0}}} }}

\newcommand{\Pij}{\ensuremath{{ \,P{_{_{i/j}}} }}}
\newcommand{\Pkk}{\ensuremath{ \,P_{_{k/k}}  }}
\newcommand{\Pkok}{\ensuremath{ \,P_{_{k+1/k}} }}
\newcommand{\Pklk}{\ensuremath{{ \,P_{_{k+l/k}} }}}
\newcommand{\Pkolk}{\ensuremath{{ \,P_{_{k+l-1/k}} }}}
\newcommand{\Pkoko}{\ensuremath{ \,P_{_{k+1/k+1}} }}
\newcommand{\Past}{\ensuremath{ \,P^{\ast} }}
\newcommand{\Pkokoast}{\ensuremath{ \Past_{_{k+1/k+1}} }}
\newcommand{\Pkkast}{\ensuremath{ \Past_{_{k/k}} }}
\newcommand{\Pz}{\ensuremath{{ \,P_{_{0}} }}}
\newcommand{\Pzz}{\ensuremath{{ \,P_{_{0/0}} }}}
\newcommand{\Ptt}{\ensuremath{ \,P_{_{\theta \theta}}  }}
\newcommand{\Pxx}{\ensuremath{{ \,P{_{_{XX}}} }}}
\newcommand{\Pxxoo}{\ensuremath{ \Pxx_{_{11}}       }}
\newcommand{\Prr}{\ensuremath{{ \,P{_{_{rr}}} }}}
\newcommand{\Prrkok}{\ensuremath{ \Prr_{_{k+1/k}} }}
\newcommand{\Prrklk}{\ensuremath{ \Prr_{_{k+l/k}} }}
\newcommand{\Pmm}{\ensuremath{{ \,P{_{_{mm}}} }}}
\newcommand{\Pko}{\ensuremath{ \,P_{_{k+1}} }}
\newcommand{\Pk}{\ensuremath{ \,P_{_{k}} }}
\newcommand{\Pbar}{\ensuremath{ \overline{P} }}
\newcommand{\Pbarzz}{\ensuremath{ \Pbar_{_{0/0}} }}
\newcommand{\Pbarkk}{\ensuremath{ \Pbar_{_{k/k}} }}
\newcommand{\Pbarkok}{\ensuremath{ \Pbar_{_{k+1/k}} }}
\newcommand{\Pbarkoko}{\ensuremath{ \Pbar_{_{k+1/k+1}} }}
\newcommand{\Pb}{\ensuremath{ \overline{P} }}
\newcommand{\Pbz}{\ensuremath{ \Pb_{_{0}} }}
\newcommand{\Pbk}{\ensuremath{ \Pb_{_{k}} }}
\newcommand{\Pbko}{\ensuremath{ \Pb_{_{k+1}} }}
\newcommand{\Pftld}{\ensuremath{ P_{_{\ftld}} }}
\newcommand{\Pyftld}{\ensuremath{ P_{_{\yv\ftld}} }}
\newcommand{\Pxtld}{\ensuremath{ P_{_{\xtld}} }}
\newcommand{\Pytld}{\ensuremath{ P_{_{\ytld}} }}

\newcommand{\Per}{\ensuremath{ Per }}
\newcommand{\Perforfor}{\ensuremath{ \Per_{4,4} }}

\newcommand{\Pq}{\ensuremath{ \,P^{\qv} }}
\newcommand{\Pqkok}{\ensuremath{ \Pq_{_{k+1/k}} }}
\newcommand{\Pqkoko}{\ensuremath{ \Pq_{_{k+1/k+1}} }}
\newcommand{\Pqkk}{\ensuremath{ \Pq_{_{k/k}} }}

\newcommand{\Pee}{\ensuremath{ P_{_{ee}} }}
\newcommand{\Pzzz}{\ensuremath{ P_{_{zz}} }}

\newcommand{\Ppr}{\ensuremath{ \,P' }}
\newcommand{\Pprkk}{\ensuremath{ \Ppr_{_{k/k}} }}
\newcommand{\Pprklk}{\ensuremath{ \Ppr_{_{k+l/k}} }}
\newcommand{\Pprkolk}{\ensuremath{ \Ppr_{_{k+l-1/k}} }}

\newcommand{\Psik}{\ensuremath{ \,\Psi_{_k} }}
\newcommand{\Psil}{\ensuremath{ \,\Psi_{_l} }}
\newcommand{\Psikr}{\ensuremath{ \Psik^r }}
\newcommand{\Psikone}{\ensuremath{ \Psik^{1} }}
\newcommand{\Psiktwo}{\ensuremath{ \Psik^{2} }}
\newcommand{\Psikthr}{\ensuremath{ \Psik^{3} }}
\newcommand{\Psikfor}{\ensuremath{ \Psik^{4} }}
\newcommand{\Psikfiv}{\ensuremath{ \Psik^{5} }}

\newcommand{\Puu}{\ensuremath{{ \,P{_{_{uu}}} }}}
\newcommand{\Puukok}{\ensuremath{ \Puu_{_{k+1/k}} }}

\newcommand{\PP}{\ensuremath{ \mathcal{P} }}
\newcommand{\PPzz}{\ensuremath{ \PP_{_{0/0}} }}
\newcommand{\PPkk}{\ensuremath{ \PP_{_{k/k}} }}
\newcommand{\PPkok}{\ensuremath{ \PP_{_{k+1/k}} }}
\newcommand{\PPkoko}{\ensuremath{ \PP_{_{k+1/k+1}} }}

\newcommand{\PPpr}{\ensuremath{ \mathcal{P'} }}

\newcommand{\Qk}{\ensuremath{ Q_{_k} }}
\newcommand{\Qok}{\ensuremath{ Q_{_{k-1}} }}
\newcommand{\QoN}{\ensuremath{ Q_{_{N-1}} }}
\newcommand{\QtwN}{\ensuremath{ Q_{_{N-2}} }}
\newcommand{\Qi}{\ensuremath{ Q_{_i} }}
\newcommand{\QQ}{\ensuremath{ \,\mathcal{Q} }}
\newcommand{\QQk}{\ensuremath{ \,\mathcal{Q}{_{_k}} }}
\newcommand{\Qq}{\ensuremath{ \,Q^{q} }}
\newcommand{\Qqk}{\ensuremath{ \Qq_{_k} }}
\newcommand{\Qkq}{\ensuremath{ \Qq_{_k} }}
\newcommand{\Qqkol}{\ensuremath{ \Qq_{_{k+l-1}} }}
\newcommand{\Qoo}{\ensuremath{{ \,\mathcal{Q}{_{_{11}}} }}}
\newcommand{\Qot}{\ensuremath{{ \,\mathcal{Q}{_{_{12}}} }}}
\newcommand{\Qto}{\ensuremath{{ \,\mathcal{Q}{_{_{21}}} }}}
\newcommand{\Qtt}{\ensuremath{{ \,\mathcal{Q}{_{_{22}}} }}}
\newcommand{\Qwk}{\ensuremath{  \,Q_{_{w_k}}   }}
\newcommand{\Qe}{\ensuremath{ \widehat{Q} }}
\newcommand{\Qekol}{\ensuremath{ \Qe_{_{k+l-1}} }}
\newcommand{\Qbar}{\ensuremath{ \overline{Q} }}
\newcommand{\Qbark}{\ensuremath{ \Qbar_{_k} }}

\newcommand{\Qeps}{\ensuremath{ Q^{\eps} }}
\newcommand{\Qkeps}{\ensuremath{ \Qeps_{_k} }}
\newcommand{\Qepsk}{\ensuremath{ \Qeps_{_k} }}
\newcommand{\Qkleps}{\ensuremath{ \Qeps_{_{k+l}} }}

\newcommand{\Qeta}{\ensuremath{ Q^{\eta} }}
\newcommand{\Qetak}{\ensuremath{ \Qeta_{_{k}} }}

\newcommand{\Rk}{\ensuremath{ R_{_k} }}
\newcommand{\Rko}{\ensuremath{ R_{_{k+1}} }}
\newcommand{\RoN}{\ensuremath{ R_{_{N-1}} }}
\newcommand{\RtwN}{\ensuremath{ R_{_{N-2}} }}
\newcommand{\Rone}{\ensuremath{ \,R_{_1} }}
\newcommand{\Rtwo}{\ensuremath{ \,R_{_2} }}
\newcommand{\RR}{\ensuremath{ \,\mathcal{R} }}
\newcommand{\RRk}{\ensuremath{{ \,\mathcal{R}{_{_k}} }}}
\newcommand{\RRko}{\ensuremath{{ \,\mathcal{R}{_{_{k+1}}} }}}
\newcommand{\Rz}{\ensuremath{{ \,\mathcal{R}{_{_0}} }}}
\newcommand{\Rb}{\ensuremath{ \,R  }}
\newcommand{\Rbi}{\ensuremath{ {\Rb^i} }}
\newcommand{\Rkob}{\ensuremath{ \Rb_{_{k+1}} }}
\newcommand{\Rkb}{\ensuremath{ \Rb_{_{k}} }}
\newcommand{\Rkobi}{\ensuremath{ {\Rkob^i} }}
\newcommand{\Rktwb}{\ensuremath{ \Rb_{_{k+2}} }}
\newcommand{\Rklb}{\ensuremath{ \Rb_{_{k+l}} }}
\newcommand{\Rq}{\ensuremath{ \,R^{q} }}
\newcommand{\Rqko}{\ensuremath{ \Rq_{_{k+1}} }}
\newcommand{\Rqkl}{\ensuremath{ \Rq_{_{k+l}} }}
\newcommand{\Rkoq}{\ensuremath{ \Rqko }}
\newcommand{\Roo}{\ensuremath{{ \,\mathcal{R}{_{_{11}}} }}}
\newcommand{\Rot}{\ensuremath{{ \,\mathcal{R}{_{_{12}}} }}}
\newcommand{\Rto}{\ensuremath{{ \,\mathcal{R}{_{_{21}}} }}}
\newcommand{\Rtt}{\ensuremath{{ \,\mathcal{R}{_{_{22}}} }}}
\newcommand{\Rest}{\ensuremath{ \widehat{R} }}
\newcommand{\Rqest}{\ensuremath{ \widehat{\Rq} }}
\newcommand{\Rekl}{\ensuremath{ \Rest_{_{k+l}} }}
\newcommand{\Rqekl}{\ensuremath{ \Rqest_{_{k+l}} }}
\newcommand{\rkocross}
{\ensuremath{ \,\left[ \rko \times \right] }}
\newcommand{\Rbar}{\ensuremath{ \overline{R} }}
\newcommand{\Rbark}{\ensuremath{ \Rbar_{_k} }}
\newcommand{\Rbarko}{\ensuremath{ \Rbar_{_{k+1}} }}
\newcommand{\Rbarz}{\ensuremath{ \Rbar_{_0} }}
\newcommand{\Rv}{\ensuremath{ \,R_{_v} }}
\newcommand{\Rsym}{\ensuremath{ R^{sym} }}
\newcommand{\Rsymko}{\ensuremath{ \Rsym_{_{k+1}} }}
\newcommand{\Ritild}{\ensuremath{ \widetilde{R}_{_{i}} }}
\newcommand{\ricross}
{\ensuremath{ \,\left[ \ri \times \right] }}
\newcommand{\Ri}{\ensuremath{ R_{_i}  }}
\newcommand{\Rcal}{\ensuremath{ \mathcal{R} }}
\newcommand{\Rcali}{\ensuremath{ \Rcal_{_i} }}
\newcommand{\Rkone}{\ensuremath{ \,R_{_{k,1}} }}
\newcommand{\Rktwo}{\ensuremath{ \,R_{_{k,2}} }}

\newcommand{\Seps}{\ensuremath{{ \,S_{_\epsilon} }}}
\newcommand{\St}{\ensuremath{ S_{t} }}
\newcommand{\Ss}{\ensuremath{ S_{s} }}
\newcommand{\ST}{\ensuremath{ S_{T} }}
\newcommand{\Sk}{\ensuremath{ S_{_{k}} }}
\newcommand{\Sko}{\ensuremath{ S_{_{k+1}} }}
\newcommand{\Si}{\ensuremath{ S_{_{i}} }}
\newcommand{\Sio}{\ensuremath{ S_{_{i+1}} }}
\newcommand{\Sj}{\ensuremath{ S_{_{j}} }}
\newcommand{\SN}{\ensuremath{ S_{_{N}} }}
\newcommand{\SoN}{\ensuremath{ S_{_{N-1}} }}
\newcommand{\Skok}{\ensuremath{ \,S_{_{k+1/k}} }}
\newcommand{\Sklk}{\ensuremath{ \,S_{_{k+l/k}} }}
\newcommand{\Skolk}{\ensuremath{ \,S_{_{k+l-1/k}} }}
\newcommand{\dSko}{\ensuremath{ \,\delta S_{_{k+1}} }}
\newcommand{\Skot}{\ensuremath{ \,S_{_{k+1}}^{o} }}
\newcommand{\Sbb}{\ensuremath{{ \,\mathcal{S}{_{_b}} }}}
\newcommand{\scross}{\ensuremath{ \,\left[ \sv \times \right] }}
\newcommand{\stcross}{\ensuremath{ \,\left[ \st \times \right] }}
\newcommand{\skocross}{\ensuremath{ \,\left[ \sko \times \right] }}
\newcommand{\Sbbi}{\ensuremath{ {\Sbb^i} }}
\newcommand{\Sbbn}{\ensuremath{ {\Sbb^n} }}
\newcommand{\Sz}{\ensuremath{ \,S^{o} }}
\newcommand{\Sy}{\ensuremath{ S_{_{y}} }}
\newcommand{\Syv}{\ensuremath{ S_{_{\yv}} }}
\newcommand{\Salf}{\ensuremath{ S_{_{\alfa}} }}

\newcommand{\Sbar}{\ensuremath{ \overline{S} }}
\newcommand{\Sbarko}{\ensuremath{ \Sbar_{_{k+1}} }}

\newcommand{\SSm}{\ensuremath{ \mathcal{S} }}
\newcommand{\SSko}{\ensuremath{ \mathcal{S}_{_{k+1}} }}

\newcommand{\Stl}{\ensuremath{ \widetilde{S} }}
\newcommand{\Stlk}{\ensuremath{ \Stl_{_{k}} }}
\newcommand{\StloN}{\ensuremath{ \Stl_{_{N-1}} }}
\newcommand{\Stlko}{\ensuremath{ \Stl_{_{k+1}} }}
\newcommand{\StlN}{\ensuremath{ \Stl_{_{N}} }}

\newcommand{\Tone}{\ensuremath{ \,T_{_{1}} }}
\newcommand{\Ttw}{\ensuremath{ \,T_{_{2}} }}
\newcommand{\Tth}{\ensuremath{ \,T_{_{3}} }}
\newcommand{\Ttwo}{\ensuremath{ \,T_{_{2}} }}
\newcommand{\Tthr}{\ensuremath{ \,T_{_{3}} }}
\newcommand{\Tfor}{\ensuremath{ \,T_{_{4}} }}

\newcommand{\Tetako}{\ensuremath{ \,\Theta_{_{k+1}}    }}
\newcommand{\Tetakok}{\ensuremath{ \,\Theta_{_{k+1/k}} }}
\newcommand{\Tetak}{\ensuremath{ \,\Theta_{_{k}}    }}
\newcommand{\Tetakr}{\ensuremath{ \Tetak^r }}
\newcommand{\Tetakone}{\ensuremath{ \Tetak^{1} }}
\newcommand{\Tetaktwo}{\ensuremath{ \Tetak^{2} }}
\newcommand{\Tetakthr}{\ensuremath{ \Tetak^{3} }}
\newcommand{\Tetakfor}{\ensuremath{ \Tetak^{4} }}
\newcommand{\Tetakfiv}{\ensuremath{ \Tetak^{5} }}
\newcommand{\Tetakt}{\ensuremath{ \Tetak^{o} }}

\newcommand{\Tetae}{\ensuremath{ \widehat{\Theta} }}
\newcommand{\Tetaeko}{\ensuremath{ \Tetae_{_{k+1}} }}

\newcommand{\Uk}{\ensuremath{{ \,U_{_k} }}}
\newcommand{\Ukok}{\ensuremath{{ \,U_{_{k+1/k}} }}}
\newcommand{\Ucovkok}{\ensuremath{ \,\mathcal{U}_{_{k+1/k}}  }}
\newcommand{\ucross}{\ensuremath{ \,\left[ \uv \times \right] }}
\newcommand{\UU}{\ensuremath{ \mathcal{U} }}
\newcommand{\UUkoN}{\ensuremath{ \UU_{k}^{N-1} }}
\newcommand{\UUoN}{\ensuremath{ \UU^{N-1} }}
\newcommand{\UUN}{\ensuremath{ \UU^{N} }}
\newcommand{\UUkooN}{\ensuremath{ \UU_{k+1}^{N-1} }}

\newcommand{\Vk}{\ensuremath{{ \,V_{_k} }}}
\newcommand{\Vi}{\ensuremath{{ \,V_{_i} }}}
\newcommand{\Vko}{\ensuremath{ \,V_{_{k+1} }}}
\newcommand{\Voj}{\ensuremath{{ \,V_{_{1j}} }}}
\newcommand{\Voi}{\ensuremath{{ \,V_{_{1i}} }}}
\newcommand{\vcross}{\ensuremath{ \,\left[ \vv \times \right] }}
\newcommand{\Vkoi}{\ensuremath{ {\,V_{_{k+1}}^i} }}
\newcommand{\Vkon}{\ensuremath{ \,V_{_{k+1}}^n }}
\newcommand{\Vki}{\ensuremath{{ \,V_{_{k+i}} }}}
\newcommand{\Vsym}{\ensuremath{  V^{sym}  }}
\newcommand{\Vsymko}{\ensuremath{  \Vsym_{_{k+1}}  }}
\newcommand{\Vort}{\ensuremath{  V^{ort}  }}
\newcommand{\Vkort}{\ensuremath{  \Vort_{_k}  }}

\newcommand{\Wk}{\ensuremath{{ W_{_{\!\!k}} }}}
\newcommand{\Wz}{\ensuremath{{ W_{_0} }}}
\newcommand{\Wzz}{\ensuremath{ \,W^{0}_{_0} }}
\newcommand{\Wzo}{\ensuremath{ \,W^{1}_{_0} }}
\newcommand{\Wo}{\ensuremath{{ \,W_{_1} }}}
\newcommand{\Wko}{\ensuremath{{ W_{_{k+1}} }}}
\newcommand{\Wki}{\ensuremath{{ W_{_{k+i}} }}}
\newcommand{\Woj}{\ensuremath{{ \,W_{_{1j}} }}}
\newcommand{\Woi}{\ensuremath{{ \,W_{_{1i}} }}}
\newcommand{\Wnk}{\ensuremath{{ \,Wn_{_k} }}}
\newcommand{\Wi}{\ensuremath{{ W_{_{\!\!i}} }}}
\newcommand{\WN}{\ensuremath{{ W_{_N} }}}
\newcommand{\WoN}{\ensuremath{{ W_{_{N-1}} }}}
\newcommand{\WNo}{\ensuremath{ W_{_{N+1}} }}
\newcommand{\Wkt}{\ensuremath{ W_{_k}^o }}

\newcommand{\xcross}{\ensuremath{ \,\left[ \xv \times \right] }}

\newcommand{\Xk}{\ensuremath{{ \,X_{_k} }}}
\newcommand{\Xko}{\ensuremath{{ \,X_{_{k+1}} }}}
\newcommand{\Xz}{\ensuremath{{ \,X_{_0} }}}
\newcommand{\Xzb}{\ensuremath{ \bar{\Xz} }}
\newcommand{\Xe}{\ensuremath{ \,\widehat{X} }}
\newcommand{\Xez}{\ensuremath{ \Xe_{_{0}} }}
\newcommand{\Xek}{\ensuremath{ \Xe_{_{k}} }}
\newcommand{\Xeko}{\ensuremath{ \Xe_{_{k+1}} }}
\newcommand{\Xkk}{\ensuremath{ \Xe_{_{k/k}} }}
\newcommand{\Xkok}{\ensuremath{ \Xe_{_{k+1/k}} }}
\newcommand{\Xkoko}{\ensuremath{ \Xe_{_{k+1/k+1}} }}
\newcommand{\Xzz}{\ensuremath{ \Xe_{_{0/0}} }}
\newcommand{\Xer}{\ensuremath{{ \,{\widetilde{X}} }}}
\newcommand{\Xb}{\ensuremath{ \overline{X} }}
\newcommand{\Xbz}{\ensuremath{ \Xb_{_{0}} }}

\newcommand{\Xik}{\ensuremath{ \Xi_{_k} }}
\newcommand{\Xikt}{\ensuremath{ \Xik^{o} }}
\newcommand{\Xiko}{\ensuremath{ \Xi_{_{k+1}} }}
\newcommand{\Xiz}{\ensuremath{{ \Xi(\qz) }}}
\newcommand{\Xio}{\ensuremath{{ \Xi(\qo) }}}
\newcommand{\Xiezz}{\ensuremath{{ \Xi(\qezz) }}}
\newcommand{\Xieoo}{\ensuremath{{ \Xi(\qeoo) }}}
\newcommand{\Xie}{\ensuremath{ \,\widehat{\Xi} }}
\newcommand{\Xiek}{\ensuremath{ \Xie_{_k} }}
\newcommand{\Xieko}{\ensuremath{ \Xie_{_{k+1}} }}
\newcommand{\Xiekok}{\ensuremath{ \Xie_{_{k+1/k}} }}
\newcommand{\Xiektwk}{\ensuremath{ \Xie_{_{k+2/k}} }}
\newcommand{\Xiekoko}{\ensuremath{ \Xie_{_{k+1/k+1}} }}
\newcommand{\Xiekolk}{\ensuremath{ \Xie_{_{k+l-1/k}} }}
\newcommand{\Xieklk}{\ensuremath{ \Xie_{_{k+l/k}} }}
\newcommand{\Xiekk}{\ensuremath{ \Xie_{_{k/k}} }}
\newcommand{\Xiekkast}{\ensuremath{ \Xiekk^{\ast} }}
\newcommand{\Xiekol}{\ensuremath{ \Xie_{_{k+l-1}} }}
\newcommand{\Xiekl}{\ensuremath{ \Xie_{_{k+l}} }}

\newcommand{\ycross}{\ensuremath{ \,\left[ \yv \times \right] }}

\newcommand{\Yk}{\ensuremath{{ \,Y_{_{k}} }}}
\newcommand{\Yko}{\ensuremath{{ \,Y_{_{k+1}} }}}
\newcommand{\Ytld}{\ensuremath{ \,\widetilde{Y} }}
\newcommand{\Ytldko}{\ensuremath{ \Ytld_{_{k+1}} }}
\newcommand{\Yl}{\ensuremath{{ \,Y_{_{l}} }}}
\newcommand{\Ysym}{\ensuremath{ Y^{sym} }}
\newcommand{\Ysymko}{\ensuremath{{ \Ysym_{_{k+1}} }}}
\newcommand{\Yz}{\ensuremath{{ \,Y_{_{0}} }}}

\newcommand{\zvcross}{\ensuremath{ \,\left[ \zv \times \right] }}
\newcommand{\zcross}{\ensuremath{ \,\left[ \zv \times \right] }}
\newcommand{\zbcross}{\ensuremath{ \,\left[ \zb \times \right] }}
\newcommand{\zbicross}{\ensuremath{ \,\left[ \zbi \times \right] }}
\newcommand{\zbncross}{\ensuremath{ \,\left[ \zbn \times \right] }}
\newcommand{\zkocross}{\ensuremath{ \,\left[ \zko \times \right] }}
\newcommand{\zepscross}{\ensuremath{ \,\left[ \zeps \times \right] }}

% OPERATORS ----------------------------------------------

\newcommand{\AD}{\ensuremath{ A^{\!\Delta} }}

\newcommand{\HHt}{\ensuremath{ H_{_{\!t}} }}
\newcommand{\HHz}{\ensuremath{ H_{_{\!0}} }}
\newcommand{\HHts}{\ensuremath{ H_{_{\!t+s}} }}
\newcommand{\HHst}{\ensuremath{ H_{_{\!s+t}} }}
\newcommand{\HHttau}{\ensuremath{ H_{_{\!t+\tau}} }}
\newcommand{\HHtau}{\ensuremath{ H_{_{\!\tau}} }}
\newcommand{\HHtmtau}{\ensuremath{ H_{_{\!t-\tau}} }}
\newcommand{\HHtmtautau}{\ensuremath{ H_{_{\!t-\tau+\tau}} }}
\newcommand{\HHs}{\ensuremath{ H_{_{\!s}} }}
\newcommand{\HHvmtz}{\ensuremath{ H_{_{\!v-\tz}} }}
\newcommand{\HHtD}{\ensuremath{ \HHt^{\!\Delta} }}
\newcommand{\HHsD}{\ensuremath{ \HHs^{\!\Delta} }}
\newcommand{\HHzD}{\ensuremath{ \HHz^{\!\Delta} }}
\newcommand{\HHtsD}{\ensuremath{ \HHts^{\!\Delta} }}
\newcommand{\HHttauD}{\ensuremath{ \HHttau^{\!\Delta} }}
\newcommand{\HHtauD}{\ensuremath{ \HHtau^{\!\Delta} }}
\newcommand{\HHtmtauD}{\ensuremath{ \HHtmtau^{\!\Delta} }}
\newcommand{\HHtmtautauD}{\ensuremath{ \HHtmtautau^{\!\Delta} }}
\newcommand{\HHvmtzD}{\ensuremath{ \HHvmtz^{\!\Delta} }}

% ABBREVIATIONS ----------------------------------------------

%\newcommand{\AD}{\ensuremath{{ \bf{AD}\,  }}}
\newcommand{\EMF}{\ensuremath{{ \bf{EMF}\,  }}}
\newcommand{\EKF}{\ensuremath{{ \bf{EKF}\,  }}}
\newcommand{\RLS}{\ensuremath{{ \bf{RLS}\,  }}}
\newcommand{\KVKF}{\ensuremath{{ \bf{KVKF}\,  }}}
\newcommand{\quatKF}{\ensuremath{{ \bf{quatKF}\,  }}}
\newcommand{\REQUEST}{ \mbox{REQUEST} }

% FACTORIZATION APPENDIX ----------------------------------------
\newcommand{\xnorm}{\ensuremath{ \, \| \xv \|  }}
\newcommand{\xonorm}{\ensuremath{ \, \| \xo \|  }}
\newcommand{\xtwnorm}{\ensuremath{ \, \| \xtw \|  }}
\newcommand{\xounit}{\ensuremath{ \,\frac{\xo}{\| \xo \|}  }}
\newcommand{\axsum}{\ensuremath{  \,\left( \ao\,\xo\;+\;\atw\,\xtw \right)  }}
\newcommand{\axsumnorm}{\ensuremath{  \|\ao\,\xo\;+\;\atw\,\xtw \| }}
\newcommand{\aoatwsum}{\ensuremath{ \,\left( \ao + \atw \right)  }}

% QUATKF SIMULATION RESULTS SYMBOLS
\newcommand{\ipo}{\ensuremath{ \,\mbox{ip}_{_1} }}
\newcommand{\iptw}{\ensuremath{ \,\mbox{ip}_{_2} }}
\newcommand{\ipmoo}{\ensuremath{ \,\mbox{ipm}_{_{11}} }}
\newcommand{\ipmtt}{\ensuremath{ \,\mbox{ipm}_{_{22}} }}

\newcommand{\pcmoo}{\ensuremath{ \,\mbox{pcm}_{_{11}} }}
\newcommand{\pcmtt}{\ensuremath{ \,\mbox{pcm}_{_{22}} }}
\newcommand{\gainoo}{\ensuremath{ \,\mbox{gain}_{_{11}} }}
\newcommand{\gaintt}{\ensuremath{ \,\mbox{gain}_{_{22}} }}

\section{Introduction}

\begin{figure}[ht]
\centering
 \includegraphics[width=0.550\textwidth]{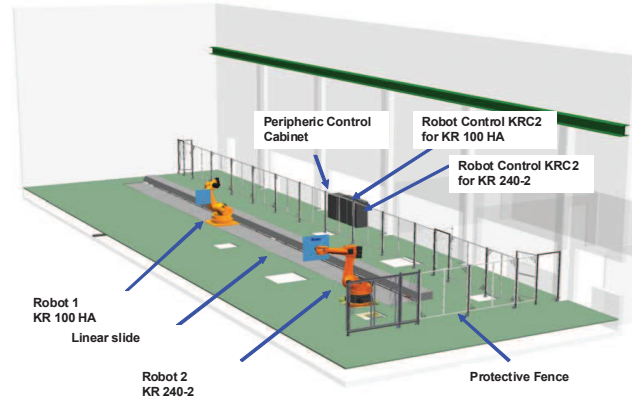}
%\caption{General view}
%\label{fig:fvepos}
  \label{s1f1}
 \caption{The EPOS facility design concept: dual-robots operations room, control room, and preparation room. The HIL simulator allows for six degrees of freedom motion of the payloads.}
\end{figure}

Simulators of satellite proximity operations have been used since the beginning of space programs~\cite{Jana2003}. Several technologies are available for testing and verification of systems' operations in a simulated micro-gravity environment, such as free-fall methods, parabolic flights, air-bearing based testbeds, neutral buoyancy, suspended systems, underwater test tanks, and robotics hardware-in-the-loop (HIL) simulators. The latter one is the preferred technology when it comes to rendezvous and docking (RvD) testing in six degrees of freedom (6DOF), i.e., for 3D translation and rotation. Examples of HIL simulators are reported in Refs.~\cite{Boge,Murphy2004,Ma1995}. The European Proximity Operations Simulator (EPOS) located at the German Space Operations Center of the German Aerospace Center belongs to this category. A schematic rendering of the facility and a picture of the dual robots system are shown in Figure~\ref{s1f1}.
%\begin{figure}[tb]
%\centering
%    \begin{tabular}{cc}
%        \subfigure[General view]
%        {\resizebox{0.5\textwidth}{0.5\textheight}{
%%        \includegraphics[width=0.450\textwidth]{fig/eposmainparts.eps}&
%	      \includegraphics{fig/eposmainparts.eps}}} &
%        \subfigure[One robot approaching along the trail]
%        {\resizebox{0.5\textwidth}{0.5\textheight}{
%         %  \includegraphics[width=0.450\textwidth]{fig/eposrobot.eps}\\
%          \includegraphics{fig/eposrobot.eps}}}  \\
%    \end{tabular}
%       \caption{The EPOS facility: dual-robots operations room, control room, and preparation room. The HIL simulator allows for six degrees of freedom motion of the payloads.}
%       \label{s1f1}
%     \end{figure}

%\begin{figure}[ht]
%\begin{minipage}[b]{0.5\linewidth}
%\centering
% \includegraphics[width=0.50\textwidth]{fig/eposmainparts.eps}
%%\caption{General view}
%%\label{fig:fvepos}
%\end{minipage}
%%\hspace{0.2cm}
%\begin{minipage}[b]{0.5\linewidth}
%\centering
%  \includegraphics[width=0.50\textwidth]{fig/eposrobot.eps}
% %\caption{One robot approaching along the trail}
%  \label{s1f1}
%\end{minipage}
%%
Particular features of EPOS are the two industrial robots, capable of positioning payload masses up to 240 kg with sub-millimeter and sub-degree accuracy. In addition to the 6DOF motion,  the facility allows for one robot to slide along a 25m long air-bearing rail at speeds varying from mm/sec to m/sec. Together with Sun illumination conditions, EPOS provides a realistic emulation of final approach and RvD scenarios. Eventually, EPOS aims at providing a truthful rendering of free-flying contact dynamics for hardware and software testing and verification of systems for future on-orbit servicing missions such as for space debris removal~\cite{Bonnal}. A detailed review of the facility is reported in~\cite{Boge,Rupp}.

\begin{figure}[h!]
	\centering
	\includegraphics[width=0.850\textwidth]{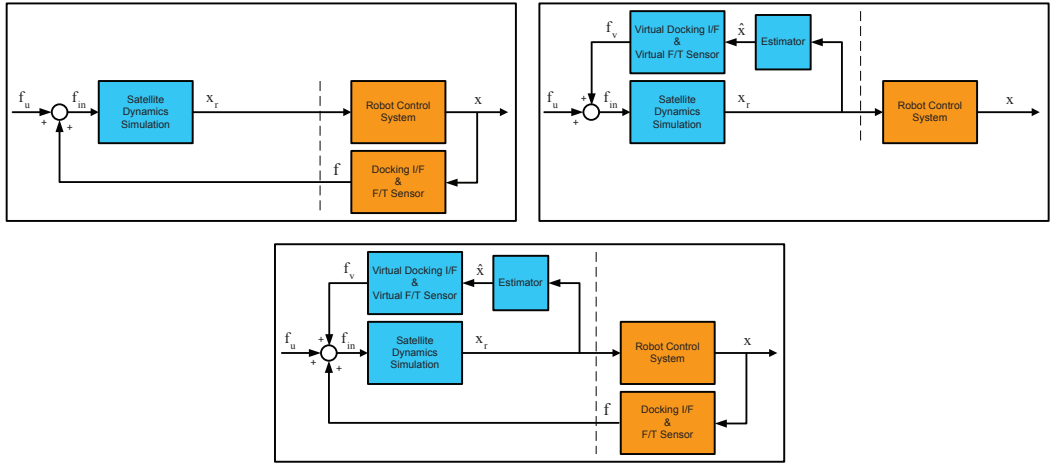}
	\caption{Block Diagram for a robotics-based docking simulator with various force feedback: pure hardware simulator (left top), pure software (right top), and hybrid (bottom). Software and hardware elements appear in blue and orange, respectively.}
	\label{s1f2}
\end{figure}

Using industrial robots for HIL docking simulation purposes is a highly challenging approach. Designed to be very accurate in position, the robotic system typically presents a very high stiffness. On the other hand, the response bandwidth is relatively small compared to the bandwidth of the dynamics between contacting surfaces. In addition, the HIL simulator relies on feedbacking the measured contact force to a numerical simulator of the satellites dynamics. That software (S/W) module calculates the satellites' positions and feeds them as command signals to the robots tracking controller. This appealing feature is also the Achille's heel of the concept. The HIL system is, in essence, a closed loop system where the robotics tracking system introduces a delay. Given the EPOS robots stiffness and delay characteristics, the loop delay system is unstable. This prompted the development of a novel concept, a \emph{hybrid contact dynamics emulator}, combining real (measured) and virtual (modeled) force feedback, as reported in~\cite{Zebenay}. Figure~\ref{s1f2} depicts simplistically different concepts of operations: pure hardware with measured force feedback (top left), pure software with virtual force feedback (top right), and the hybrid approach using a combination of both (bottom).

This work, continuing earlier efforts reported in~\cite{Zebenay,Ma_Melak,Melak2013may} presents an analytical and experimental investigation that verifies and validates the proposed hybrid concept for contact in one dimension. The main contributions are the stability analysis and the extensive experimental results. The stability analysis offers closed-form results developed using the pole location method~\cite{Marshall,Mirkin2006}. In addition, graphical tools for characterization of the stability boundaries are presented both via the pole location method and via the classical Root-Locus method based on a first-order Pade approximation of the delay.  Extensive testing was performed with the EPOS system in 1D, where a probe installed on the chaser robot came in contact with a target metallic sheet. Empirical data-driven stability criteria, such as the coefficient of restitution~\cite{Gilardi, Nakanishi} and the observed energy~\cite{Slotine,988969,Rainer} were applied and compared to the stability prediction from the analysis. The test results exhibit a very good agreement with the analytical results, in particular, with respect to the stability boundaries and the contact duration. In addition, the proposed hybrid simulator is successfully implemented in order to emulate test results from an experiment conducted on an air-floating testbed at the Space Robotics Laboratory at Tohoku University.

Section~2 presents the mathematical modeling and stability analysis of the loop delay system. Section~4 describes the EPOS experimental setup, tests and results, and the air-floating testbed results emulation. Conclusions are drawn in the last section.

\section{Modeling and stability analysis of the docking simulator}
\label{subsec:1A}

\subsection{1D Mathematical modeling}
\begin{figure}[htb]
      \centering
      \includegraphics[width=0.70\textwidth]{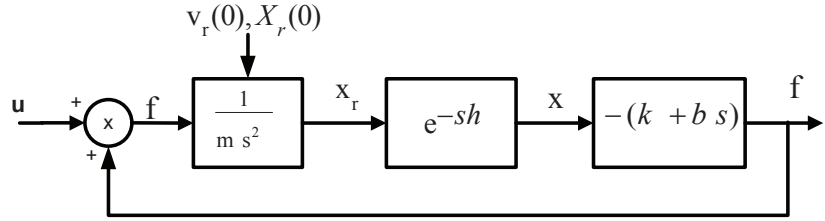}
      \caption{Block diagram of the HIL docking simulator for single-dimensional analysis}
       \label{fig:1dofhil}
\end{figure}

The present modeling is single-dimensional. The chaser and target satellites are simulated as point masses, $m_{_C}$ and $m_{_T}$, respectively, moving along a straight line at a relative distance $x_r(t)$. The masses move towards each other with an initial relative velocity as free-floating masses with respect to an inertial frame. During and following the impact, the contact force is assumed to be the single cause of motion. The applied force is transmitted to the chaser via a compliant device that is modeled as a lightly damped spring with known stiffness and damping coefficients, $k$ and $b$. Assuming small relative displacements and relative velocities, the linear spring-dashpot model is adopted for the contact force modeling~\cite{Gilardi,Nakanishi,Kraus}. The stiffness of the compliant device is lower than that of the materials in contact. The applied force is assumed to be perfectly sensed. The underlying assumption is that calibration of the force sensor has been performed and that the level of noises is negligible compared to the force magnitude. That force is fed back to the satellites numerical simulator, which provides the calculated motion as a command to the robots. The robots are commanded in position and reach the required position, i.e. $x_r(t)$, with no steady-state error after a known delay $h$. Let $x(t)$ denote the true relative position of the target with respect to the chaser. The delay, masses, stiffness and damping coefficients are assumed to be time-invariant. As a result, the equations governing the dynamics of the loop delay system are as follows:
\begin{align}
\label{s2eq02}
& m \ddot{x}_r(t) =  f(t)  \\
\label{s2eq01}
& f(t)= - k x(t) - b \dot{x}(t)  \\
\label{s2eq00}
& x(t) = x_r(t-h) \
\end{align}
where
\begin{align}
& m = \frac{m_{_C} m_{_T} }{m_{_C} + m_{_T}}\
\end{align}
and $m$ denotes the equivalent mass in the relative motion dynamics. This design model is captured by the block diagram in Figure~\ref{fig:1dofhil}.

\subsection{Stability analysis}
\subsubsection{Pole location method}
\begin{figure}[h!]
    \centering
    \begin{tabular}{ccc}
        \subfigure[$b$ vs $h$, m=60 kg, k=1000 N/m]
        {\resizebox{0.3\textwidth}{0.17\textheight}{
	      \includegraphics{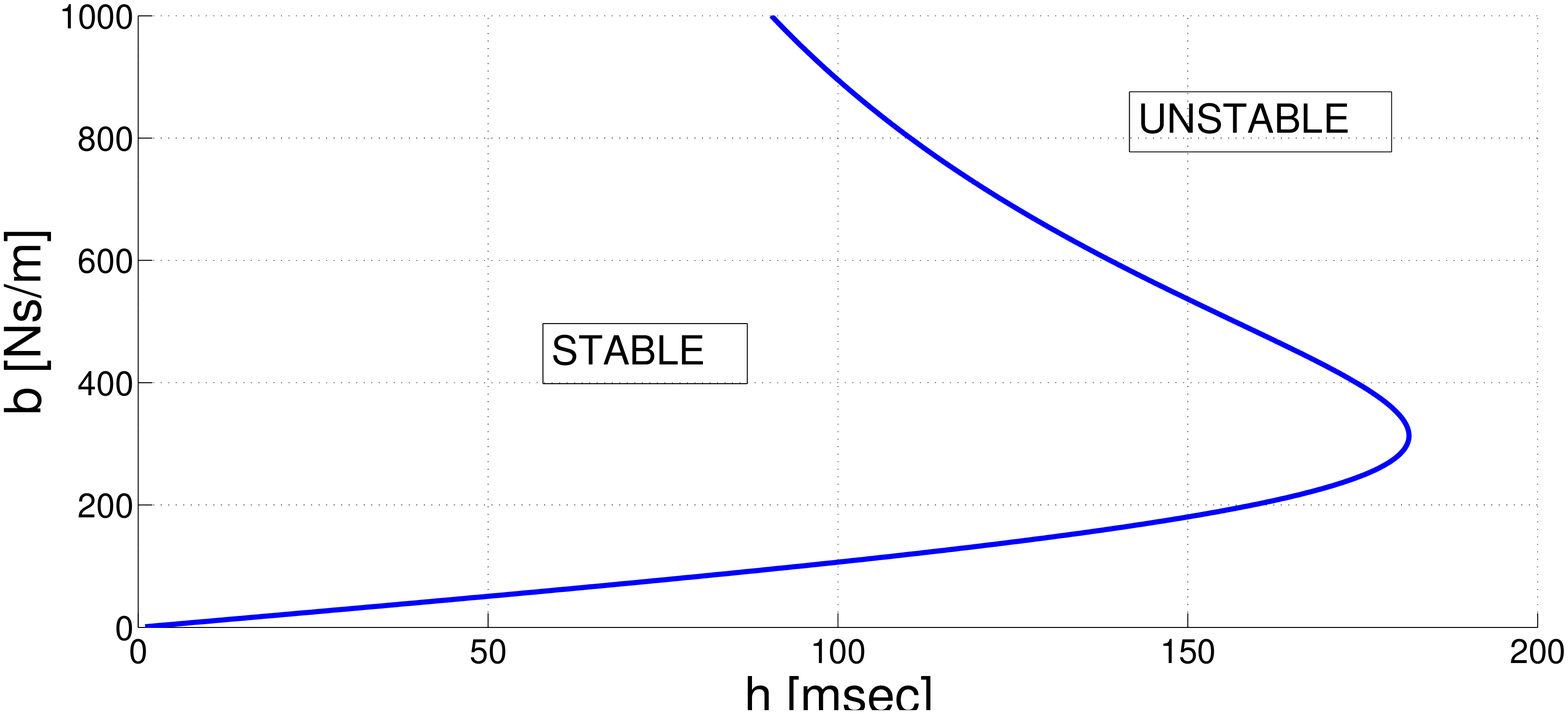}}} &
        \subfigure[$k$ vs $h$, m=60 kg, b=50 Ns/m]
        {\resizebox{0.3\textwidth}{0.17\textheight}{
          \includegraphics{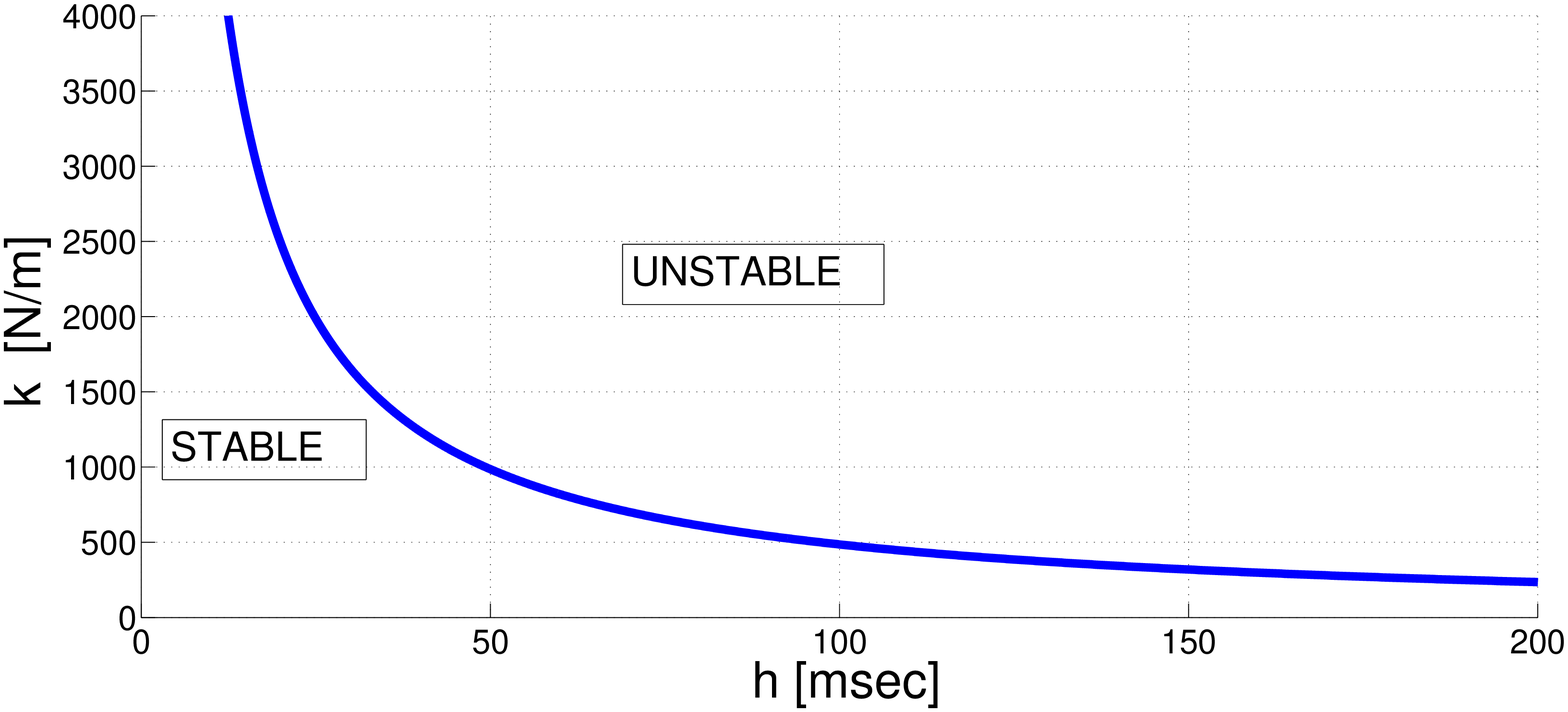}}}  &
        \subfigure[$m$ vs $h$, b=50 Ns/m, k=1000 N/m]
        {\resizebox{0.3\textwidth}{0.17\textheight}{
	      \includegraphics{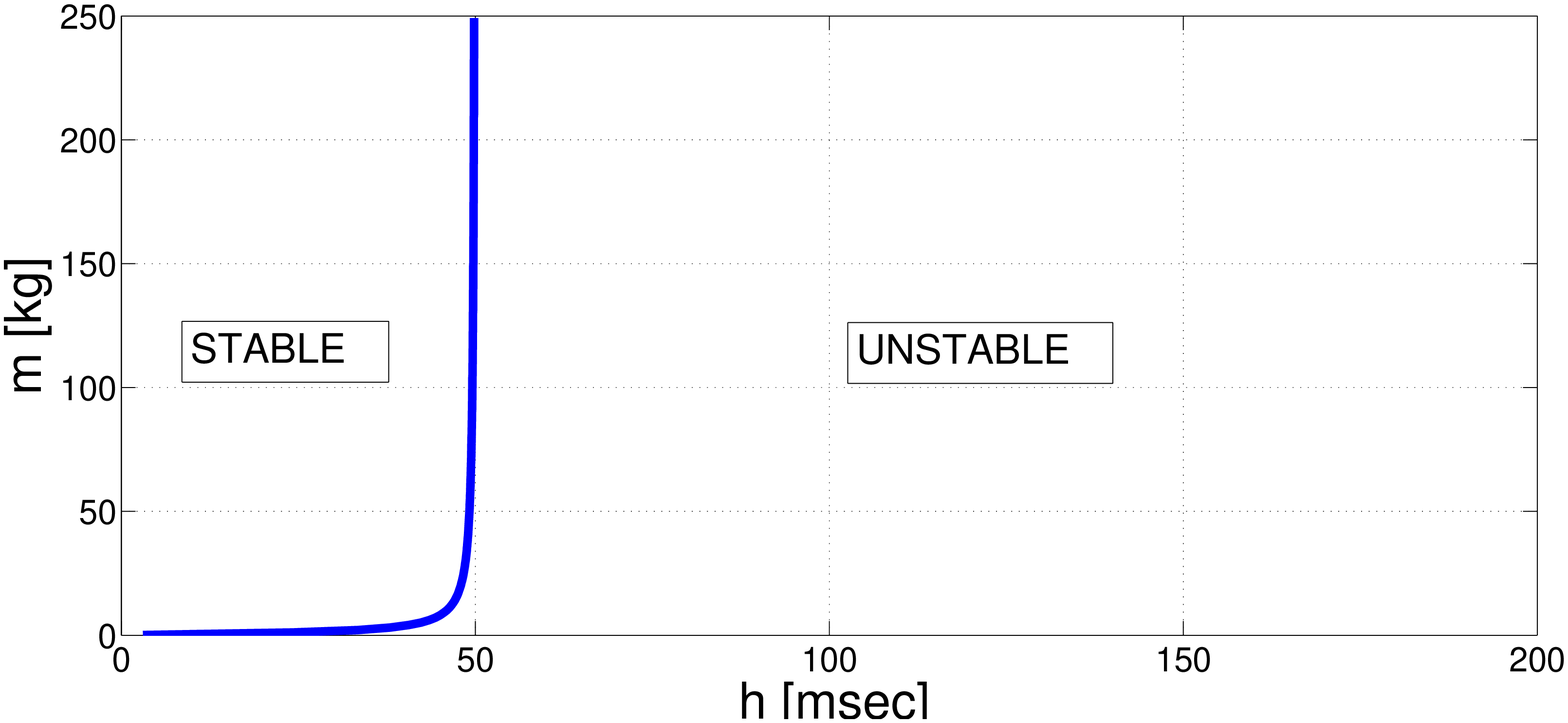}}} \\
    \end{tabular}
       \caption{Stability domains for a typical operational point m=60 kg, k=1000 N/m, b=50 Ns/m.}
       \label{f2}
     \end{figure}

\begin{figure}[h!]
    \centering
    \begin{tabular}{ccc}
        \subfigure[$b$ vs $h$, $m=$60 kg, varying stiffness]
        {\resizebox{0.3\textwidth}{0.17\textheight}{
	      \includegraphics{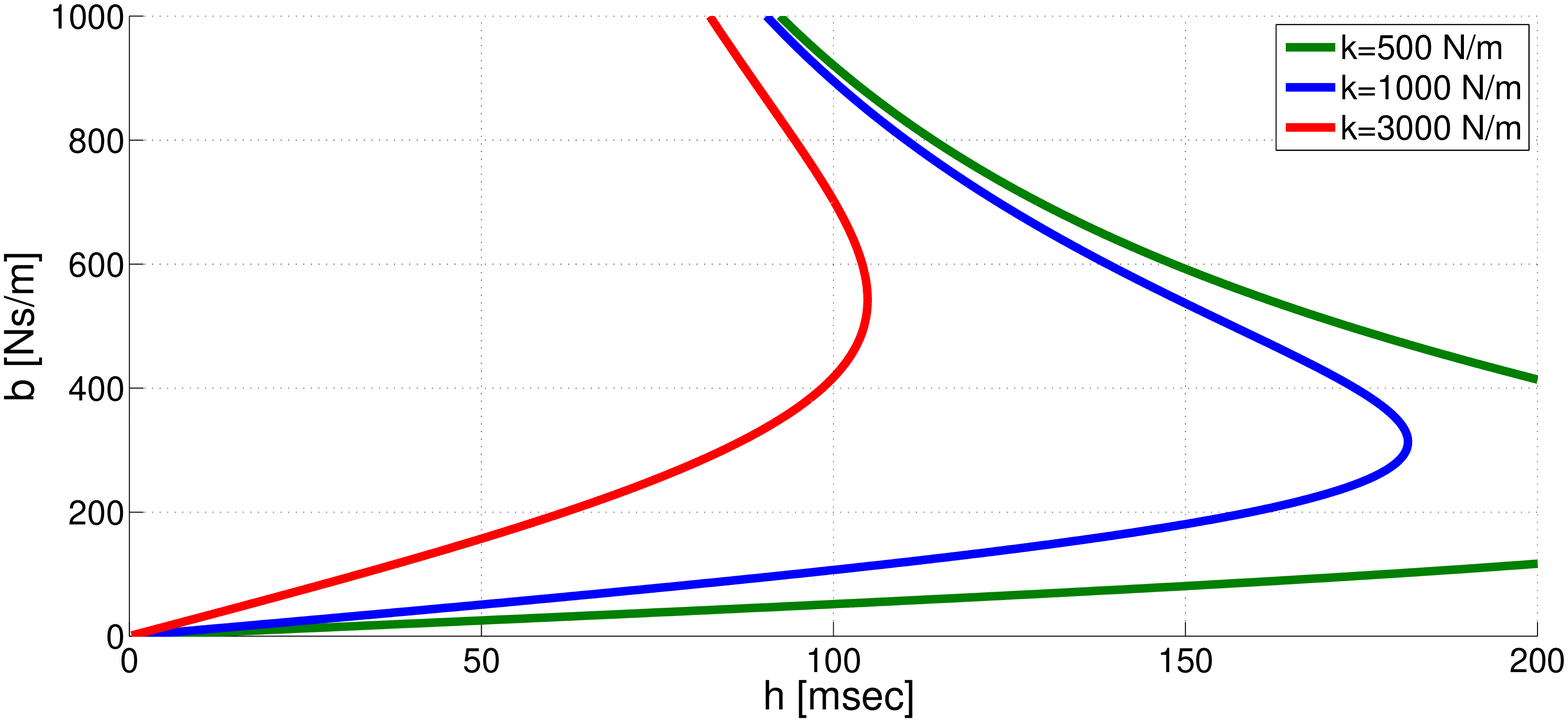}}} &
        \subfigure[$k$ vs $h$, m=60 kg, varying damping]
        {\resizebox{0.3\textwidth}{0.17\textheight}{
          \includegraphics{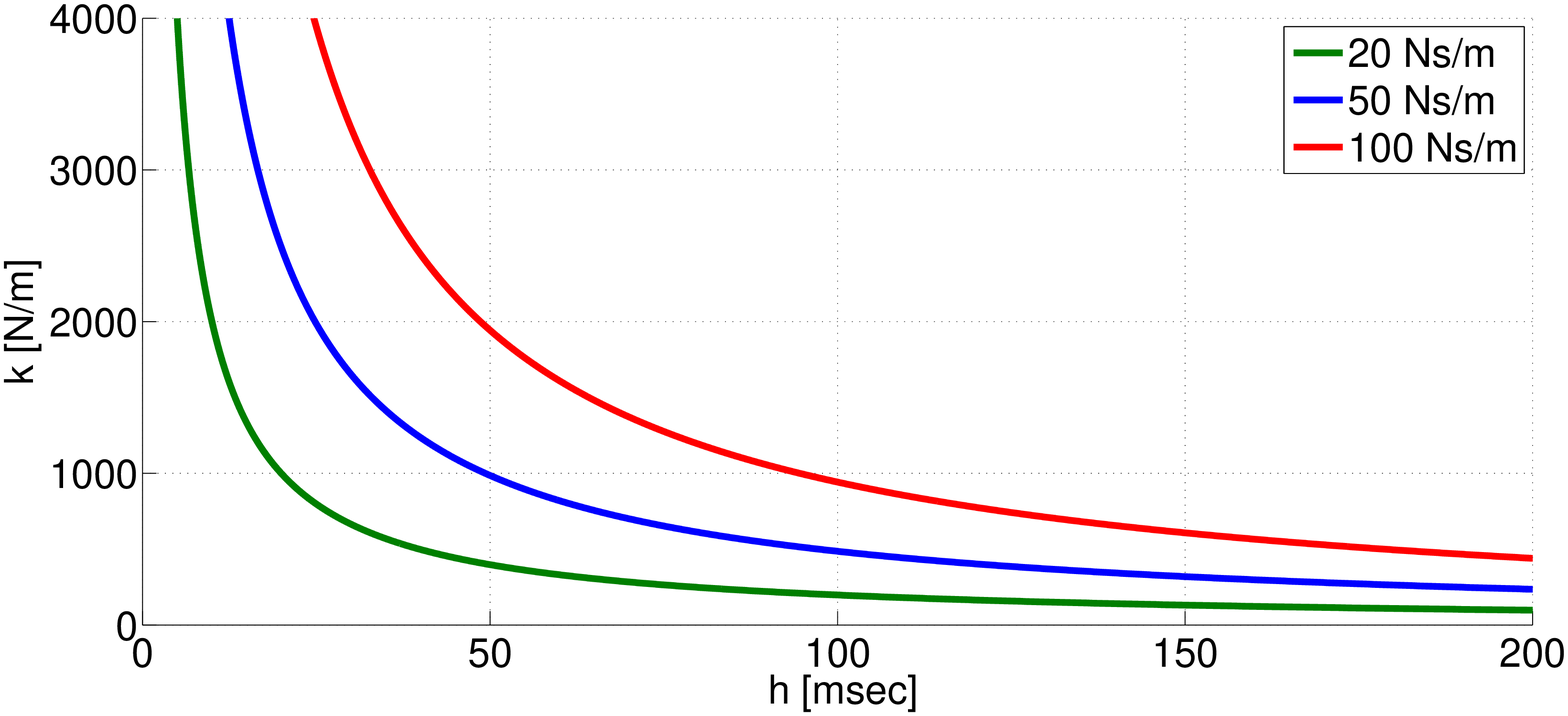}}}  &
        \subfigure[$m$ vs $h$, $k=$1000 N/m, varying damping]
        {\resizebox{0.3\textwidth}{0.17\textheight}{
	      \includegraphics{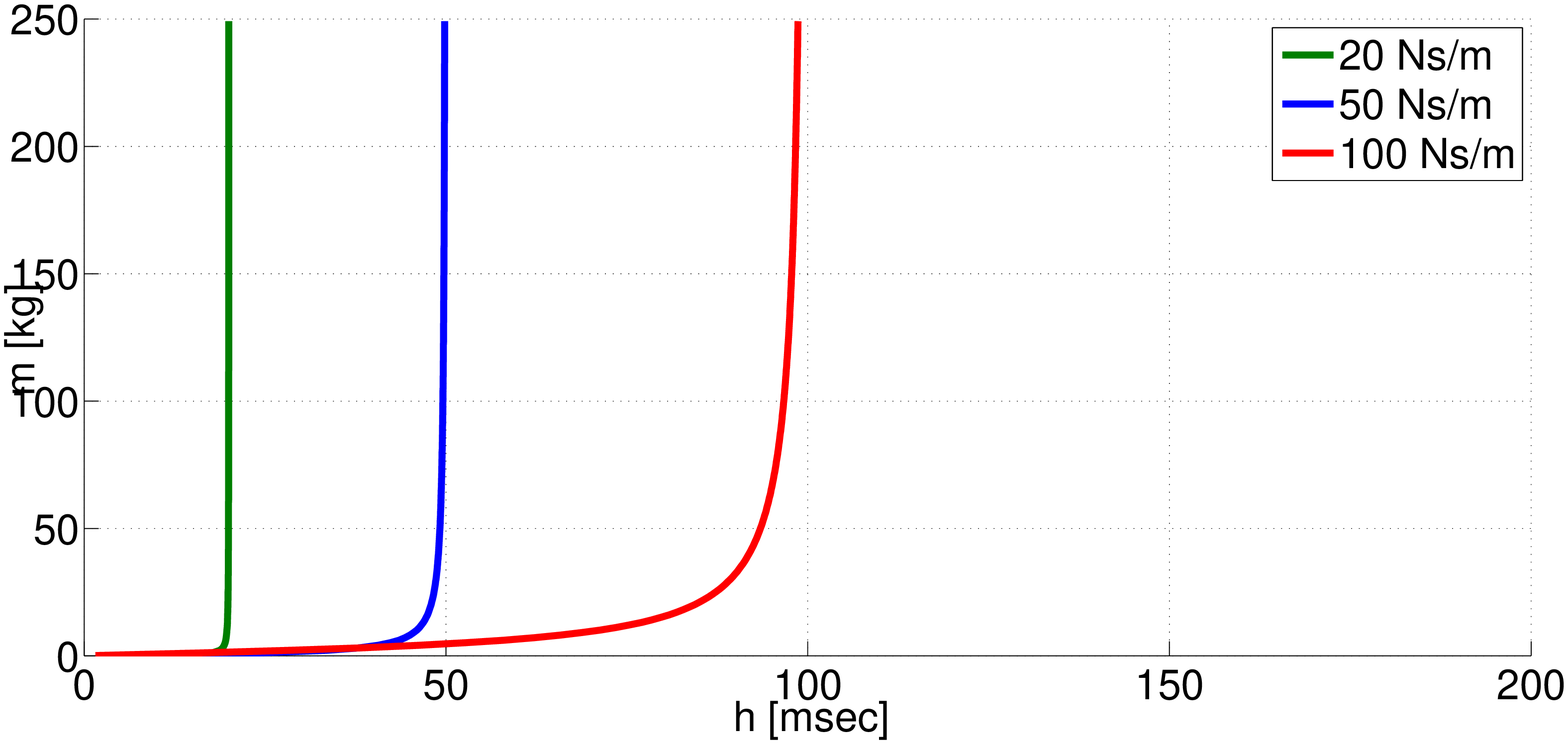}}} \\
        \subfigure[$b$ vs $h$, k=1000 N/m, varying mass]
        {\resizebox{0.3\textwidth}{0.17\textheight}{
	      \includegraphics{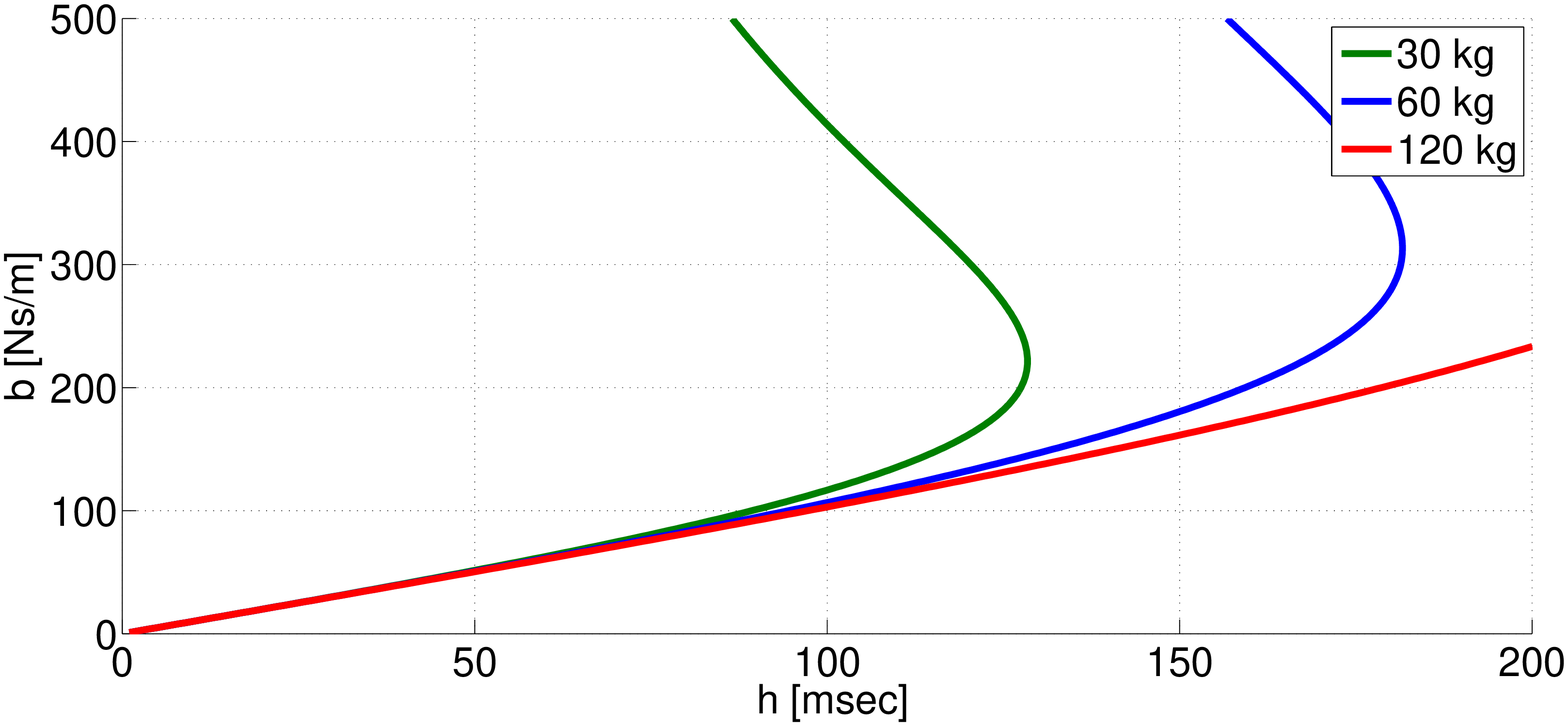}}} &
        \subfigure[$k$ vs $h$, $b=$50 Ns/m, varying mass]
        {\resizebox{0.3\textwidth}{0.17\textheight}{
          \includegraphics{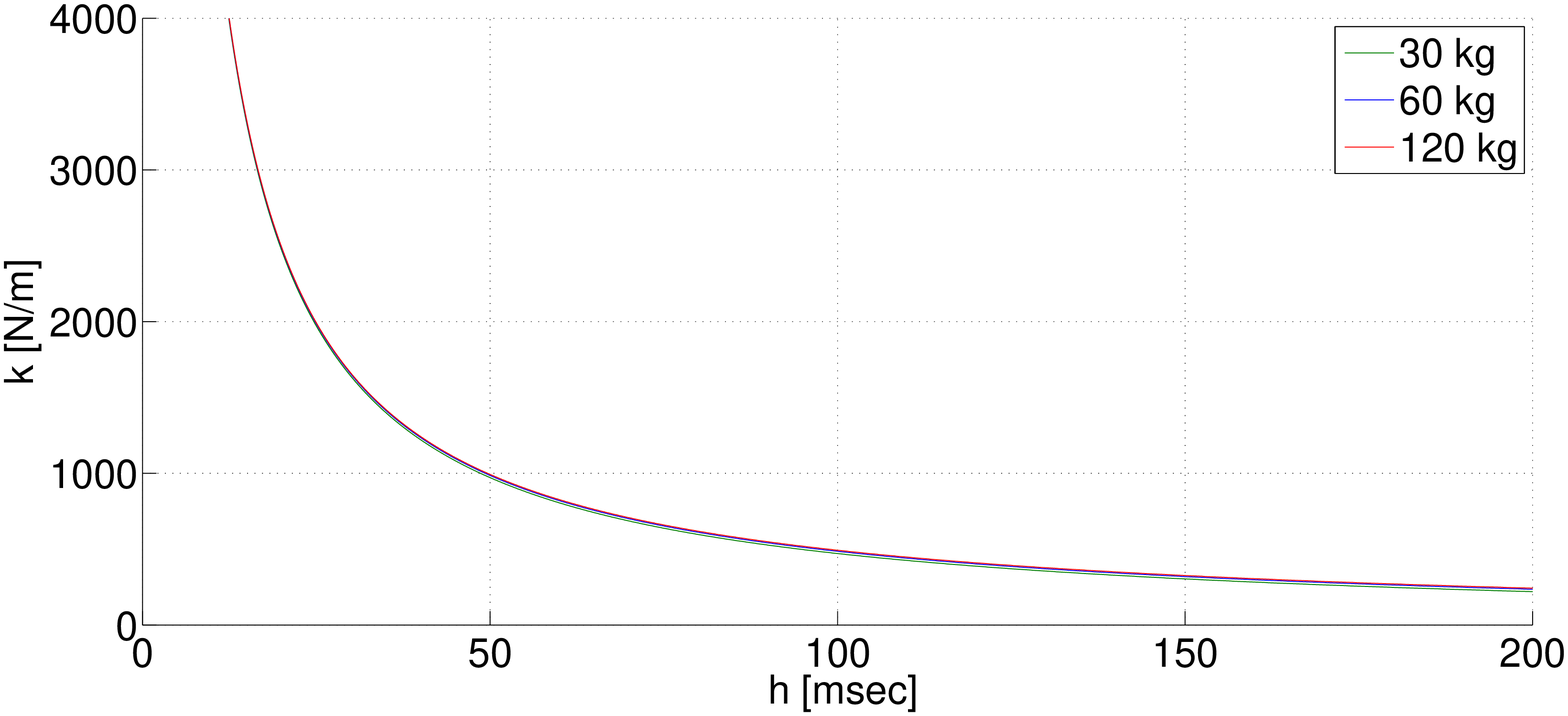}}}  &
        \subfigure[$m$ vs $h$, $b=$50 N/m, varying stiffness]
        {\resizebox{0.3\textwidth}{0.17\textheight}{
	      \includegraphics{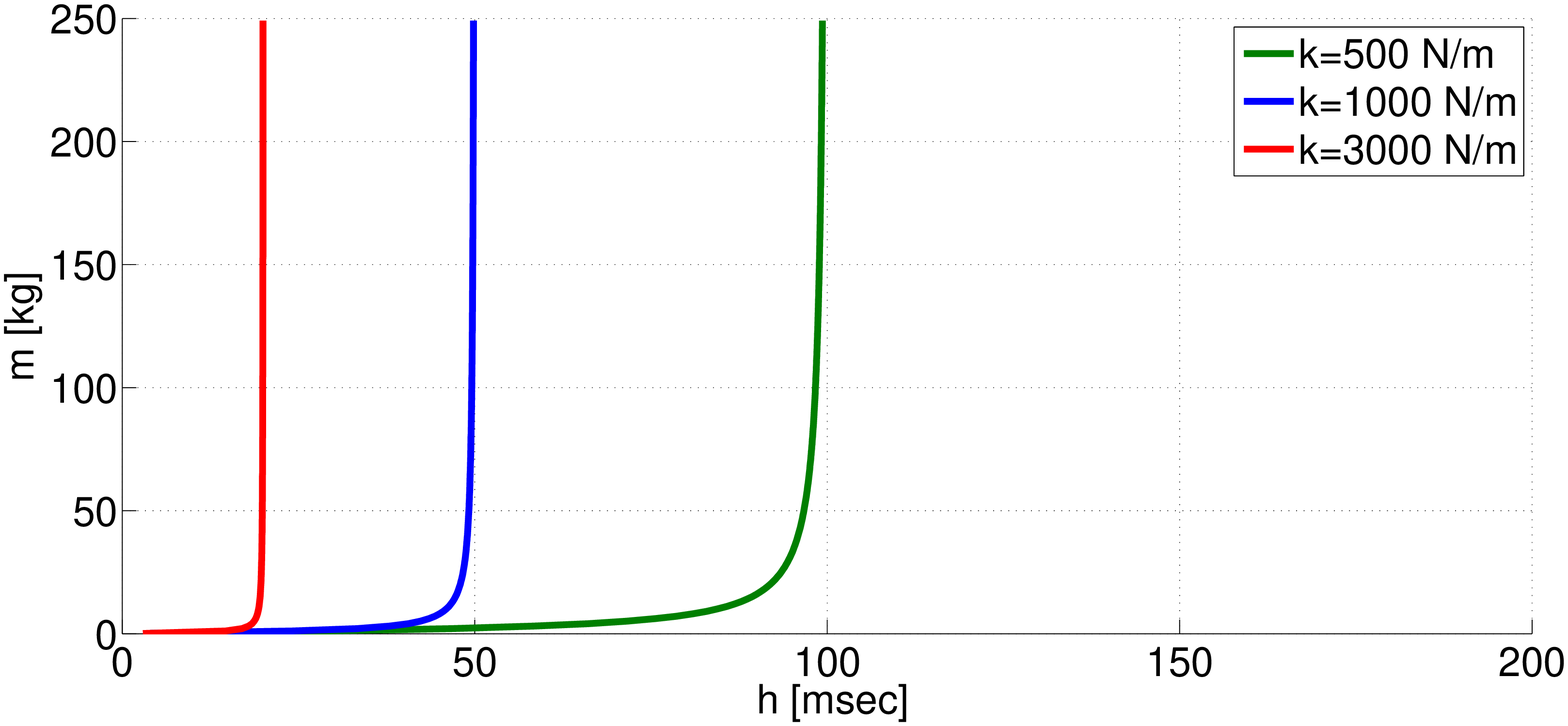}}} \\
    \end{tabular}
       \caption{Stability domains variations for an operational point: 60kg, 1000 N/m, 50 Ns/m.}
       \label{f3}
     \end{figure}

The proposed analysis follows the pole location method. A description of the method can be found in \cite{Marshall}. The characteristic equation of the loop delay system given in Fig.~\ref{fig:1dofhil} is as follows:
 \begin{align}
 \label{eq:cheq}
&    \chi_h(s)=  m s^2+ e^{-sh} (k + b s) \
\end{align}
The stability of this loop delay system is analyzed by studying the behavior of roots of Eq.~\eqref{eq:cheq} as $h$ increases from zero. This approach is based on the continuous property of the roots of $\chi_h(s)$ as functions of $h$, and on the fact that the number of unstable roots can only change when some roots cross the imaginary axis. The condition for stability is that all the roots of $\chi_h(s)$ lie in the open left half-plane (OLHP) of the complex plane. The pole location method provides an analytical mean to determine the value(s) of the delay $h$, as a function of the system's parameters $m$, $k$, and $b$, such that some roots of $\chi_h(s)$ lie on the imaginary axis.  The first step consists in examining the delay-free characteristic polynomial, which is
\begin{equation}
	\chi(s)= m s^2+ b s + k
	\label{eq:transhzero}
\end{equation}
IN this case, necessary and sufficient conditions of stability are that all coefficients are positive: the delay-free loop system is stable as long as there is stiffness and damping in the feedback force. The second step consists in analyzing the roots as $h$ increases from zero. The number of roots becomes infinite and some of them will cross the imaginary axis for a critical value of $h$. Let $D(s)$ and $N(s)$ be defined as follows:
\begin{align}
\label{s2eq10}
& D(s) = m s^2 \\
\label{s2eq11}
& N(s) = bs + k\
\end{align}
The condition for $\chi_h(s)$ to have roots on $(j\omega)$ is expressed as follows:
\begin{align}
\label{s2eq12}
& \chi_h(j\omega) = 0
\;\;\Leftrightarrow \;\;\left\{
\begin{tabular}{l}
$| \frac{N(j \omega)}{D(j \omega)}| = 1$ \\
$\arg[\frac{N(j \omega)}{D(j \omega)}] = -\omega h \pm 2 \pi n$ \\
\end{tabular}
\right.
\
\end{align}
where $n = 0,1,2,\ldots$ This reminds of the gain-phase relationships of the Root-Locus method for rational transfer functions, with the difference stemming from the phase delay. Using Eqs.~\eqref{s2eq10} and \eqref{s2eq11} in Eq.~\eqref{s2eq12} yields
\begin{align}
\label{s2eq13a}
&  m^2 \omega^4 - b^2 \omega^2 - k^2 = 0 \\
\label{s2eq13b}
& \omega h  = \arctan(\frac{\omega b}{k}) \pm 2 \pi n \
\end{align}
Selecting the positive root for $\omega$ yields the following final expressions:
\begin{align}
\label{s2eq14a}
&\omega = \sqrt{\frac{b^2}{2 m^2} + \sqrt{ \frac{b^4}{4 m^4} + \frac{k^2}{m^2}}} \\
\label{s2eq14b}
& h_n  = \frac{1}{\omega}\arctan(\frac{\omega b}{k}) \pm \frac{2 \pi}{\omega} n \
\end{align}
Equations~\eqref{s2eq14a}, \eqref{s2eq14b} show that, as the delay increases, poles are crossing the imaginary axis each time $h$ reaches one of the values $h_n$ of the described set. The first value, denoted by $h_c$, is computed at $n=0$, thus
\begin{align}
\label{s2eq15}
& h_c = \frac{1}{\omega_c}\arctan(\frac{\omega_c b}{k}) \
\end{align}
where the natural frequency at the $j\omega$-crossing, $\omega_c$, is expressed from Eq.~\eqref{s2eq14a}. Notice that the value of $\omega_c$ is independent of the delay.\footnote{This general result stems from the fact that a pure delay is a unitary operator that does not change the loop gain.} According to \cite{Marshall}, the criteria that determines whether poles are crossing on their way out of the OLHP (switch) or on their way into the OLHP (reversal) is the sign of the following quantity, $\sigma(\omega)$:
\begin{align}
\label{eq:gainqu1}
& \sigma (\omega_c) \eqdef \left[\frac{d}{d\omega}(\mid D(j\omega)\mid^2-\mid N(j\omega)\mid^2)\right]_{\omega= \omega_c}
=  \sqrt{\frac{b^4}{4 m^4}+\frac{k ^2}{m^2}} \
\end{align}
where the second equation in Eq.~\eqref{eq:gainqu1} results from Eqs.~\eqref{s2eq10}, \eqref{s2eq11}, and \eqref{s2eq14a}. A switch occurs if $\sigma(\omega_c) > 0$, a reversal occurs if $\sigma(\omega_c)  < 0$, and no crossing occurs if $\sigma(\omega_c) = 0$. Obviously, in the present case, only switches occur, i.e., poles successively leave the OLHP as $h$ takes on the values $h_n$. A particular limit case of the analysis consists of the absence of damping, i.e. $b=0$. It is straightforward to check that the critical delay is simply $0$, and the associated crossing frequency is $\omega_c = \sqrt{\frac{k}{m}}$, which are the expected values. Notice that for relative large values of the stiffness $k$, the frequency $\omega_c$ is of order $\Order(\sqrt{k})$ (see Eq.~\ref{s2eq14a}), which yields an order $\Order(\frac{1}{k})$ for the critical delay $h_c$ (Eq.~\ref{s2eq15}). This illustrates the known phenomenon that higher values of a proportional feedback gain - here $k$ - are adverse to stability in presence of delays. Further, since typical values of the stiffness $k$ yield low ratios $\frac{\omega_c b}{k}$, and using the equivalence $\arctan(x) \sim x$ for small $x$, it appears from Eq.~\eqref{s2eq15} that the critical delay $h_c$ is equivalent to $\frac{b}{k}$, independently from the frequency $\omega_c$.  In other words, if $\omega_c b << k$, one can use the following approximation formula in order to compute the critical delay:
\begin{align}
\label{s2eq16}
& h_c = \frac{b}{k} \
\end{align}
As a conclusion, Eqs.~\eqref{s2eq15} and \eqref{s2eq14a} provide analytical expressions for the critical delay that will destabilize the closed-loop system, and for the natural frequency at which this happens, as functions of the system's parameters, $m$, $k$, and $b$. \\
As an example, Fig.~\ref{f2} depicts the stability regions for typical values of the delay, the mass, the stiffness, and the damping coefficient. The plot in Fig.~\ref{f2}-(a) illustrates the existence of a minimum required damping for a given delay. It also shows that there exists a limit for a system time delay beyond which there is no stability for any damping. Figure~\ref{f2}-(b) provides the curve of the critical stiffness $k$ as a function of the delay. Figure~\ref{f2}-(c) illustrates the existence of a region of $m$ in which the critical delay becomes independent of $m$.
A numerical sensitivity investigation of the stability regions with respect the parameters $m$, $k$, and $b$ was performed. The results are summarized in Figure~\ref{f3}. Figures~\ref{f3}-(a)(d) depict the variations of Figure~\ref{f2}-(a) when $k$ and $m$ are modified, respectively, while holding the other parameter constant. It appears that an increase in the stiffness $k$ reduces the stability region, Fig.~\ref{f3}-(a), while an increase in the mass $m$ increases it Fig.~\ref{f3}-(d). Henceforth, for a higher stiffness the HIL simulator will require more damping to guarantee stability. Notice that for small delays and damping values, the curve $b$ vs $h$ is approximately insensitive to the mass, as expected (see Eq.~\ref{s2eq16}). Figures \ref{f3}-(b)(e) illustrate the sensitivity of the curve $k$ vs $h$ of Fig. ~\ref{f2}-(b) when $b$ and $m$ are varied. Increasing the damping coefficient has the effect of increasing the stability region. The increase in the mass also has the effect of increasing the stability domain, albeit by a small amount. Figures~\ref{f3}-(c)(f) depict the sensitivity of Figure~\ref{f2}-(c) to changes in $b$ and in $k$. The increase in $b$ enlarges the domain of stability. It also shows that the maximum allowed delay becomes more mass-dependent for higher damping values. Notice that for the value of 20 Ns/m and a stiffness of 1000 N/m the plot depicts a critical delay of 20 ms, which validates the approximation of Eq.~\eqref{s2eq15}. The increase in $k$ has the inverse effect with a similar factor.

\subsubsection{Root-locus analysis via Pade approximation}
\begin{figure}[h]
\centering
\hspace{-5ex}
    \begin{tabular}{cc}
        \subfigure[$h\geq0$]
        {\resizebox{0.4\textwidth}{0.3\textheight}{
	      \includegraphics{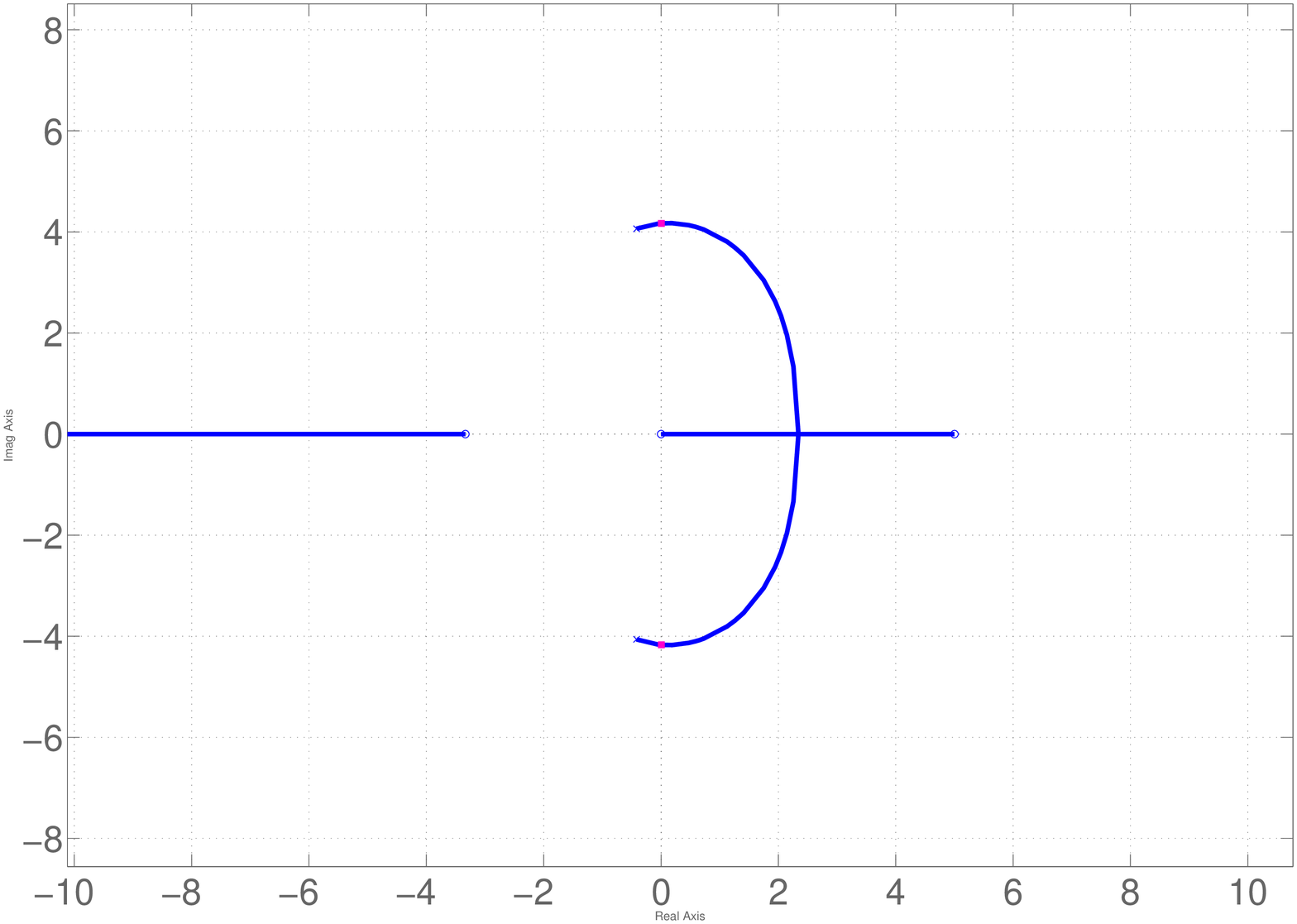}}} &
        \subfigure[$b\geq0$]
        {\resizebox{0.4\textwidth}{0.3\textheight}{
          \includegraphics{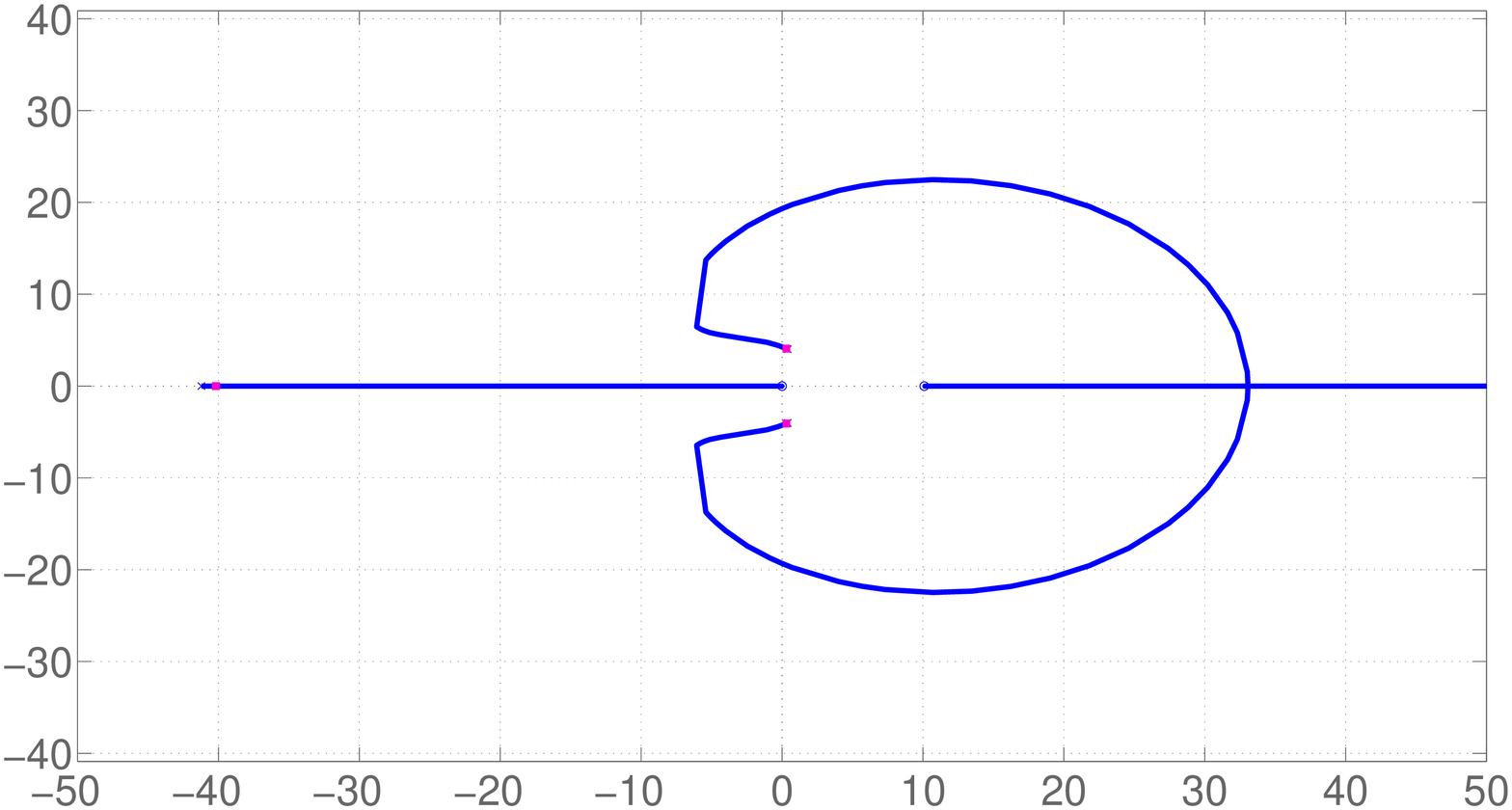}}}
         \end{tabular}
      \caption{Root-loci using Eq~\eqref{s2eq30} and  \eqref{s2eq31}}
       \label{f4}
     \end{figure}

\begin{figure}[h]
     \centering
\hspace{-5ex}
    \begin{tabular}{cc}
        \subfigure[$k\geq0$]
        {\resizebox{0.4\textwidth}{0.3\textheight}{
	      \includegraphics{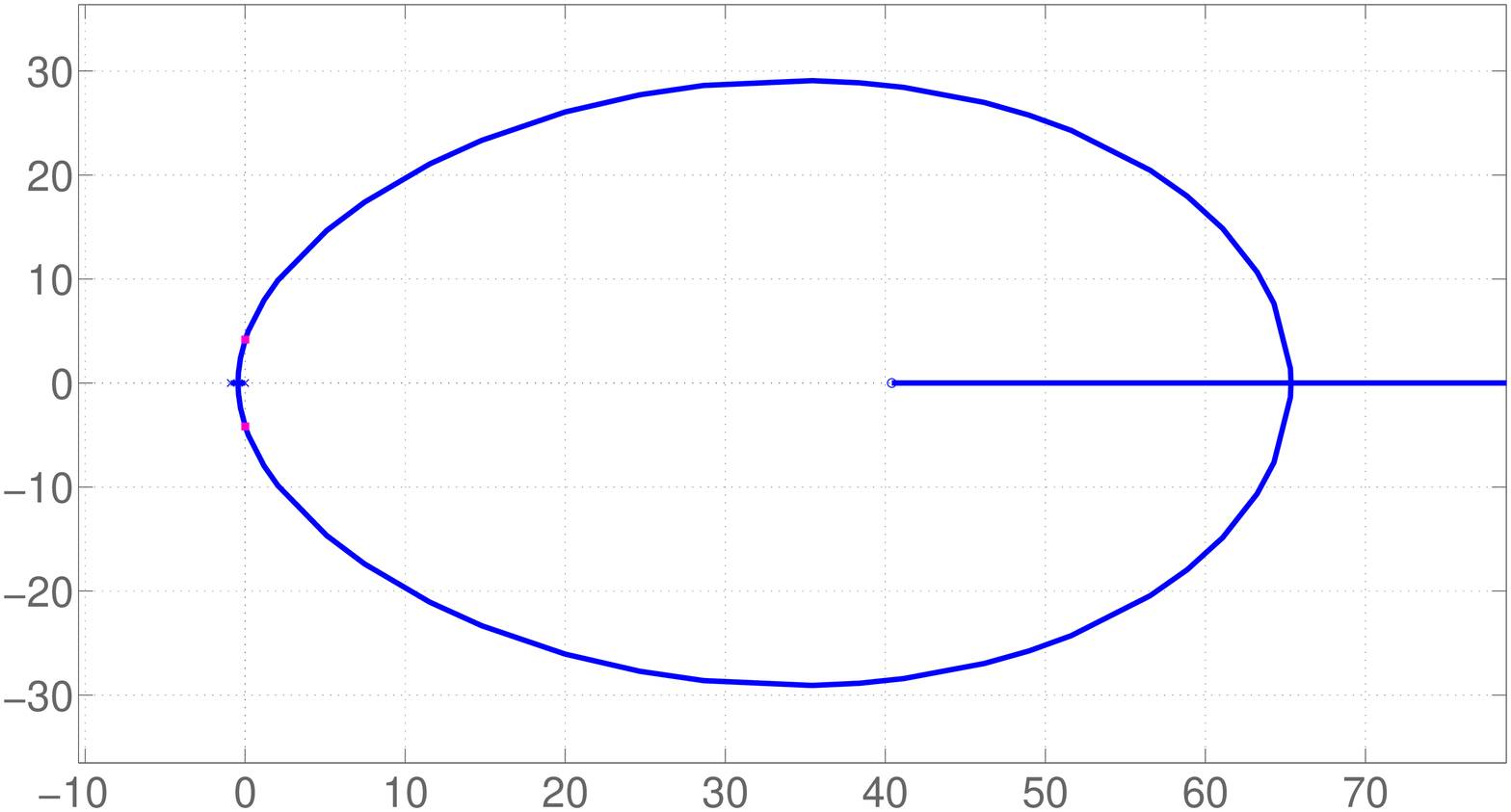}}} &
        \subfigure[$m\geq0$]
        {\resizebox{0.4\textwidth}{0.3\textheight}{
	      \includegraphics{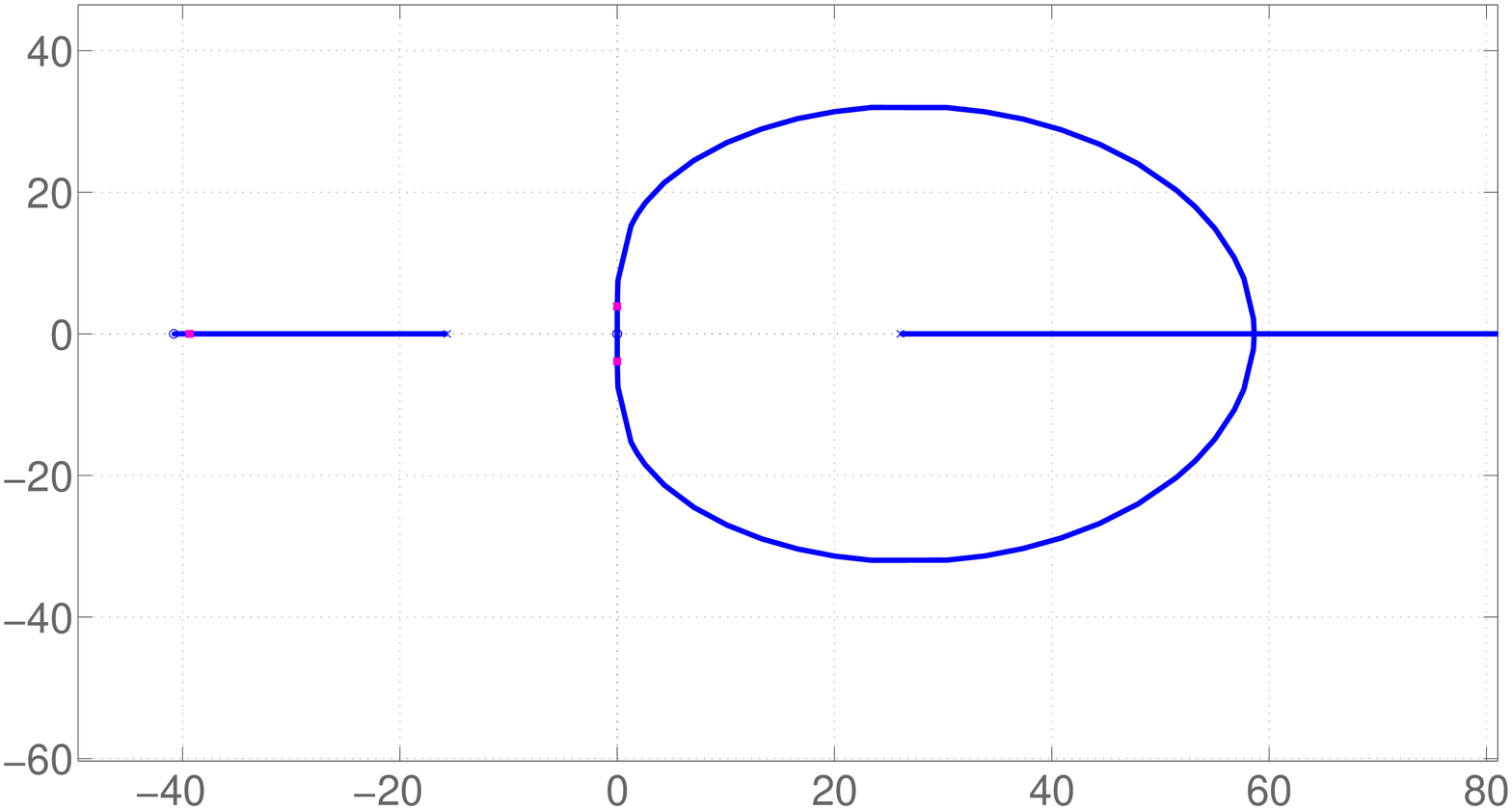}}} \\
    \end{tabular}
       \caption{Root-loci using Eq~\eqref{s2eq32} and  \eqref{s2eq33} }
       \label{f4b}
     \end{figure}

The above analysis and graphical results can be advantageously used, but lack the flexibility of classical tools like the Root-Locus. As shown next via an example, a simple first-order Pade approximation for the delay transfer function provides very good agreements with the general pole location method. It has the advantage of easily investigating stability while modifying independently the system parameters $h$, $k$, $b$, and $m$. Using the following expression for the delay:
\begin{align}
& e^{-sh} = \frac{2+sh}{2-sh} \
\end{align}
in the characteristic polynomial of the closed loop system, $\chi_h(s)$, in Eq.~\eqref{eq:cheq}, yields the following expanded expression
\begin{align}
 & \chi_h(s) = m h s^3 + 2 (m- b h) s^2 +(2 b - k h ) s + 2 k \
\end{align}
The Evans forms of $\chi_h(s)$ with respect to $h$, $b$, $k$, and $m$, respectively, as free parameters, are easily developed as follows:
\begin{align}
\label{s2eq30}
& \chi_h(s) = (m s^2 + b s + k ) + h \;[\frac{1}{2} s (m s^2 - 2 b s +k)] \\
\label{s2eq31}
 & \chi_h(s) = (m h s^3 + 2 m s^2 + - k h s + 2 k)+ b \; [2 s (-2 h s + 1)] \\
 \label{s2eq32}
& \chi_h(s) = [m h s^3 + 2 (m - b h ) s^2 + 2 b s] + k\;(-h s + 2) \\
 \label{s2eq33}
& \chi_h(s) = (- 2 b h s^2 + 2 b s - k h s + 2 k) +m\;(h s^3 + 2 s^2) \
\end{align}
Henceforth applying the Routh-Hurwitz rules for necessary and sufficient conditions is straightforward. This yields for instance the following inequalities:
\begin{align}
\label{s2eq34}
& h^2 - 2 \left(\frac{b}{k} + \frac{2m}{b}\right) h + \frac{4m}{k} > 0 \\
\label{s2eq35}
& m - b h > 0 \\
\label{s2eq36}
& 2 b - k h >0 \
\end{align}
where Eq.~\eqref{s2eq34} is typically the active constraint. As a numerical example, consider the following values: $m=60$ kg, $b=50$ Ns/m, $k=1000$ N/m. For reference, according to Eqs.~\eqref{s2eq15} and \eqref{s2eq14a}, the critical delay is $49.3$ ms and the associated crossing frequency is 4.19 rad/s. Using these numerical values, the root-loci are drawn and pictured in Fig.~\ref{f4}(a) for $h\geq0$, Fig.~\ref{f4}(b) for $b\geq0$, Fig.~\ref{f4b}(a) for $k\geq0$, and Fig.~\ref{f4b}(b) for $m\geq0$.
     Figure~\ref{f4}(a) shows that the system becomes unstable when the delay grows beyond a critical value. The later can be computed from the following expression, which directly stems from Eq.~\eqref{s2eq34}:
\begin{align}
 \label{s2eq36a}
 & h_c = \left(\frac{b}{k} + \frac{2m}{b}\right) - \sqrt{\left(\frac{b}{k} + \frac{2m}{b}\right)^2 - \frac{4 m}{k}} \
\end{align}
 \FloatBarrier
With the given mass, stiffness, and damping, the value is 49.5 ms, which is in good approximation with 49.3 ms obtained via the pole location method. The root-locus method provides a crossing frequency of 4.15 rad/s in good agreement also. The impact of varying the damping parameter $b$ is shown on Fig.~\ref{f4}(b), and supports the findings in Fig.~\ref{f2}(a). The Root-locus enters the OLHP for a minimal value of $b$, here 50 Ns/m, before exiting it for a higher value. Figure~\ref{f4b}(a) depicts how the poles stay in the OLHP until $k$ reaches a critical value, here 1000 N/m, which agrees with the plot in Fig.~\ref{f2}(b). In Fig.~\ref{f4b}(b), the Root-locus never enters the OLHP, for any value of the mass when the damping is 50Ns/m and time delay assumed 50ms. When $m$ takes on high values, the poles asymptotically approach the origin. This concurs with the asymptote of the plot in Fig.~\ref{f2}(c) which appears for a delay of 50 ms, showing that the system remains unstable as the mass increases although the distance to the stable domain diminishes.
    \FloatBarrier

\section{Experimental results}
This section presents results from 1D tests using the EPOS system. These tests serve as a proof-of-concept of the proposed hybrid simulator, albeit in 1D, showing its ability in achieving stability by introducing a virtual damping. They also provide a validation of the 1D stability analysis. In addition, a series of tests were performed that emulate results from experiments carried on an air-bearing testbed at the Space Laboratory of Tohoku University in the Fall of 2012. This paper provides an account of one of these tests.

\subsection{Experimental setup}
\begin{figure}[h]
	\centering
%{\resizebox{0.7\textwidth}{0.45\textheight}{\includegraphics{fig/3Dsetup.eps}}}
	\includegraphics[width=0.7\textwidth]{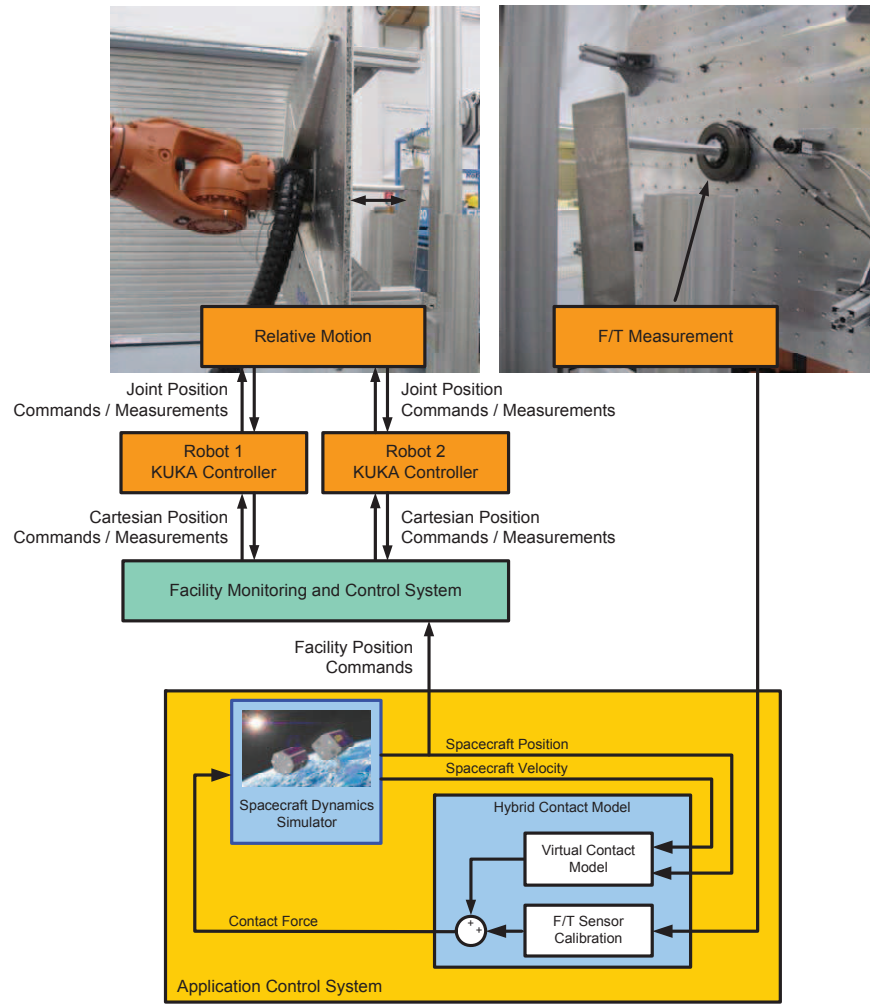}
	\caption{1D Experimental setup of the DLR European Proximity Operations System (EPOS)}
	\label{fig:HILsetup}
\end{figure}

Figure~\ref{fig:HILsetup} conceptually pictures the HIL EPOS test setup. The hardware module consists of the chaser robot, its tracking controller, a target element, the force sensor, and a compliance device that has the function of a docking interface. The force sensor is attached to a tool plate that is fixed at the chaser's end-effector. The docking interface is also attached to the tool plate which has a stiff shaft (probe) with a pin-like head. The probe makes thus contact with the target element in a pin-pointed manner. The target element is a metal sheet at rest with respect to the room's referential. This was done for the sake of simplicity and does not limit the validity of the tests since they are conducted in 1D only. The software module of the HIL simulator includes the numerical simulation of the chaser and target satellites, an estimator of the current relative displacement of the target with respect to the chaser, the computation of a virtual contact force according to specified damping and stiffness coefficients, and the gravity compensation in the measured force. The latter is using a calibration procedure which will be described in a different work. The robotics tracking system operates at a frequency of 250 Hz. It provides a sub-millimeter accuracy positioning of the probe tip with a time-invariant delay of 16ms after receiving the command signal. The force sensor provides readings at a frequency of 1 kHz, with calibrated errors of order 0.25 N.

\subsection{Stability analysis validation using the coefficient of restitution}
Several series of tests were performed in order to validate the single-dimension stability analysis. In order to compare with the theoretical results each series consisted in varying $b$ from 0 to 100 Ns/m, while keeping the other parameters constant. The value of the damping and of the mass were precisely set in the software part of the simulator. The delay was dictated by the robotics (16 ms). No virtual stiffness feedback force was added to the sensed force: the only stiffness element stems from the hardware. Once calibrated to provide (approximately) a stiffness of 1000 N/m, the experimental set-up was unchanged. For verification, the measured force and displacement were processed in order to estimate the actual stiffness values for each test. Each test consisted of an initial acceleration of the chaser robot to a velocity of approximately 20 mm/s, followed by a phase of constant velocity, where the sensor calibration was improved, until contact. Data were recorded until the chaser robot reached a steady velocity after bouncing back from the target element. The simulated target mass value was several orders of magnitude heavier than that of the chaser in order to avoid motion of the target with respect to the room.

\begin{figure}[h]
\hspace{-5ex}
    \begin{tabular}{ccc}
            \subfigure[m=5 kg]
        {\resizebox{0.32\textwidth}{0.25\textheight}{
          \includegraphics{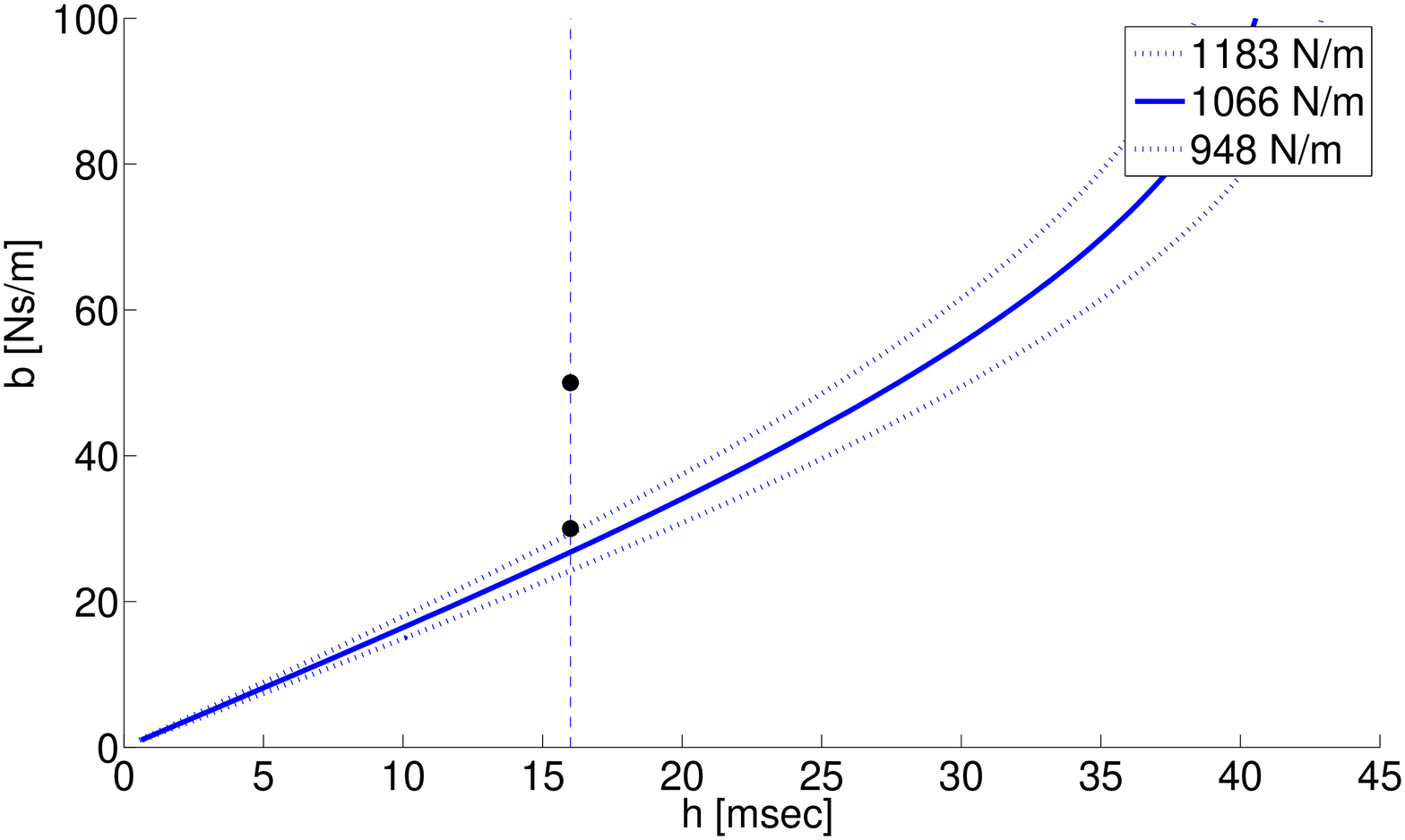}}}  &
        \subfigure[m=63.2 kg]
        {\resizebox{0.40\textwidth}{0.25\textheight}{
	      \includegraphics{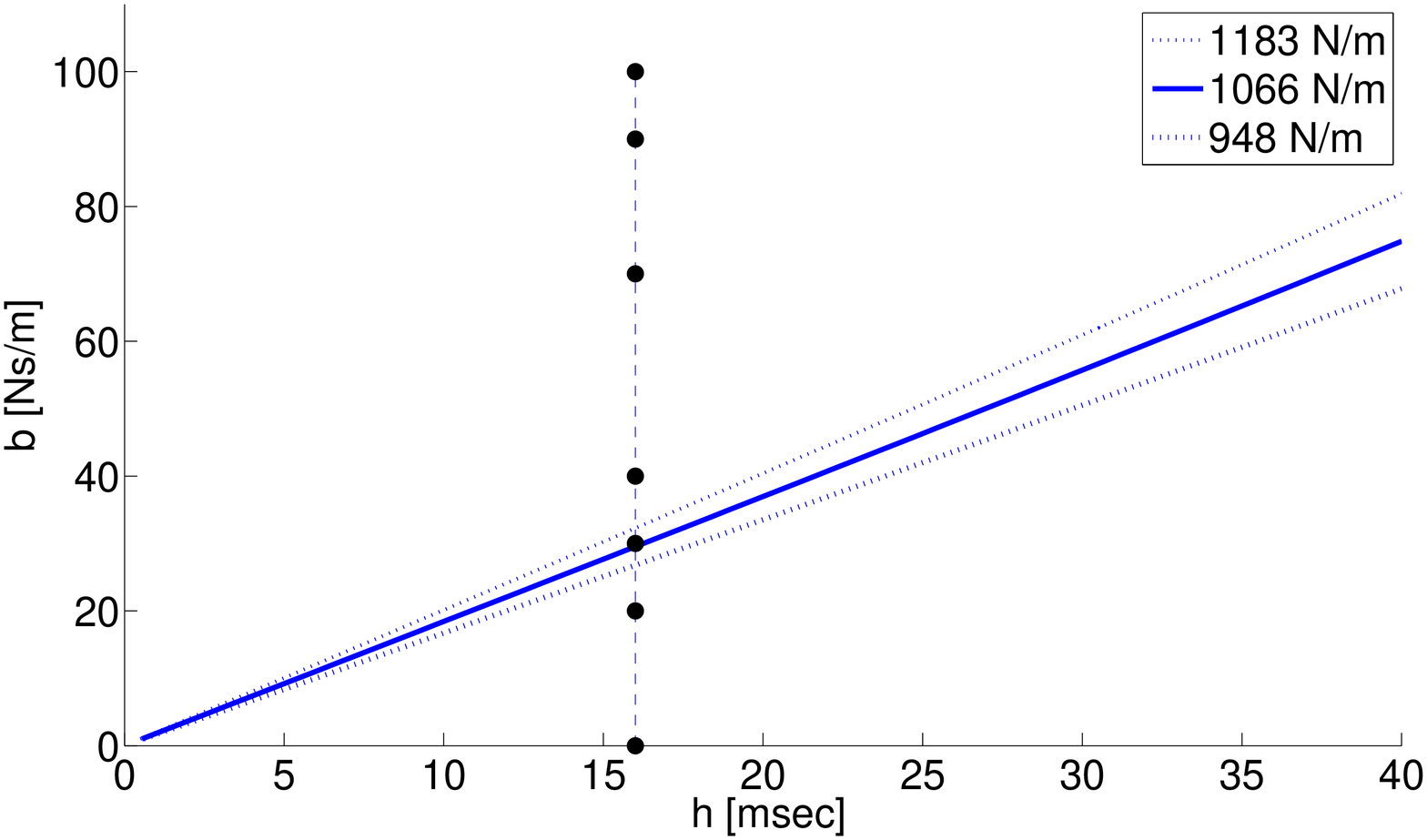}}} &
        \subfigure[m=500 kg]
        {\resizebox{0.32\textwidth}{0.25\textheight}{
	      \includegraphics{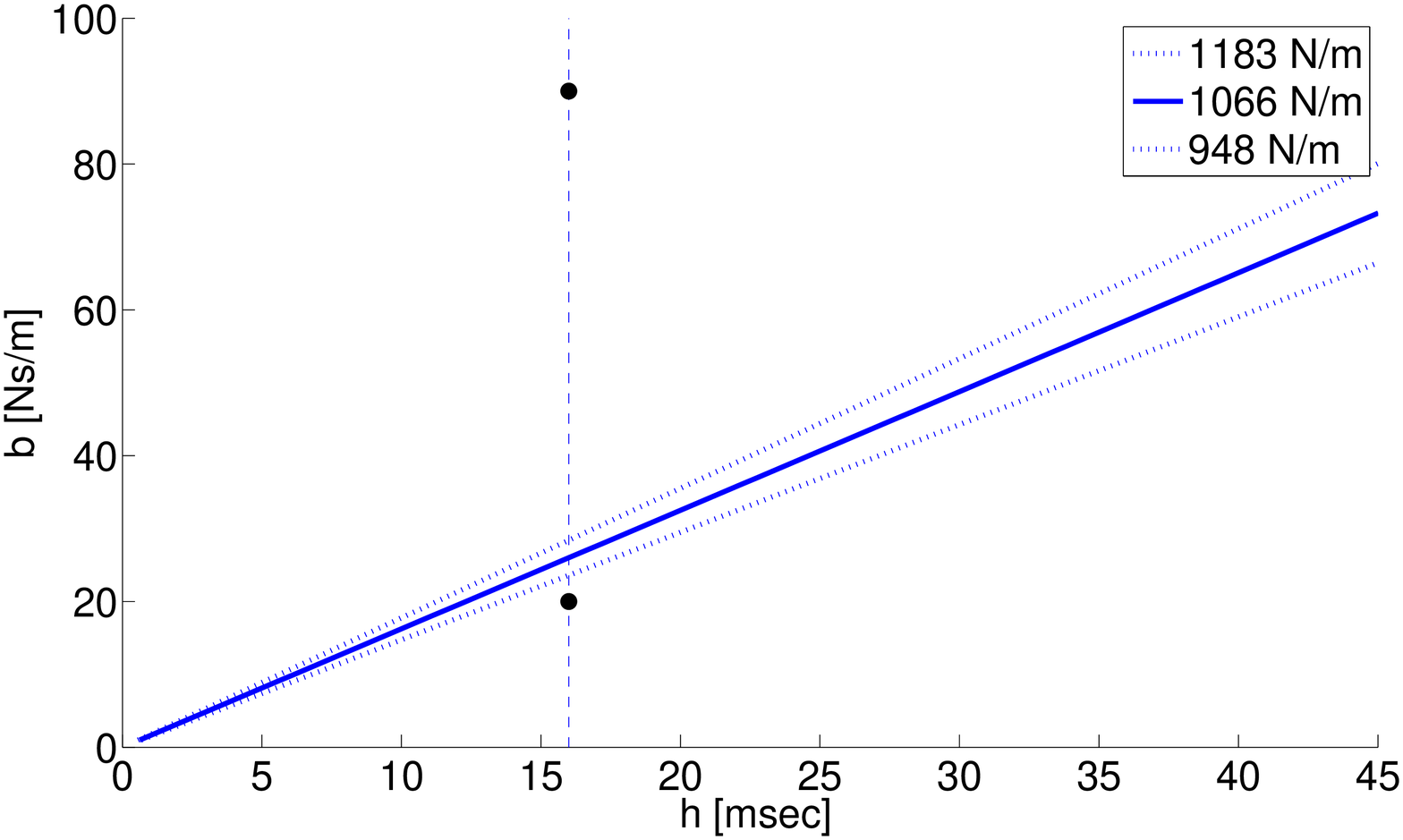}}} \\
    \end{tabular}
       \caption{Experimental validation of the stability analysis. The black dots indicate the various test points for various damping values and a 16 ms delay. The stability regions are above the critical envelopes.}
       \label{f6}
     \end{figure}

The relative velocities before and after impact were recorded and averaged over several seconds. These averages are denoted by $v^{-}$ and $v^{+}$, respectively. The experimental criteria stability considered here is the ratio between $v^{+}$ and $v^{-}$, known as coefficient of restitution~\cite{Gilardi,Nakanishi}, i.e.,
\begin{align}
 \label{s3eq00}
 & \epsilon = \frac{v^{+}}{v^{-}} \
\end{align}
The system is stable if $\epsilon < 1$, neutrally stable if $\epsilon= 1$, and unstable otherwise. Notice that $\epsilon$ is identical to the overshoot, for delay-free second order systems, i.e., $\epsilon = e^{\frac{-\pi \zeta}{\sqrt{1-\zeta^2}}}$, where $\zeta$ is the non-dimensional damping coefficient. Table~\ref{tab1} summarizes the results for the first series of tests, where the mass was 63.2 kg.
\begin{table}[t]
 \centering
 \caption{Tests results for varying values of the damping $b$. The mass is 63.2 kg, the delay is 16 ms, and the stiffness is around 1066 N/m.}
 \label{tab1}
 \vspace{10mm}
\begin{tabular}{@{\extracolsep{10mm}}ccccccc}
 \hline\hline\\
$b$     & $v^{-}$    & $v^{+}$    & $\epsilon$ & $\kes$  & $\toes$ & $\tau$ \\
$[Ns/m]$  &  [mm/s] & [mm/s]    &            & [N/m]   &  [ms] & [ms] \\\\
\hline\\
$0$   & $21.0$  & $26.5$  & $1.26$  & $1280$  & $698$ & $700$ \\
$0$   & $21.0$  & $26.5$  & $1.26$  & $1100$  & $753$ & $756$ \\
$0$   & $21.0$  & $26.0$  & $1.24$  & $955$  & $878$ & $810$ \\
$0$   & $21.0$  & $23.4$  & $1.11$  & $977$  & $799$ & $800$ \\
$20$   & $18.5$  & $20.0$  & $1.08$  & $1020$  & $782$ & $780$ \\
$\bf30$   & $18.0$  & $18.0$  & $\bf1.00$  & $1150$  & $736$ & $735$ \\
$\bf30$   & $18.0$  & $18.0$  & $\bf1.00$  & $975$  & $800$ & $800$ \\
$\bf40$   & $17.5$  & $17.0$  & $\bf0.97$  & $1270$  & $700$ & $700$ \\
$\bf40$   & $17.0$  & $16.5$  & $\bf0.97$  & $1140$  & $739$ & $740$ \\
$\bf40$   & $17.5$  & $16.7$  & $\bf0.98$  & $1080$  & $760$ & $760$ \\
$\bf40$   & $17.5$  & $17.0$  & $\bf0.97$  & $1050$  & $771$ & $772$ \\
$70$   & $18.0$  & $15.0$  & $0.83$  & $1095$  & $755$ & $756$ \\
$70$   & $20.0$  & $17.0$  & $0.85$  & $1030$  & $778$ & $780$ \\
$90$   & $20.0$  & $15.0$  & $0.75$  & $1040$  & $774$ & $776$ \\
$100$   & $21.0$  & $15.0$  & $0.71$  & $822$  & $871$ & N/A \\\\
\hline\hline
\end{tabular}
\end{table}
When the damping coefficient is zero, the system is, as expected, unstable, as evidenced by the fact that $\epsilon$ is greater than one. Incremental increases of the value of $b$, up to 30-40 Ns/m in the software, produce stronger damping forces, which results in a decrease of $\epsilon$ down to unity. The test was repeated several times (in bold in Table~\ref{tab1}), consistently yielding values of $\epsilon$ between between $0.97$ and $1$. The system has thus become neutrally stable. Further increasing the coefficient $b$ to 70, 90, and 100 Ns/m, results in a consistent reduction of $\epsilon$. Comparison with the analytical results of the previous section is done as follows. For each test, an empirical value of the stiffness, $\kes$, was determined (assuming a linear spring model). Using the values for $\kes$ as given in Table~\ref{tab1}, the sample average $\bar{k}$ and standard deviation $\sigma_k$ are computed, yielding 1066 N/m and 118 N/m, respectively. This is consistent with the levels of accuracy of 0.25 N and 1 mm in the force and position knowledge, respectively. This shows that the experiment was well calibrated. Using the values for the mass (63.2 kg), the delay (16 ms), and the three stiffness values $\bar{k}$, $\bar{k}\pm\sigma_{k}$, three plots of the curve $b$ vs $h$ are depicted in Fig.~\ref{f6}(b). The black points represent the experimental data. It appears that the points corresponding to neutral stability (i.e. $b$ at 30 and 40 Ns/m) lie inside or are close to the critical envelope (in dotted lines). There is thus a good agreement between the tests and the analysis. Additional comparison is done based on the predicted and observed contact durations, denoted by $\toes$ and $\tau$, respectively. The value of $\toes$ is computed as
\begin{align}
\label{s3eq02}
 & \toes= \frac{\pi}{\widehat{\omega}_c} \
\end{align}
where $\widehat{\omega}_c$ is an estimated crossing frequency as given from Eq.~\eqref{s2eq14a} with $\kes$ instead of $k$. Considering $k$ as the only uncertain parameter in the expression for $\omega_c$, and given the values of the mass and the damping from Table~\ref{tab1}, the estimated duration $\toes$ has got an accuracy of 1 ms. The values of the contact durations are direct observations from experimental data, here the force profiles. The accuracy is here limited by the HIL controller sample time of 4 ms. Table~\ref{tab1} shows that there is an excellent agreement between the predicted and the actual contact durations when $\epsilon$ is equal or close to $1$, around marginal stability. Other tests were performed with higher and lower values of the mass. Table~\ref{tab2} summarizes the results for two cases: 500 kg and 5 kg.
\begin{table}[t]
 \centering
 \caption{Tests results for two mass and damping values. The delay is 16 ms and the stiffness is around 1066 N/m.}
 \label{tab2}
 \vspace{10mm}
\begin{tabular}{@{\extracolsep{10mm}}rcccc}
 \hline\hline
$m$ & $b$     & $v^{-}$    & $v^{+}$    & $\epsilon$  \\
$[kg]$ & [Ns/m]  &  [mm/sec] & [mm/sec]  & \\
\hline\\
$500$ & $20$   & $20.0$  & $21.0$  & $1.05$ \\
  &     $90$   & $20.0$  & $18.0$  & $0.90$ \\
  \hline\\
$5$ &  $30$   & $20.0$  & $20.5$  & $1.03$ \\
  &    $50$   & $20.0$  & $18.7$  & $0.94$ \\
\hline\hline
\end{tabular}
\end{table}
Figures~\ref{f6}(a)(c) depict the experimental points for $m=500$ kg and $m=5$ kg, respectively. It clearly validates the stability/instability prediction.
\FloatBarrier
\subsection{Stability analysis validation using the observed energy}
\par
\begin{figure}[h]
\centering
    \begin{tabular}{cc}
        \subfigure[$b=0$ Ns/m]
        {\resizebox{0.43\textwidth}{0.33\textheight}{
	      \includegraphics{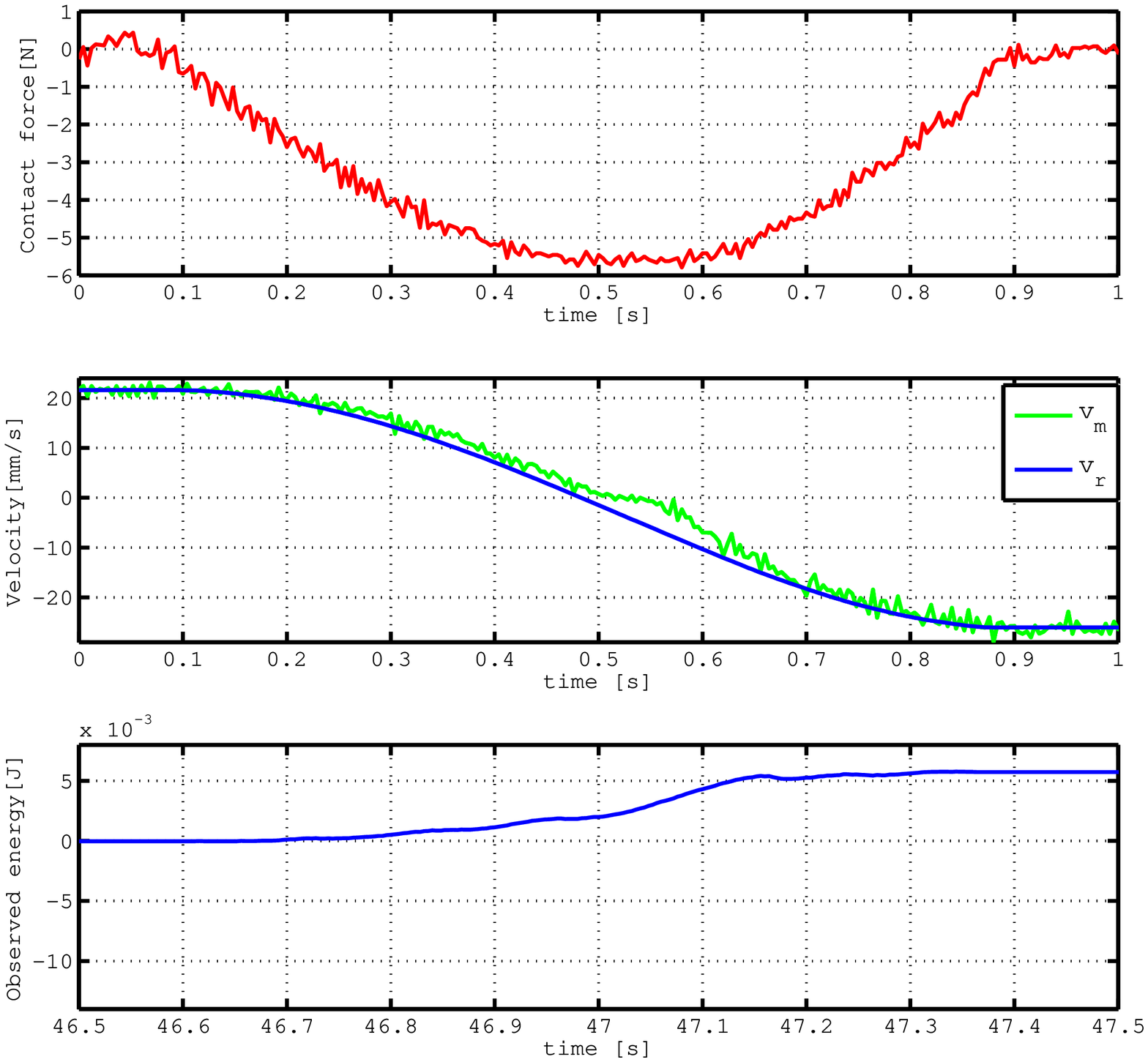}}} &
        \subfigure[$b=20$ Ns/m]
        {\resizebox{0.43\textwidth}{0.33\textheight}{
	      \includegraphics{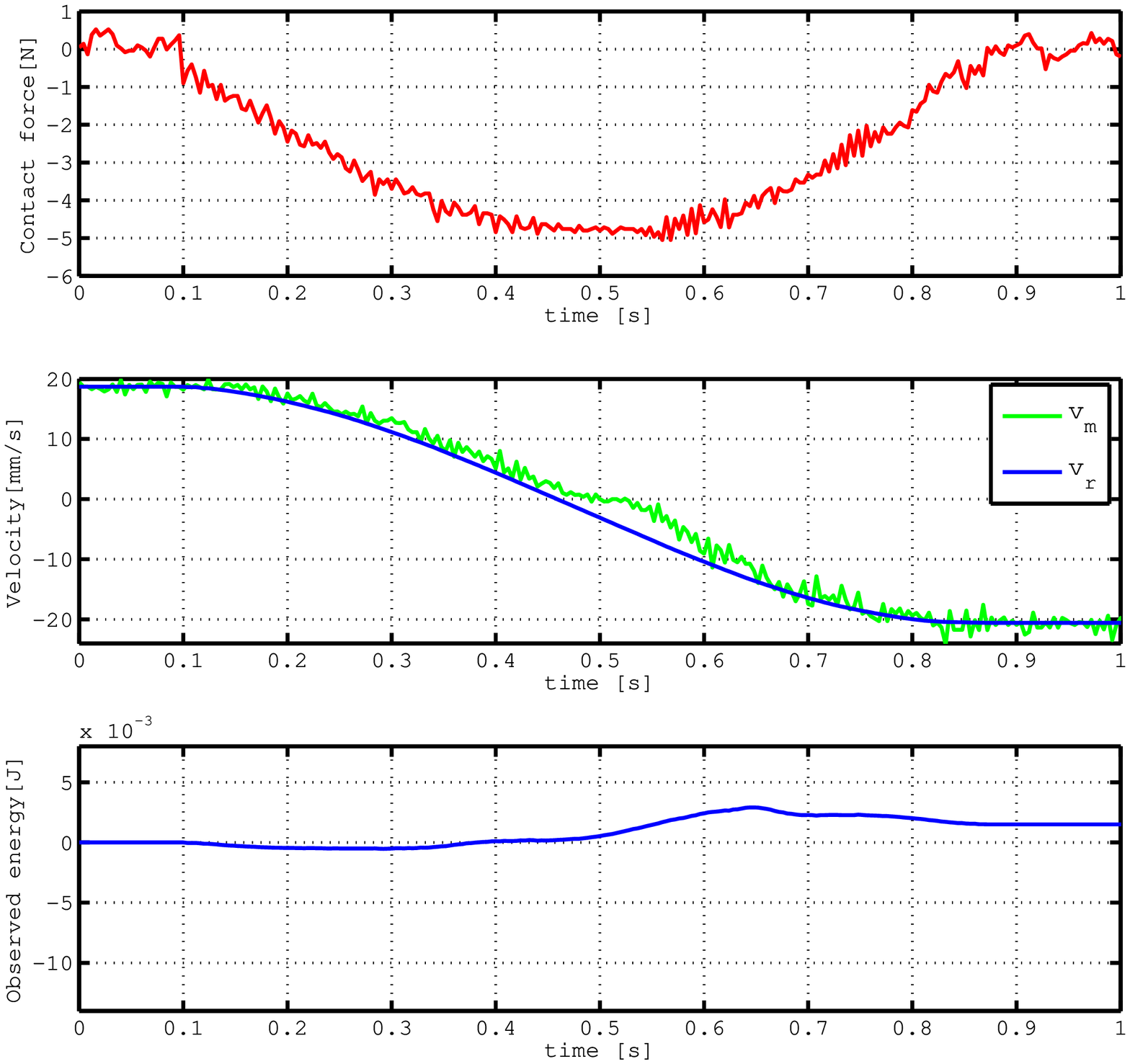}}} \\
    \end{tabular}
       \caption{The simulator system is active both for $b$ = 0 and 20 Ns/m.}
       \label{s3f1}
\end{figure}

\begin{figure}[h]
\centering
    \begin{tabular}{cc}
        \subfigure[$b=40$ Ns/m]
        {\resizebox{0.43\textwidth}{0.33\textheight}{
	      \includegraphics{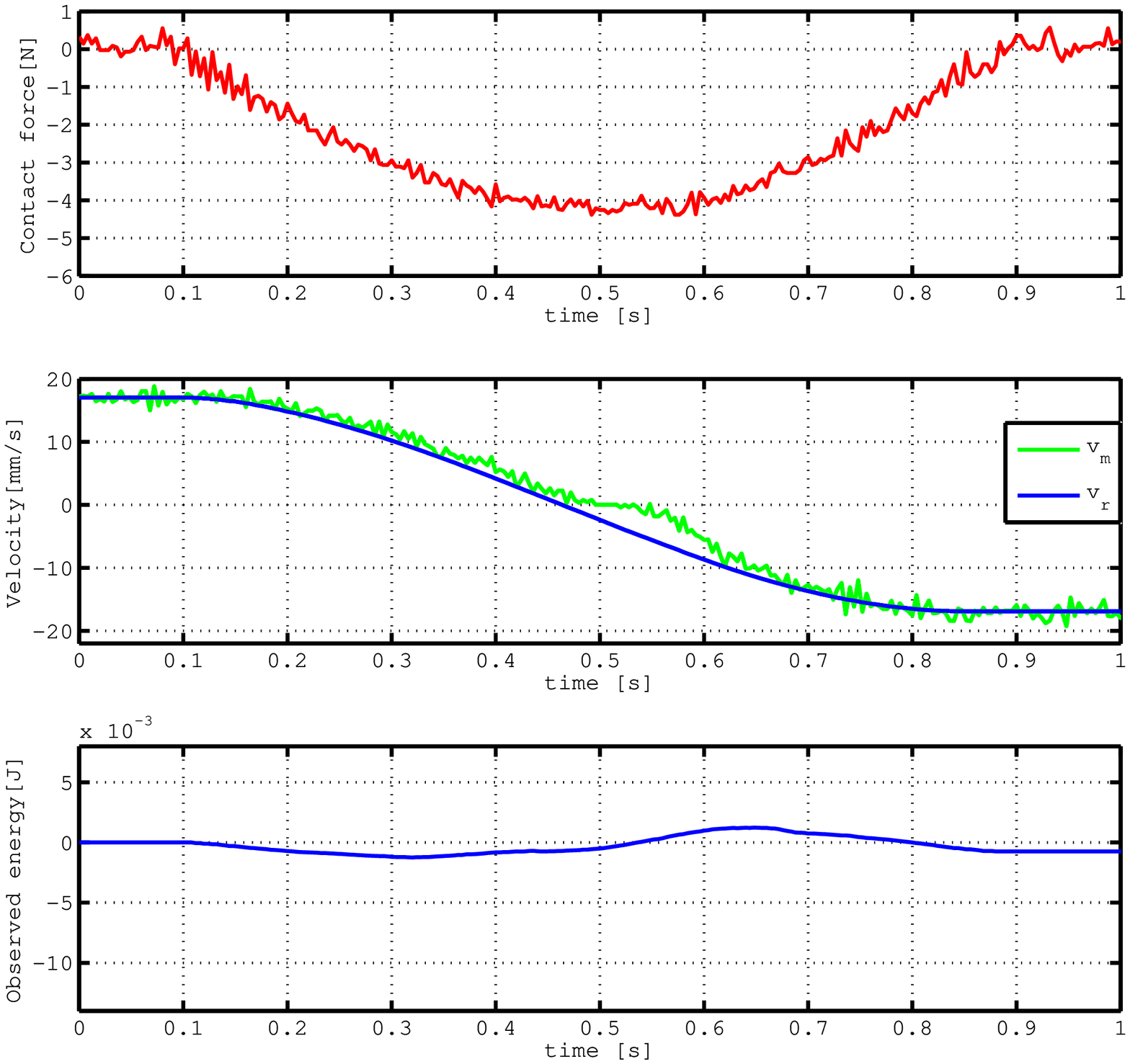}}} &
        \subfigure[$b=70$ Ns/m]
        {\resizebox{0.43\textwidth}{0.33\textheight}{
	      \includegraphics{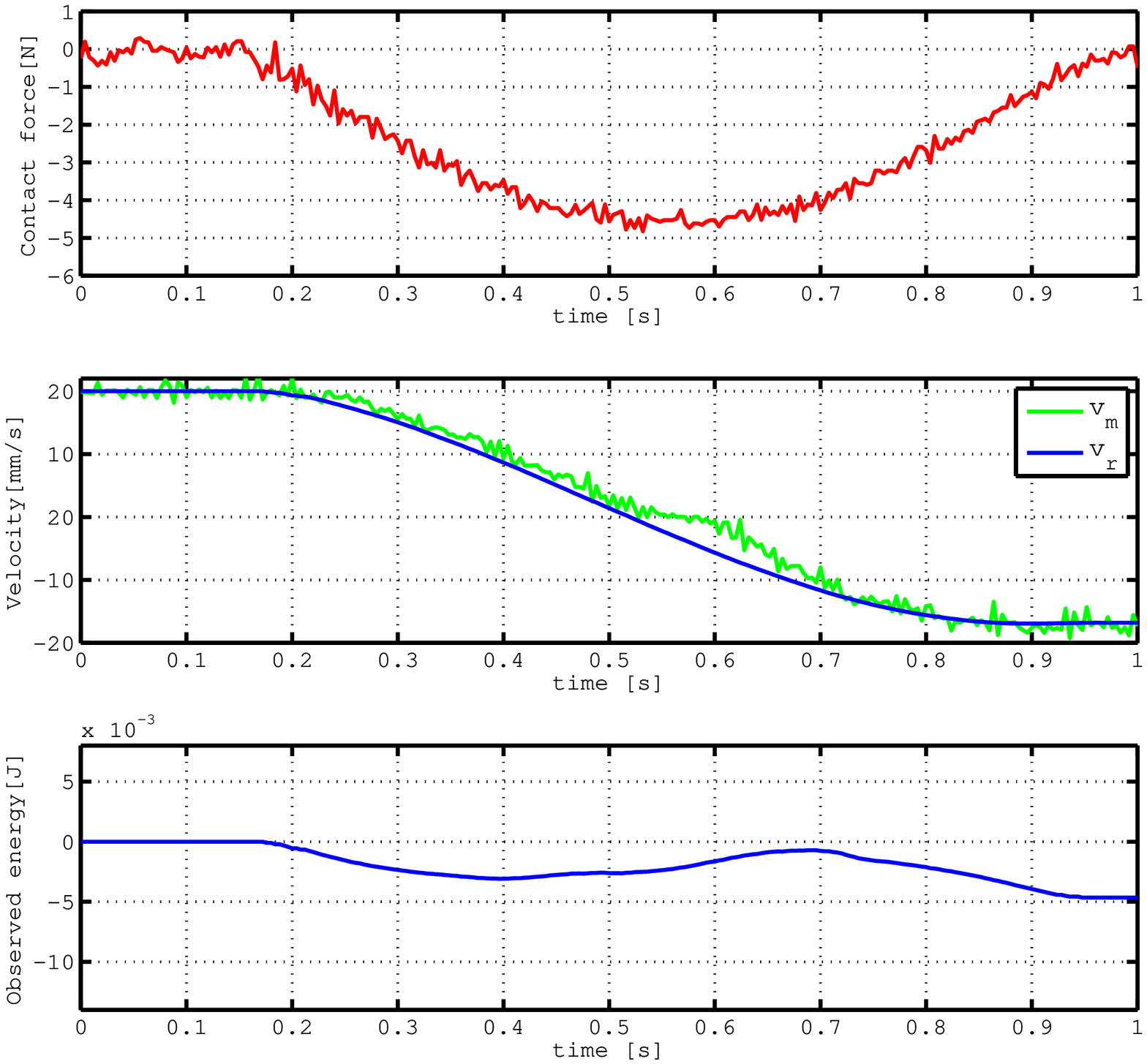}}} \\
    \end{tabular}
       \caption{The simulator system is almost neutrally stable for $b$ = 40 Ns/m. It is passive at $b$ = 70 Ns/m.}
       \label{s3f2}
\end{figure}

 The stability-related property of passivity was used in previous experimental works~\cite{988969, Rainer} because it lends itself to an empirical method where only input-output signals are required. As opposed to the pole location method it is not model-dependent and it applicable in the time domain. It also has the advantage to provide an insight on the passivity of the system \emph{during} the contact, as opposed to the method using the coefficient of restitution which determines the stability after the contact ending. The algorithm, called passivity observer~\cite{Rainer}, consists in computing the following performance measure, a.k.a the added energy:
\begin{equation}
\label{eq:obstwo1}
\Delta \EE = \Delta t \sum_{i=1}^N [f(i) v_m(i) - f_{in}(i) v_r(i)]
\end{equation}
where $f(i)$, $f_{in}(i)$, $v_m(i)$ and $v_r(i)$ are the sampled signals of the measured force, the force input to the HIL system, the measured velocity, and the command velocity, respectively, $\Delta t$ is the sample time (4 ms), and $N=1,2,\ldots$ until the end of the contact. The HIL simulator is passive if $\Delta\EE < 0$, lossless if $\Delta\EE = 0$, and active if $\Delta\EE >0$ for any $N$. Notice that, looking at the whole contact duration, this approach is equivalent to the $\epsilon$-approach. Indeed, considering the change of energy of a point mass before and after an impact yields $\Delta \EE = \EE^{-} \,(\epsilon - 1) $ where $\epsilon$ is defined in Eq.~\eqref{s3eq00}, showing that the sign of $\Delta \EE$ and of $(\epsilon-1)$ are identical. Four cases, taken from Table~\ref{tab1}, were analyzed, corresponding to damping 0, 20, 40 and 70 Ns/m.   Figures~\eqref{s3f1}(a)-(b) show, for each case, the time variations during contact of the sensed force (upper plots, red lines), of the required and measured relative velocities, $v_r$ and $v_m$ (middle plots, black and green lines), and of the observed energy, $\Delta\EE$ (bottom plots, blue lines). In the absence of virtual damping ($b=0$), it can be seen from Fig.~\ref{s3f1}(a) that $\DEE$ only increases with time, accumulating about 6 mJ during the contact. The maximum rate of increase in $\DEE$ occurs at half the contact duration, when the chaser robot stops and inverts its motion direction. It can be seen from the velocities plots that the maximum discrepancy between $v_m$ and $v_r$ happens at that time. This ``deadzone'' lasts around 100 ms and stems from a combination of robotic controller delay, non-linearities, and inertia. The increase in $\DEE$ and its settling at a positive value is an evidence for the HIL system to be active. This concurs with the fact that the final velocity (26.5 mm/s) is greater than the initial velocity (21 mm/s), and $\epsilon$ is henceforth greater than one. Figure~\ref{s3f1}(b) depicts the case of a lightly damped system. The energy profile shows a small dip before increasing and settling around 1.5 mJ. The system is thus slightly active over the contact duration. In Fig.~\ref{s3f2}(a), which depicts the case $b=40$ Ns/m, the added energy over the contact duration is almost zero. The system is here approximately neutrally stable, as supported by the value $0.98$ for $\epsilon$. Interestingly, the plot of $\DEE$ shows an oscillation with negative values in the first contact half followed by positive values in the second half. The results for the case $b=70$ Ns/m are depicted in Fig.~\ref{s3f2}(b). The energy dip is here lower than in previous cases, such that the following pick does not reach the positive values, and $\DEE$ settles at -5 mJ. The system is thus passive and energy is being dissipated, in good agreement with the value of $\epsilon$ (0.85).
\FloatBarrier

\subsection{Emulation of an air-floating table test}
\begin{figure}[h]
    \centering
	   \includegraphics[width=0.5\textwidth]{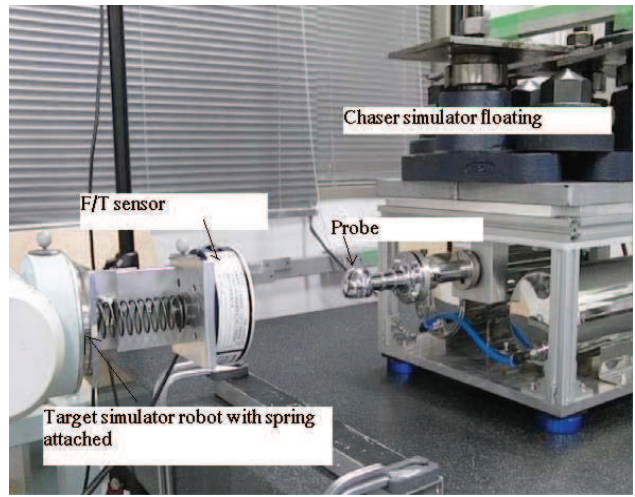}
       \caption{Air-floating test bed setup at Tohoku University}
       \label{fig:airfloating}
     \end{figure}

The HIL docking simulator concept was validated experimentally by emulating the results from the air-floating table test conducted at the Space Robotics Laboratory at the Tohoku University. Figure~\ref{fig:airfloating} depicts the experimental setup.

 The chaser body is equipped with four air pads. Each pad is connected to on-board air-tanks enabling the pads to float via static air-pressure. A spring with a stiffness of 1000 N/m  is attached at the end-effector of the target body which has a viscous friction with an equivalent damping of 47.5 Ns/m. The damping value is estimated using the experiment data. The force sensor is attached at the head of the spring. The motion measurement system used infrared (IR) cameras tracking the motion of IR reflecting balls fixed to the floating bodies. The force and position accuracies were of order 0.1 N and 1 mm, respectively.  The target body was actuated in order to reach an absolute velocity of 20 mm/sec and then to float with constant speed towards the chaser body, initially at rest. During and after the impact the target speed was maintained at 20 mm/sec while the chaser body was set in motion due to the impact. The mass of the chaser body was 63.2kg, which is the maximum allowed on this testbed. Much care was given to correctly align the chaser center of mass with the axis of the contact force in order to avoid torques. The relative velocity changed from 20 mm/sec before impact to -15mm/sec after impact. This corresponds to a value of 0.75 for $\epsilon$. These conditions were successfully emulated using the EPOS testbed. For that purpose, the virtual damping in the HIL simulator was set at 90 Ns/m. Compared to the damping value at the air-floating test bed, the HIL simulator required a 52.5 Ns/m higher value because the delay added energy, which needed to be dissipated.  As a result, the relative velocities, and $\epsilon$, were identical to those of the air-floating test. Further, the maximum force magnitudes were similar (5 N and 4.6 N). Table~\ref{tab3} summarizes the comparison between both tests.

\begin{figure}[h]
\centering
    \begin{tabular}{cc}
        \subfigure[HIL EPOS test]
        {\resizebox{0.45\textwidth}{0.35\textheight}{
	      \includegraphics{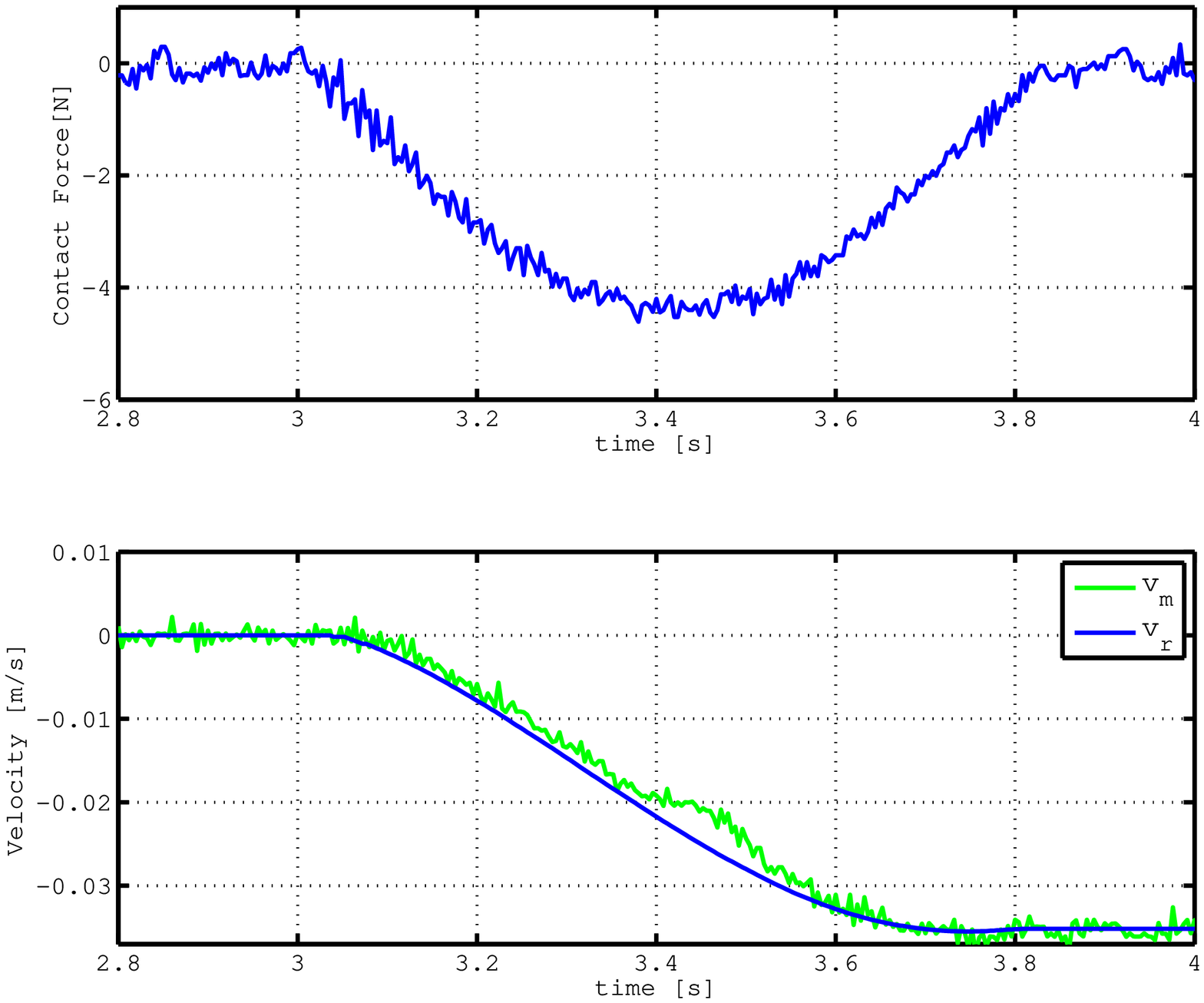}}} &
        \subfigure[Air-floating table test]
        {\resizebox{0.45\textwidth}{0.35\textheight}{
	      \includegraphics{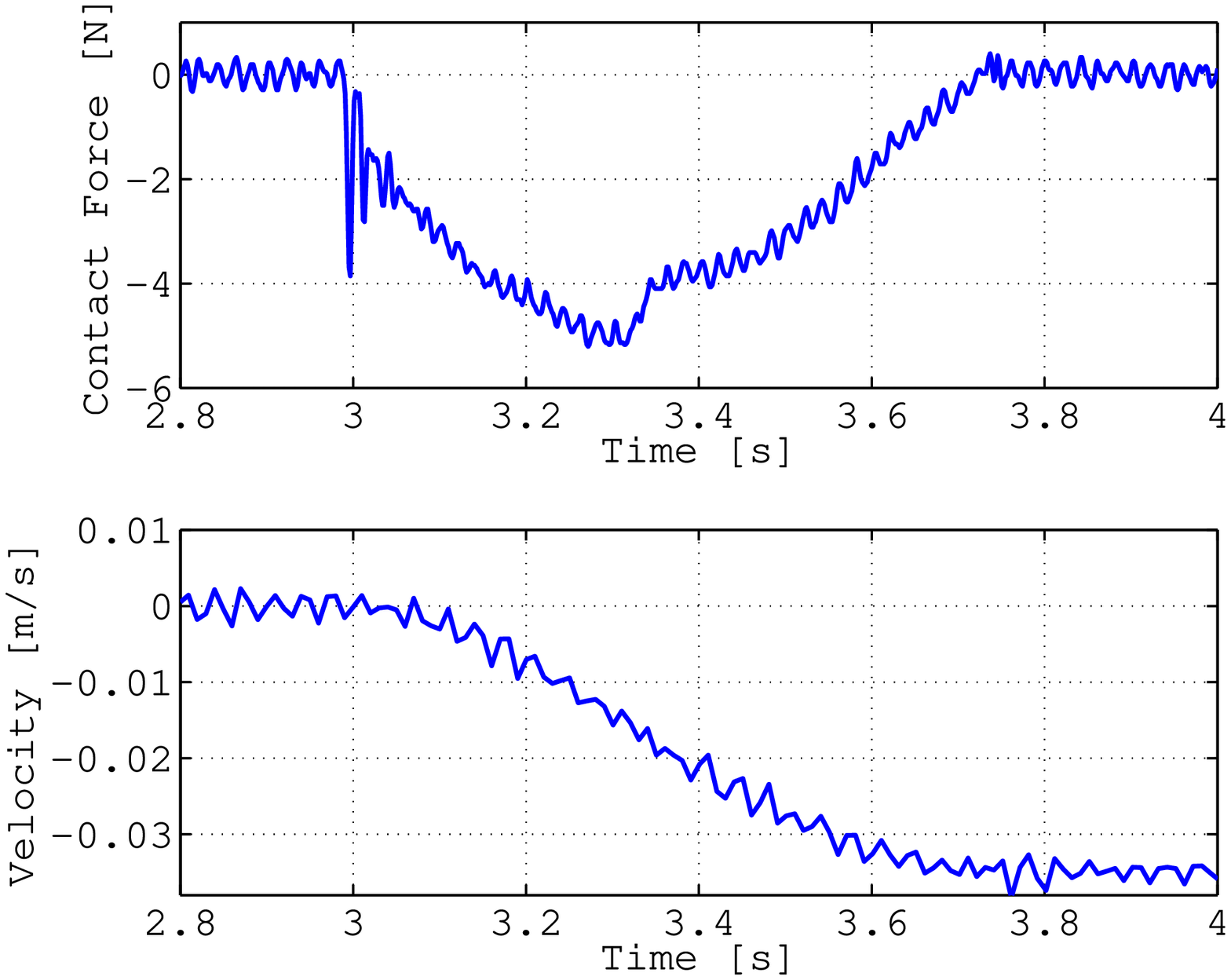}}} \\
    \end{tabular}
       \caption{Emulation of the air-floating test using HIL EPOS system. Contact force and relative velocity profiles are similar. The velocity of HIL EPOS test is shifted from 0.02m/s to zero for visualization purpose. Both experiments showed a 35mm/s final relative velocity.}
       \label{s3f5}
\end{figure}

%%%%
%%
Figure~\ref{s3f5}(a) and (b) show the time variations of the measured forces and velocities for the EPOS test and the air-floating table test, respectively. There is overall a very good similarity in the profiles of the forces and velocities. Some dissimilarities appear. The ``deadzone'' in the velocity profile at half the contact duration is characteristics of the robotics testbed. The force profile in Fig.~\ref{s3f5}(b) shows a jump at half the contact due a dry friction phenomenon experienced by the compressed spring. Also, the early phase of the contact depicts high frequencies force variations: they stem from the high stiffness encountered at contact, and recorded by the force sensor since it is attached between the spring and the contact point.

\begin{table}[h]
 \centering
 \caption{Emulation of the air-floating table test using the HIL EPOS}
 \label{tab3}
 \vspace{10mm}
\begin{tabular}{@{\extracolsep{5mm}}rccccccc}
 \hline\hline
              & $m$   &   $k$   &   $b$     &   $h$    & $\epsilon$  &$f_{max}$  \\
              & [kg] &   [N/m]   &  [Ns/m]  &  [ms]    &            & [N]       \\
\hline\\
Air-floating & $63.2$   & $1000\;$   & $47.5^{\ast}$     & N/A      & $0.75$      &   $5.0$    \\
HIL EPOS     & $63.2$  & $1030^{\ast}$   & $90$     & $16$    & $0.75$      &   $4.6$  \\
\hline\hline
 $\ast$ \small{estimated from experiment}  & & & & & &
\end{tabular}
\end{table}%%

\FloatBarrier

\section{Conclusion} \par
This work presented a numerical and experimental investigation of the DLR robotics-based hardware-in-the-loop simulator EPOS for single-dimensional contact. For stability analysis, the robotics tracking system was represented as a pure delay and the force feedback as a linear spring-dashpot device. The pole location method and a first order Pade approximation were applied providing complementary analytical and graphical tools for this analysis. The simulator was operated using a hybrid method for contact dynamics emulation. Extensive tests were performed showing good agreement between analysis and tests. The ability to modify parameters in the software, like the delay, the mass, the damping, or the stiffness, without changing the hardware elements, demonstrates the powerful flexibility of the proposed hybrid force feedback concept. The successful emulation of the experiment performed at Tohoku University on an air-floating table was achieved. The usefulness of data-driven stability indexes, such at the coefficient $\epsilon$ or the added energy $\DEE$, were validated via experiments.

\begin{figure}[h]
    \centering
	   \includegraphics[width=0.5\textwidth]{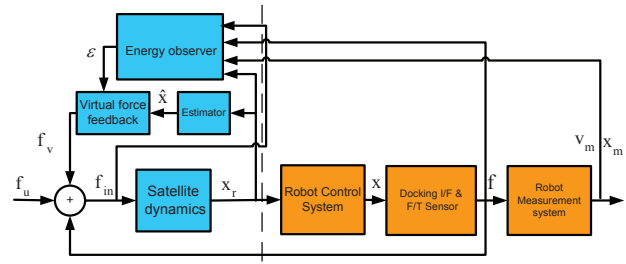}
       \caption{Block diagram of a damping adaptation system using the observed energy}
       \label{fig:conceptpass}
     \end{figure}

There is a good agreement between the $\epsilon$-based approach and the $\EE$-based approach. These approaches present the advantage of only relying on input-output data, and do not depend on the model. In addition, the $\EE$-approach can shed light on the systems passivity property during contact, not only after contact. Hence, properly monitored in real time, the energy signal provides a clue on the passivity property of the HIL system during operations. As suggested in~\cite{988969, Rainer}, applying a feedback force such as to prevent the system from becoming active appears as a promising direction for future work. The concept is visualized in Fig.~\ref{fig:conceptpass}. Additional work will focus on the extension of the HIL system to 3D, on the stability analysis extension, and on the performances of 3D tests. Further directions include automated tuning of the virtual feedback force for various purposes, e.g. ensuring passivity, emulating desired force-velocity profiles, or enabling safe operations of the simulator with guaranteed stability margins. Potential challenges to address include uncertain and time-varying delay, stiffness, and damping, randomness in the force measurement, and non-linearity of the contact dynamics. %%

  \section*{\large{\textbf {Acknowledgments}}}
The authors acknowledge fruitful discussions with Mr. Roberto Lamparello of the DLR Robotics and Mechatronics Institute. The first author acknowledges many fruitful discussions with Prof. Kazuya Yoshida and Dr. Naohiro Uyama from Tohoku University of Japan and their kind support during the experiment.
\FloatBarrier

\bibliographystyle{elsarticle-num}

\bibliography{bibliography}

\end{document}

%%
%% End of file `ecrc-template.tex'. 